\documentclass{emulateapj}

\def \htwo{H$_{2}$}
\def \as{\hbox{$^{\prime\prime}$}}
\def \deg{\hbox{$^\circ$}}

\def \hi{H\,I}
\def \ha{H$\alpha$}

\defcitealias{hic13}{Paper~I}

\shorttitle{Fueling AGN II: Spatially Resolved Molecular Inflows and Outflows}
\shortauthors{Davies et al.}

\begin{document}


\title{Fueling Active Galactic Nuclei II:\\Spatially Resolved Molecular Inflows and Outflows}


\author{
R.I.~Davies\altaffilmark{1},
W.~Maciejewski\altaffilmark{2},
E.K.S.~Hicks\altaffilmark{3,4},
E.~Emsellem\altaffilmark{5},
P.~Erwin\altaffilmark{1,6},
L.~Burtscher\altaffilmark{1},
G.~Dumas\altaffilmark{7},
M.~Lin\altaffilmark{1},
M.A.~Malkan\altaffilmark{8},
F.~M\"uller-S\'anchez\altaffilmark{9},
G.~Orban de Xivry\altaffilmark{1},
D.J.~Rosario\altaffilmark{1},
A.~Schnorr-M\"uller\altaffilmark{1}
and A.~Tran\altaffilmark{4}}

\altaffiltext{1}{Max-Planck-Institute f\"ur extraterrestrische Physik, Postfach 1312, 85741, Garching, Germany}
\altaffiltext{2}{Astrophysics Research Institute, Liverpool John Moores University, IC2 Liverpool Science Park, 146 Brownlow Hill, L3 5RF, UK}
\altaffiltext{3}{Astronomy Department, University of Alaska, Anchorage, USA}
\altaffiltext{4}{Department of Astronomy, University of Washington Seattle, WA 98195, USA}
\altaffiltext{5}{European Southern Observatory, Karl-Schwarzschild Str. 1, 85748 Garching, Germany}
\altaffiltext{6}{Universit\"{a}ts-Sternwarte M\"{u}nchen, Scheinerstrasse 1, D-81679 M\"{u}nchen, Germany}
\altaffiltext{7}{Institut de Radio Astronomie Millim\'etrique (IRAM), 300 Rue de la Piscine, Domaine Universitaire, F-38406 Saint Martin d'Heres, France}
\altaffiltext{8}{Astronomy Division, University of California, Los Angeles, CA 90095-1562, USA}
\altaffiltext{9}{Center for Astrophysics and Space Astronomy, University of Colorado, Boulder, CO 80309-0389, USA}

\begin{abstract}
We analyse the 2-dimensional distribution and kinematics of the stars as well as molecular and ionised gas in the central few hundred parsecs of 5 active and 5 matched inactive galaxies.
The equivalent widths of the Br$\gamma$ line indicate there is no on-going star formation in their nuclei, although recent (terminated) starbursts are possible in the active galaxies.
The stellar velocity fields show no signs of non-circular motions, while the 1-0\,S(1) H$_2$ kinematics exhibit significant deviations from simple circular rotation.
In the active galaxies the H$_2$ kinematics reveal inflow and outflow superimposed on disk rotation.
Steady-state circumnuclear inflow is seen in three AGN, and hydrodynamical models indicate it can be driven by a large scale bar.
In three of the five AGN, molecular outflows are spatially resolved.
The outflows are oriented such that they intersect, or have an edge close to, the disk -- which may be the source of molecular gas in the outflow.
The relatively low speeds imply the gas will fall back onto the disk; and with moderate outflow rates, they will have only a local impact on the host galaxy.
H$_2$ was detected in two inactive galaxies.
These exhibit chaotic circumnuclear dust morphologies and have molecular structures that are counter-rotating with respect to the main gas component, which could lead to gas inflow in the near future.
In our sample, all four galaxies with chaotic dust morphology in the circumnuclear region exist in moderately dense groups with 10--15 members where accretion of stripped gas can easily occur.
\end{abstract}

\keywords{galaxies: active
-- galaxies: ISM
-- galaxies: kinematics and dynamics
-- galaxies: nuclei
-- galaxies: Seyfert
-- infrared: galaxies}

\section{Introduction}
\label{sec:intro}

This paper is the second part of a project to study the molecular gas and stellar properties in the circumnuclear region of five matched pairs of nearby active and inactive galaxies.
The rationale for embarking on this project was that, despite the statistical studies of correlations between AGN and host galaxy properties \citep[e.g.][see also other references below]{kau03,cid04,ho08araa,sch10,kos11}, there is no consensus on the mechanisms that drive gas to the nuclear region.
We aim to identify which structural and kinematic properties of the stars and molecular gas are present in active, but absent in inactive galaxies, and hence may be related to fueling AGN.
Hicks et al. (\citeyear{hic13}; hereafter \citetalias{hic13}) report on systematic differences within the central few hundred parsecs:
with respect to inactive galaxies, hosts of Seyfert nuclei have more centrally concentrated surface brightness profiles for both stellar continuum and H$_2$ 1-0\,S(1) emission, as well as lower stellar velocity dispersions and elevated H$_2$ 1-0\,S(1) luminosity.
These results suggested that Seyfert galaxies have a nuclear structure that is dynamically colder than the bulge, and comprises a significant gas reservoir together with a relatively young stellar population.
In this paper we focus on the spatially resolved stellar and molecular gas kinematics, with a goal to trace inflow mechanisms working on these scales.
As such, we distinguish between the following mechanisms that might lead to gas inflow and ultimately to accretion onto the central massive black hole:
(i) major merger, the coalescence of two approximately equally massive galaxies;
(ii) minor merger, the accretion of a small galaxy such as a dwarf into a larger one (i.e. coalescence of very unequal mass galaxies);
(iii) accretion of gas streamers (i.e. intergalactic atomic or molecular gas, perhaps in the form of spurs or bridges), which might originally have been produced during a merger or interaction, perhaps between other galaxies;
(iv) secular evolution, which is inflow of gas due to long-lasting disk driven processes, perhaps (but not necessarily) stimulated originally by an interaction.

There have been a large number of studies addressing at least some of these issues, focussing in particular on the role of major mergers.
While there are differences in the selection (e.g. hard X-ray or mid-infrared), the luminosity range (typically from $10^{41-42}$\,erg\,s$^{-1}$ to $10^{44-45}$\,erg\,s$^{-1}$), the redshift range (from $z$=0 to 3), and the technique used to identify mergers (e.g. close pairs or disturbed hosts), a clear consensus is emerging that major mergers appear to play a relatively minor role in triggering AGN activity \citep[e.g.][]{koc12,sch12,kar14,vil14}.
A similar conclusion is reached when looking at the star formation rates: typical QSOs at $z\sim2$ lie in galaxies with normal, rather than enhanced, star formation rates \citep{ros13}.
Only above L$_{AGN}\sim10^{45}$\,erg\,s$^{-1}$ is there some observational evidence that major mergers may begin to dominate AGN triggering \citep{tre12,ros12}.
With respect to low and intermediate luminosity AGN, \cite{nei14} argue that they may be triggered mostly by minor mergers; and, at least for early type galaxies, \cite{sim07} and \cite{mar13} suggest that external accretion, perhaps in the form of minor mergers, may be fueling the nuclear activity.
Beyond finding that about half of AGN are in disky hosts and also that about half of AGN have hosts that appear undisturbed \citep{koc12}, the observational studies above cannot probe further into the relative roles of minor mergers, gas accretion, or secular evolution.
Clues to their roles may lie in the local environment of the AGN.
For example, \cite{arn09} found that the fraction of X-ray selected AGN in groups is a factor 2 higher than in clusters.
And \cite{geo08} showed that X-ray AGN are more frequently found in groups than in the field.

A difficulty with all statistical studies, expecially in the context of a control sample, is the transient nature of AGN activity.
The timescale on which it occurs is expected to be short, characteristically of order 100\,Myr, and during this time it can flicker on and off with a timescale of 1--10\,Myr due to stochasticity at small scales \citep{hae93,ulr97,nov11,hic14,nei14}.
Indeed, \cite{kee12} and \cite{schi13} report AGN light echoes that suggest variability on even shorter timescales is possible.
In these galaxies, illumination of the very extended narrow line region requires a recent AGN luminosity significantly greater than that now inferred, implying that the AGN has decreased in luminosity by 1--2 orders of magnitude on a timescale of 0.1\,Myr.
In comparison, the timescale on which a merger occurs can be measured in Gyr \citep{lot08}.
And even for compact groups that evolve rapidly, the timescale over which the group survives (i.e. the galaxies have not yet all merged together) is measurable in Gyr \citep{bar89}.
A similar mismatch in timescales is also an issue when investigating links
between AGN and host galaxy morphological structures.
For example, despite much effort, there is at best only marginal evidence for a direct link between bars and AGN \citep{shl00,lai02,lau04,cis13} -- although there is a strong indirect link for at least one subset of AGN \citep{orb11}.
Looking in more detail for links between Seyferts and their host galaxies, \cite{hun04} found no significant differences in the presence of circumnuclear bars, boxy and disky isophotes, or other non-axisymmetric features in a comparison of matched samples of Seyferts and inactive galaxies.
They do, however, report that Seyferts are more likely to show isophotal twisting, suggesting a potential increase in the disturbance of the kinematics in these active galaxies (and noting that this is driven by the Seyfert~2s in their sample).
When looking at their data, we find that this difference is occurring equally in both early and late type hosts.
We also find that there is a higher fraction of late type inactive galaxies than late type Seyferts with circumnuclear bars, the implications of which are not clear.

In this work, we attempt to overcome the difficulties due to mismatched timescales by focusing on the mechanisms operating in the central
few hundred parsecs of active and inactive galaxies.
Here, where the orbital velocities are 100--150\,km\,s$^{-1}$ at a radius of 100\,pc \citepalias[see][]{hic13}, the dynamical timescales are 2--3\,Myr, comparable to the AGN duty cycle (i.e. the timescale for order-of-magnitude variability in accretion rate) mentioned above.

Previous studies of matched active and inactive galaxy samples based on optical integral field spectroscopy, such as \cite{dum07} and \cite{wes12} which we discuss in Sec.~\ref{sec:perturb}, have tended to focus on larger scales, from a few kpc down to their resolution limit of 100--200\,pc (or more typically 600\,pc in the latter study).
The seeing under which our data were taken have enabled us to achieve resolutions of $\sim50$\,pc, and we probe specifically the circumnuclear region out to radii of only a few hundred parsecs.
With respect to the optical regime, observing in the near-infrared around 2\,$\mu$m provides two advantages.
It enables us to probe to greater optical depth ($A_K \sim 0.1 A_V$);
and our spectral range includes the H$_2$ 1-0\,S(1) line which directly traces molecular gas at $\sim2000$\,K, in addition to the Br$\gamma$ line that probes the ionised phase.
We combine this with the stellar distribution and kinematics traced through the CO\,2-0 bandhead.

This paper is organised as follows, beginning with a short recap about the sample in Section~\ref{sec:sample}.
In Section~\ref{sec:environ} we look at the environment of the host galaxies since this is fundamental to the later discussion.
Using the galaxy orientations adopted in the Appendix, Section~\ref{sec:stellardiskfits} assesses whether a simple dynamical model is an appropriate prescription for, and if there is evidence for perturbations in, the circumnuclear stellar kinematics.
Then in Section~\ref{sec:h2vel} we make a detailed analysis of the H$_2$ 1-0\,S(1) distribution and kinematics.
We bring the results on the individual galaxies together in Sections~\ref{sec:starform} and~\ref{sec:outflow} where we discuss the evidence for and against nuclear star formation, and the properties of the observed molecular outflows.
Finally, in Section~\ref{sec:perturb} we broaden our discussion to address evidence for internal secular driven inflow versus external accretion, and the role of environment and host galaxy type. We do this not only in the context of our sample, but also other samples with spatially resolved stellar and gas kinematics, and also with reference to H\,I studies.
We summarise our conclusions in Section~\ref{sec:conc}.

\begin{deluxetable}{cllcccc}
\tabletypesize{\scriptsize}
\tablecaption{Galaxy Sample\label{tab:sample}} 
\tablewidth{0pt}
\tablehead{
\colhead{Pair} &
\colhead{Galaxy} &
\colhead{Type\tablenotemark{a}} &
\colhead{AGN} &
\colhead{D} &
\multicolumn{2}{c}{PSF FWHM} \\
\colhead{} &
\colhead{} &
\colhead{} &
\colhead{} &
\colhead{(Mpc)} &
\colhead{(\as)} &
\colhead{(pc)} \\
}
\startdata

1 & NGC 3227 & SABa  & Sey 1.5 & 21 & 0.55 & 56 \\
2 & NGC 5643 & SABc  & Sey 2   & 17 & 0.49 & 40 \\
3 & NGC 6300 & SBb   & Sey 2   & 17 & 0.48 & 40 \\
4 & NGC 6814 & SABbc & Sey 1.5 & 23 & 0.51 & 57 \\
5 & NGC 7743 & SB0   & Sey 2   & 19 & 0.54 & 50 \\
\\
1 & IC 5267  & SA0/a & no      & 30 & 0.61 & 90 \\
2 & NGC 4030 & SAbc  & no      & 27 & 0.66 & 87 \\
3 & NGC 3368 & SABab & no      & 11 & 0.58 & 30 \\
4 & NGC  628 & SAc   & no      & 10 & 0.59 & 28 \\
5 & NGC  357 & SB0/a & no      & 32 & 0.62 & 97 \\

\enddata

\tablecomments{The data here are repeated from Tables~1--3 of \cite{hic13}, which should be consulted for original sources, details about the quantities given, and other sample properties.}

\tablenotetext{a}{Abbreviated classification taken from the NASA/IPAC Extragalactic Database.}

\end{deluxetable}


\begin{deluxetable}{lllll}
\tabletypesize{\scriptsize}
\tablecaption{Line and Continuum Luminosities (2\as\ aperture)\label{tab:lum}} 
\tablewidth{0pt}
\tablehead{
\colhead{Galaxy} &
\colhead{\htwo} &
\colhead{Br$\gamma$} & 
\colhead{L$_K$} & 
\colhead{EW$_{Br\gamma}$} \\
\colhead{} &
\colhead{10$^4$ L$_\sun$} &
\colhead{10$^4$ L$_\sun$} &
\colhead{10$^4$ L$_\sun$} &
\colhead{\AA} \\
}
\startdata

NGC 3227 & \phm{$<$}31.1 & \phm{$<$}13.5 & 23.9 & \phm{$<$}1.4 \\
NGC 5643 & \phm{$<$}16.2 & \phm{$<$}\phn6.0 & \phn4.8 & \phm{$<$}1.3 \\
NGC 6300 & \phm{$<$}\phn6.4 & \phm{$<$}\phn0.3 & \phn4.7 & \phm{$<$}0.1 \\ 
NGC 6814 & \phm{$<$}\phn4.5 & \phm{$<$}\phn2.3 & \phn8.7 & \phm{$<$}0.5 \\ 
NGC 7743 & \phm{$<$}\phn6.9 & \phn$<$0.2 & \phn7.0 & $<$0.03\\ 
\\
IC 5267  & \phm{$<$}\phn5.9 & \phn$<$0.1 & 12.9 & $<$.004 \\
NGC 4030 & \phn$<$1.4 & \phn$<$0.1 & \phn7.3 & $<$0.007 \\ 
NGC 3368 & \phm{$<$}\phn1.3 & \phn$<$0.1 & \phn2.8 & $<$0.003 \\ 
NGC  628 & \phn$<$0.05 & \phn$<$0.01 & \phn0.2 & $<$0.01 \\ 
NGC  357 & \phn$<$4.7 & \phn$<$0.1 & \phn8.4 & $<$0.01 \\ 

\enddata

\tablecomments{Br$\gamma$ equivalent widths are determined using the stellar luminosities after correcting the total {\em K}-band luminosity for any non-stellar AGN contribution
(i.e. $L_{stellar}$ = L$_{K}$ $\times$ f$_{dilution}$).  See text for details.}

\end{deluxetable}

\section{Sample}
\label{sec:sample}

This paper focusses on a small sample that was selected as matched pairs of active and inactive galaxies.
However, our analysis does not treat them as such.
We assess them first as an active sample and an inactive sample; and then as a combined sample of which some members are active and some are inactive.
The targets were already discussed in \citetalias{hic13}, where a full description of the sample selection and properties is given, together with a discussion of the observations and data reduction.
Here, Table~\ref{tab:sample} summarises the galaxy properties and Table~\ref{tab:lum} provides an overview of the measured line and continuum luminosities.
Below we discuss a few key points about the targets and sample.

The observations were obtained with SINFONI, performing K-band integral field spectroscopy in seeing (at the same wavelength) of $0.57\pm0.06$\arcsec.
The targets were a subset of 5 matched pairs of active and inactive galaxies at distances of 10--32\,Mpc, taken from the sample of \cite{mar03} who published $V-H$ dust structure maps of the circumnuclear region of all the objects.
The original matching was based on the host galaxy Hubble type, B-band luminosity, heliocentric velocity, inclination, and angular size.
In our subsample, we performed an additional comparison of the H-band luminosity (as a proxy for stellar mass) and physical size of the disks.
For both of these quantities, values for the active galaxies as a whole are $\sim25$\% smaller, although this difference is only at the 1$\sigma$ level of significance.

The nuclear properties of the active and inactive samples were originally based on optical emission line ratios.
The difference between our two subsamples was confirmed through their 2--10\,keV X-ray luminosities, which are listed in \citetalias{hic13}.
The active galaxies are all in the range 5--$200\times10^{40}$\,erg\,s$^{-1}$;
only two of the inactive galaxies have measured hard X-ray luminosities, and these are $<2\times10^{39}$\,erg\,s$^{-1}$.
In addition, \citetalias{hic13} showed that 3 out of 5 of the active galaxies exhibit a strong spatially unresolved non-stellar continuum in the K-band, associated with hot dust heated by the AGN.
This is not seen in any of the inactive galaxies.

\section{Environment}
\label{sec:environ}

Before embarking on our analysis, we first look at the environment of the host galaxies, since this can help with understanding the galaxy structure and also shed light on the origin of the gas feeding the AGN.

The barred galaxy NGC\,3227 is in an overdense environment, being part of a group of 13 \citep{gar93} or 14 \citep{cro07} galaxies with a virial mass of $10^{12.8}\,M_\odot$.
The latter authors also include it in a high density contrast (HDC) group of 4 galaxies, and it is interacting with the dwarf elliptical NGC\,3226 which is only 13\,kpc away and has the same heliocentric velocity. 
In addition, \cite{mun95} detected large H\,I plumes extending 70\,kpc north and 30\,kpc south of NGC\,3227, although NGC\,3226 itself was undetected.

With respect to the other active galaxies, NGC\,5643, NGC\,6814, and NGC\,7743 all have undisturbed disks, and none are listed as being in groups by either \cite{cro07} or \cite{gar93}.
However, \cite{kat11} point out that, based on the observations of \cite{dup96}, NGC\,7743 has two massive ($10^{7.8}$ and $10^{8.7}$\,M$_\odot$) H\,I rich companions at distances of 46 and 61\,kpc with systemic velocities differing by only 100 and 200\,km\,s$^{-1}$ respectively from NGC\,7743.
Thus, while NGC\,7743 appears undisturbed, it is not isolated.
NGC\,6300 also does not show obvious signs of disturbance. 
Although it is reported by \cite{cro07} as being part of a low density contrast (LDC) group with 9 members, \cite{kar10} classify it not only as an isolated galaxy based on their analysis of 2MASS sources but without significant neighbours identified in optical images.

Turning to the inactive galaxies, IC\,5267 was marked as one of the brightest members of the Grus group of 15 bright galaxies by \cite{san75}, and is included in a low density contrast (LDC) group of 11 galaxies with a virial mass of $10^{12.7}\,M_\odot$ by \cite{cro07}.
Within this, it is part of a HDC group of 3 \citep{cro07} or 4 \citep{gar93} galaxies.
It has been suggested that it may be interacting with IC\,5267A which lies at a projected distance of 125\,kpc, and for which the H\,I line profiles overlap in velocity range \citep{dri88}.
While IC\,5267A has hints that it may have been perturbed (although available imaging is too shallow and low resolution to be certain either way), the star forming rings in IC\,5267 \citep{gil07,gro10} are very symmetric.
Since the primary effect of an interaction is to disturb the outer gas disk, this argues against such a scenario.
Instead, it suggests a perturbation propagating outwards from the centre, perhaps triggered by a companion passing through the centre perpendicular to the disk.
If that companion was IC\,5267A, one might expect IC\,5267A to be more disrupted and IC\,5267 itself to look more like the Cartwheel galaxy \citep{str93,hor01}.
Another cause might be a stream of gas feeding into the central regions of IC\,5267, a scenario that is consistent with the group environment but does not necessarily involve IC\,5267A.
Although the H\,I image of \cite{dri88} does not show evidence for spurs or other intergalactic gas, it does reveal a significant asymmetric extension of the disk to the south-east.

In contrast, NGC\,3368 does show clear evidence for intergalactic H\,I.
It is the brightest member in the Leo group of 12 galaxies \citep{giu00}, and has also been included in a HDC group of 14 galaxies (within a LDC group of 45 galaxies having a virial mass of $10^{14.0}\,M_\odot$) by \cite{cro07}, and a group of 9 galaxies by \cite{gar93}. 
It appears to be lying at the end of a gas bridge extending from a large H\,I structure \citep{mic10}.
Thus, while the outer star forming ring \citep{gil07} may be associated with the outer Lindblad resonance of the bar, it is also possible that the disk has been perturbed by the intergalactic atomic gas.

The other 3 inactive galaxies all appear to be undisturbed disk galaxies and, although not isolated, do not reside in massive groups.
NGC\,357 is not listed as being in a group by either \cite{cro07} or \cite{gar93}, although \cite{ber02} suggests that, despite it showing no signs of interaction, it is part of a group with at least 6 other members.
NGC\,628 and NGC\,4030 are reported as being in groups by \cite{gar93} with 7 and 4 members respectively (the latter being a small group in a southern extension to the Virgo cluster), but do not appear in the listings of \cite{cro07}.
The HI map of NGC\,628 also suggests it is undisturbed \citep{wal08}.

\begin{figure}
\epsscale{1.05}
\plotone{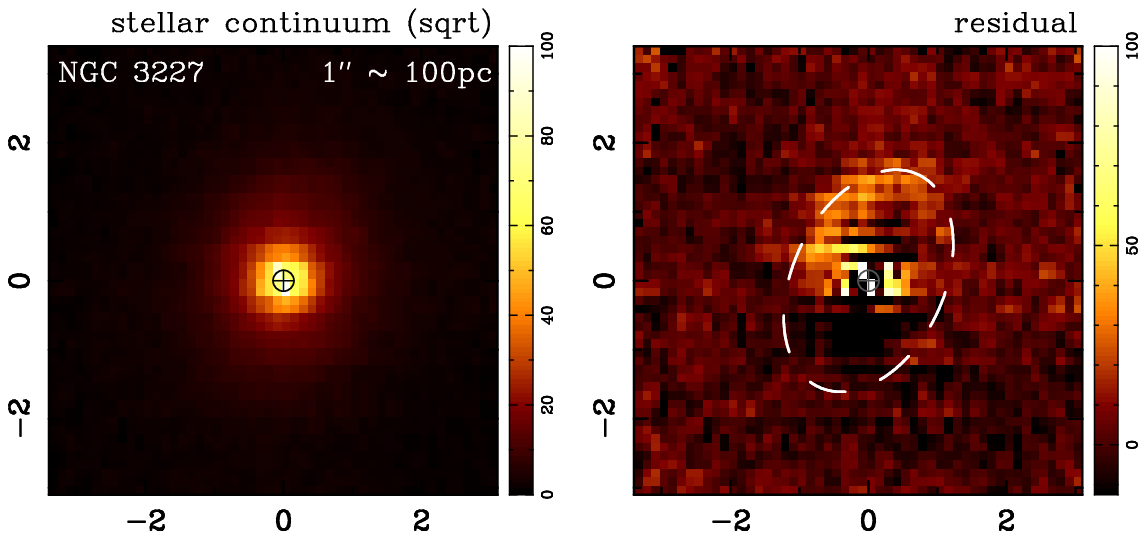}\\
\vspace{2mm}
\plotone{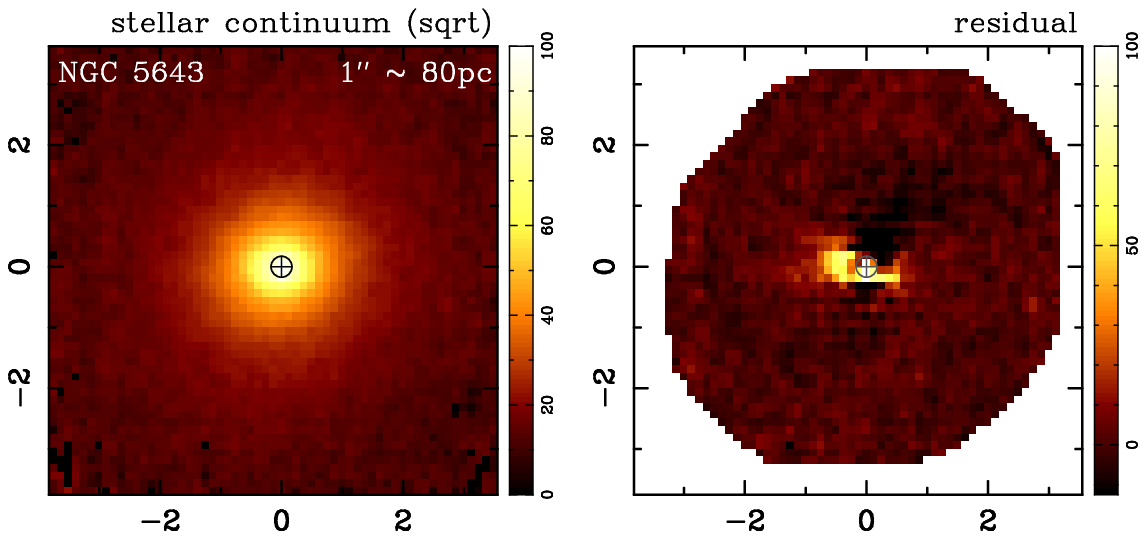}\\
\vspace{2mm}
\plotone{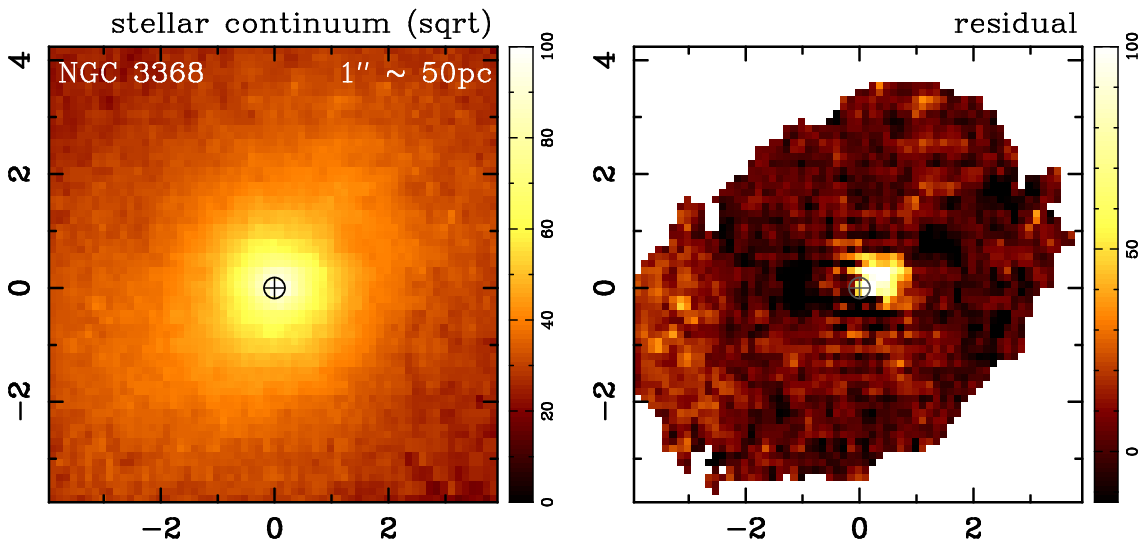}
\caption{\label{fig:starcont}
Stellar continuum of 3 galaxies, together with the residuals after subtracting elliptical isophotes fitted at a single position angle and with a single axis ratio (based on the isophotal values in Tables~\ref{tab:pa} and~\ref{tab:inc}).
These 3 galaxies have notable circumnuclear structure, which is discussed in Sections~\ref{sec:stellardiskfits} and~\ref{sec:h2vel}, and also in the Appendix.
Axis scales are in arcsec, with a conversion to parsec given for each object; north is up and east is left.}
\end{figure}

\begin{deluxetable}{lllll}
\tabletypesize{\scriptsize}
\tablecaption{Position Angle\label{tab:pa}} 
\tablehead{
\colhead{} &
\colhead{} &
\multicolumn{2}{c}{CO bandheads} &
\colhead{} \\
\cline{3-4} \\
\colhead{Galaxy} &
\colhead{adopted\tablenotemark{a}} &
\colhead{kinemetry\tablenotemark{b}} &
\colhead{isophotes\tablenotemark{c}} &
\colhead{large scale\tablenotemark{d}} \\
\colhead{} &
\colhead{(\deg)} &
\colhead{(\deg)} &
\colhead{(\deg)} &
\colhead{(\deg)} \\
}
\startdata

NGC 3227 & \phn155      & \phn155                  & \phn168  & \phn 159 \\
NGC 5643 & \phn\,-39    & \phn\,-39                  & \phn\,-64  & \phn\,-51 \\
NGC 6300 & \phn114      & \phn114                  & \phn105  & \phn 109 \\
NGC 6814 & \phn\phn\,-6 & \phn\phn\,-6                   & \phn\phn54   & \phn\phn 15  \\
NGC 7743 & \phn\,-65    & \phn\,-65                  & \phn\,-82  & \phn\,-65 \\
\\
IC 5267  & \phn\,-29    & \phn\phn19\tablenotemark{e}  & \phn\,-29  & \phn\,-41 \\
NGC 4030 & \phn\phn40   & \phn\phn40                   & \phn\phn38   & \phn\phn 32 \\
NGC 3368 & \phn173      & \phn173                  & \phn144  & \phn 172  \\
NGC  628 & \phn\phn---\tablenotemark{f} & \phn\phn42\tablenotemark{d}  & \phn\phn85   & \phn\phn 20   \\
NGC  357 & \,-159       & \,-159                 & \,-135 & \,-160 \\

\enddata

\tablecomments{All PAs given are the angle east of north for the side with positive velocities (if known).  Large scale PAs are typically based on photometry, as described for each galaxy in the Appendix. Where appropriate, the PA has been switched by 180\degr\ so that it can be compared more easily to the small scale kinematic PA.}
\tablenotetext{a}{This refers to the PA adopted on small scales when fitting disk models; the large scale values are used to address the uncertainties in direct estimates of the PA derived from the small scale data.}
\tablenotetext{b}{Kinemetry fit of the 2-D CO bandhead velocity and dispersion.}
\tablenotetext{c}{Mean PA fit to the isophotes of the CO bandhead flux}
\tablenotetext{d}{See the Appendix for details.}
\tablenotetext{e}{There is little stellar rotation detected and so the kinemetry is poorly constrained.}
\tablenotetext{f}{Due to the weak stellar velocity gradient and non-detection of 1-0\,S(1) emission, we do not model this galaxy and so have not adopted a PA.}

\end{deluxetable}


\begin{deluxetable}{lllll}
\tabletypesize{\scriptsize}
\tablecaption{Inclination Angle (Axis Ratio)\label{tab:inc}} 
\tablehead{
\colhead{} &
\colhead{} &
\multicolumn{2}{c}{CO bandheads} &
\colhead{} \\
\cline{3-4} \\
\colhead{Galaxy} &
\colhead{adopted\tablenotemark{a}} &
\colhead{kinemetry\tablenotemark{b}} &
\colhead{isophotes\tablenotemark{c}} &
\colhead{large scale\tablenotemark{d}} \\
\colhead{} &
\colhead{(\deg)} &
\colhead{(\deg)} &
\colhead{(\deg)} &
\colhead{(\deg)} \\
}
\startdata

NGC 3227 & 47\tablenotemark{e} & 41 & 35 & 56 \\ 
NGC 5643 & 34 & 34                  & 19 & 24 \\
NGC 6300 & 44 & 44                  & 44 & 50 \\ 
NGC 6814 & 21 & 43                  & 17 & 21 \\ 
NGC 7743 & 37 & 37                  & 25 & 40 \\ 
\\
IC 5267  & 40 & 59\tablenotemark{f} & 20 & 40 \\
NGC 4030 & 38 & 38                  & 25 & 40 \\ 
NGC 3368 & 53 & 22                  & 35 & 53 \\ 
NGC  628 & ---\tablenotemark{g} & 57\tablenotemark{f} & 28 & \phn7 \\ 
NGC  357 & 37 & 20                  & 22 & 37 \\ 

\enddata

\tablenotetext{a}{This refers to $i$ adopted on small scales when fitting disk models; the large scale values are used to address the uncertainties in direct estimates $i$ derived from the small scale data.}
\tablenotetext{b}{Kinemetry fit of the 2-D CO bandhead velocity and dispersion; these values are associated with a large uncertainty.}
\tablenotetext{c}{Mean PA fit to the isophotes of the CO bandhead flux.}
\tablenotetext{d}{See the Appendix for details.}
\tablenotetext{e}{We have adopted an inclination matching that implied by the circumnuclear ring, as discused in the Appendix.}
\tablenotetext{f}{There is little stellar rotation detected and so the kinemetry is poorly constrained.}
\tablenotetext{g}{Due to the weak stellar velocity gradient and non-detection of 1-0\,S(1) emission, we do not model this galaxy and so have not adopted any $i$.} 

\end{deluxetable}


\begin{figure*}
\epsscale{0.75}
\plotone{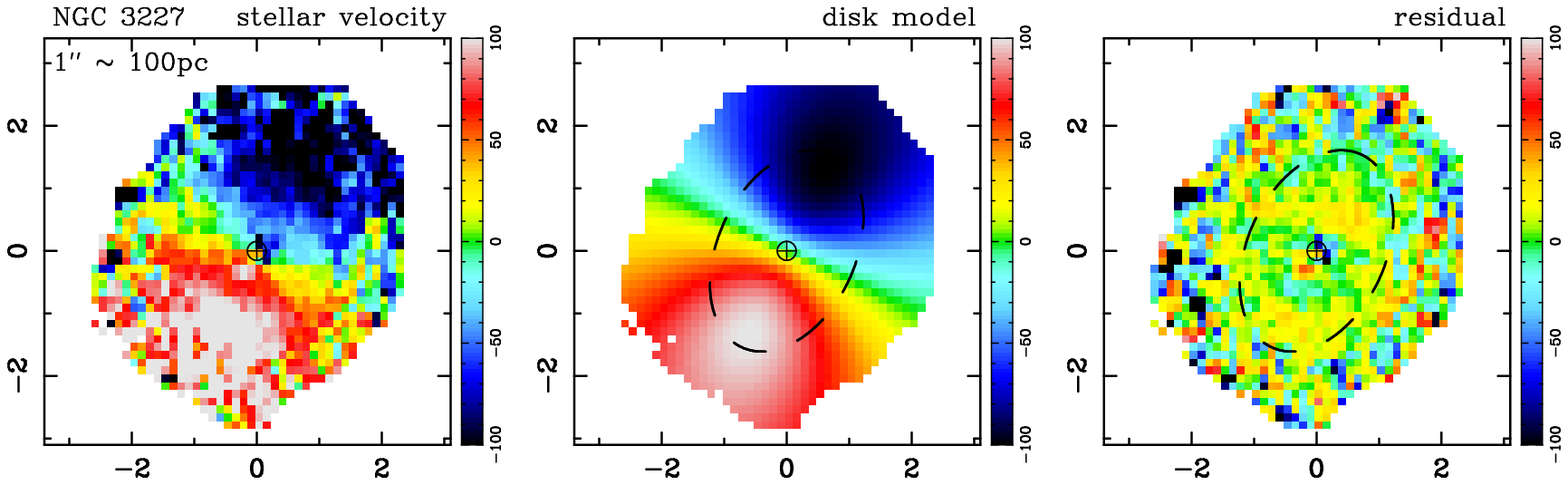}
\vspace{2mm}
\plotone{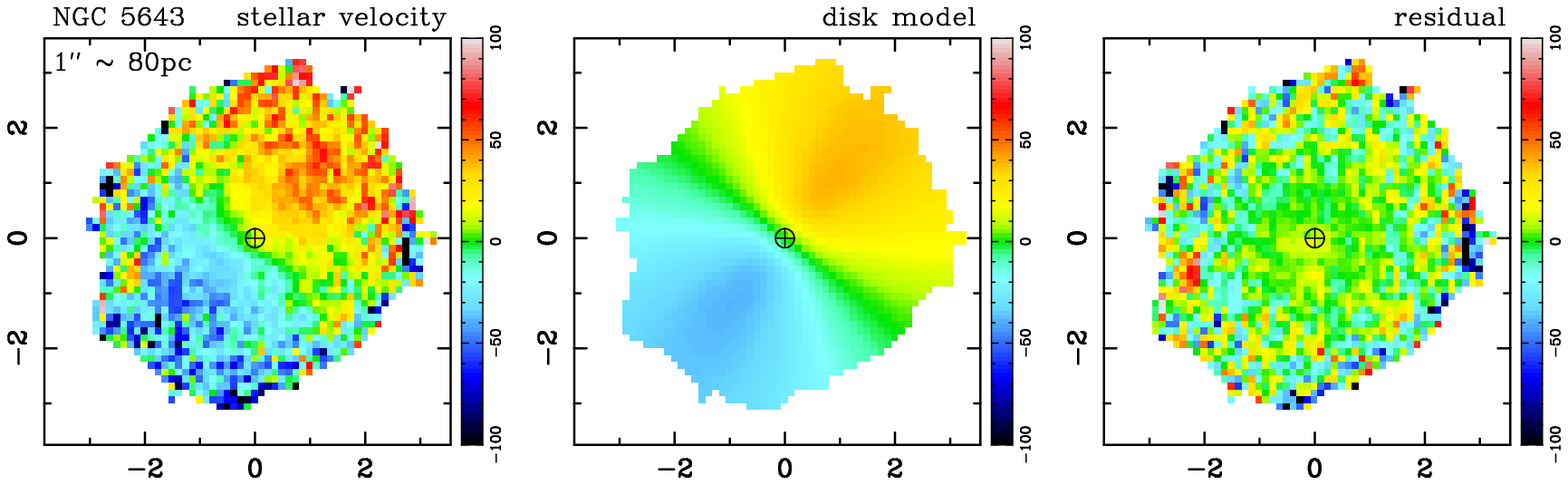}
\vspace{2mm}
\plotone{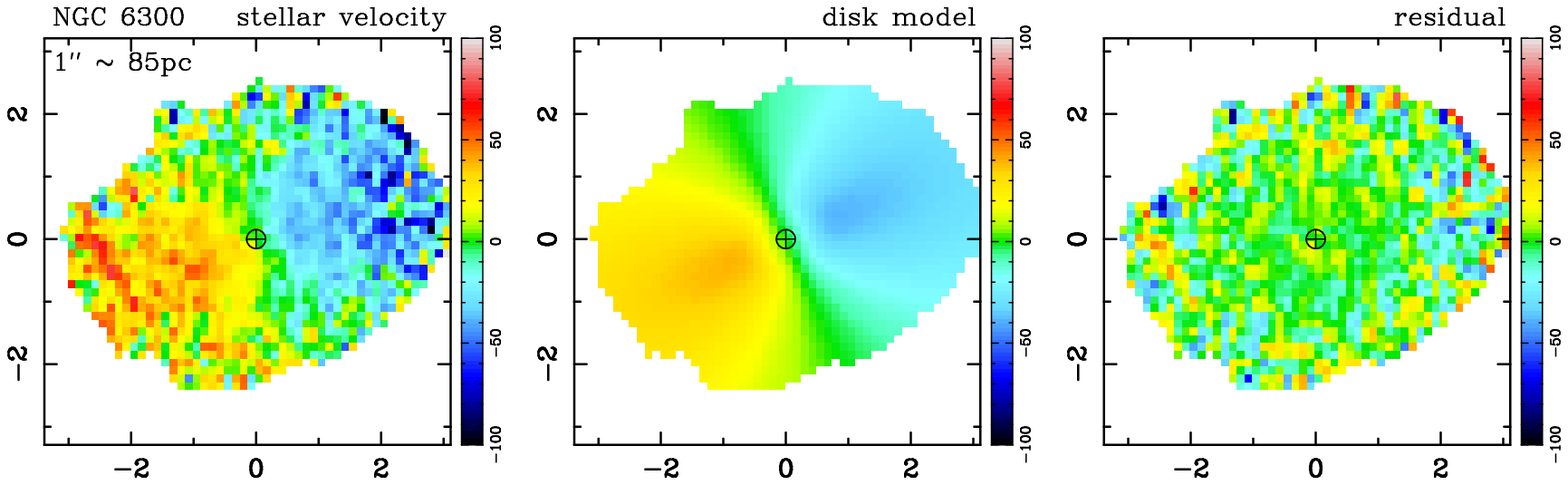}
\vspace{2mm}
\plotone{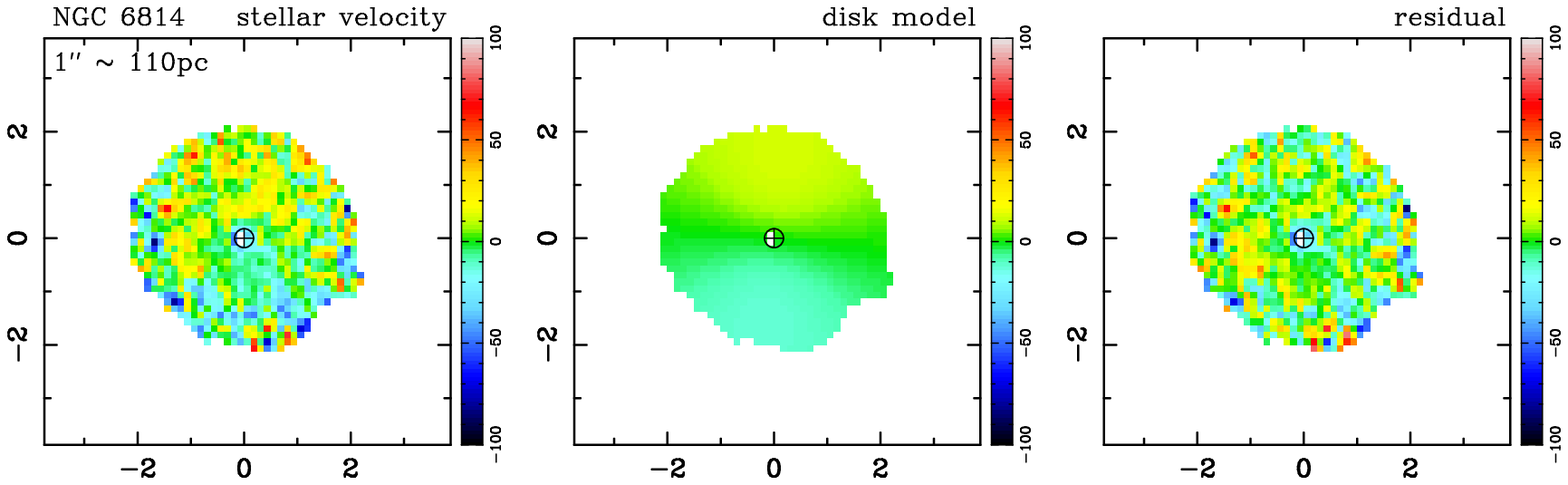}
\vspace{2mm}
\plotone{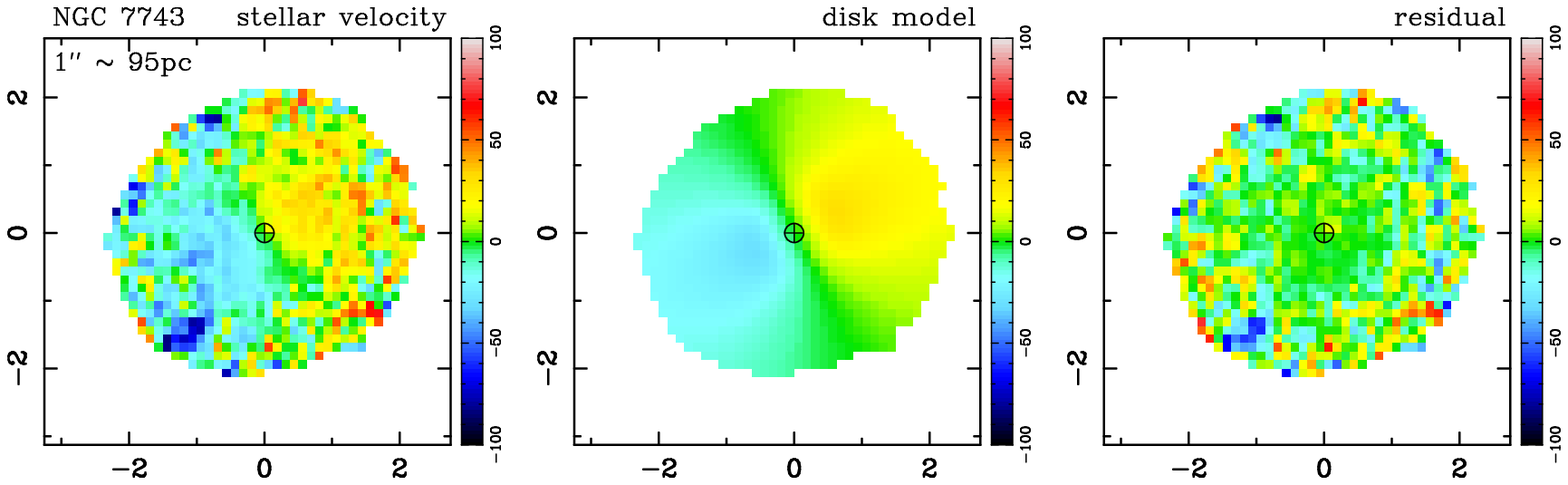}
\caption{\label{fig:starvelagn}
Stellar velocity fields of the active galaxies, together with the very simple disk models and residuals (all plotted on the same velocity scale from $-$100 to $+$100\,km\,s$^{-1}$). The model used a rotation curve defined solely by a mass distribution described by a single S\'ersic function (with the exception of NGC\,7743 for which we included an additional unresolved central mass). The aims were to confirm the orientation adopted for each galaxy, assess whether the stellar velocity fields can be well matched in this way; and hence whether we can apply this method to the more complex H$_2$ velocity fields in order to determine the non-circular motions of the molecular gas.
The dashed ellipse in NGC\,3227 traces the circumnuclear ring \citep{sch00,dav06}.
Axis scales are in arcsec, with a conversion to parsec given for each object; north is up and east is left.}
\end{figure*}

\begin{figure*}
\epsscale{0.75}
\plotone{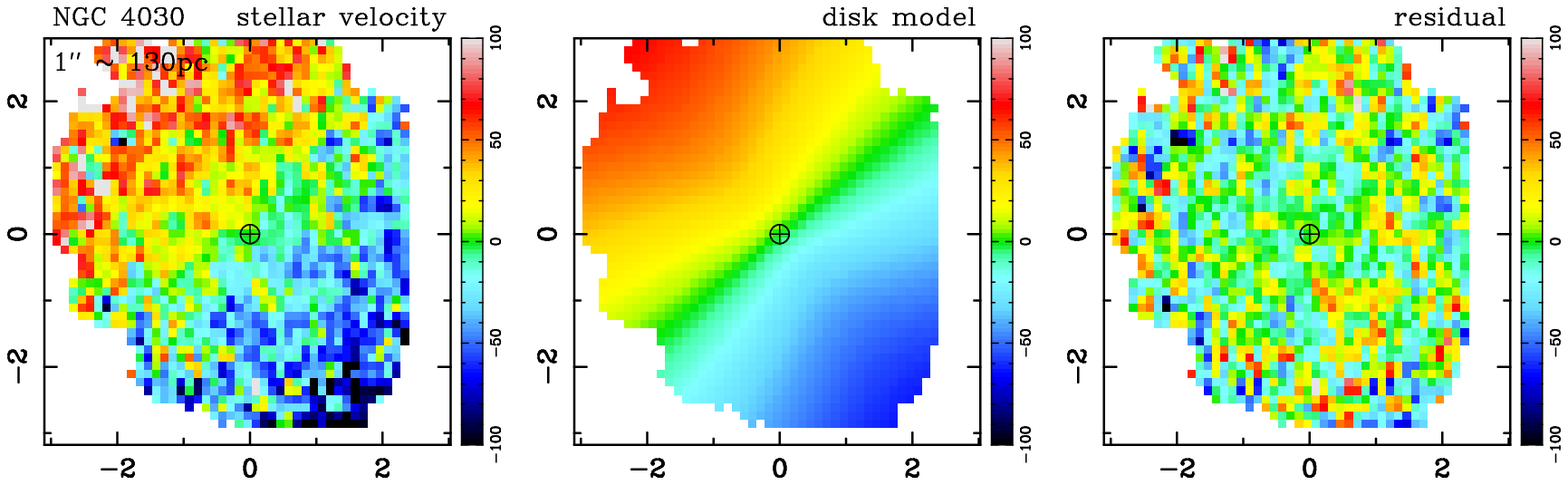}
\vspace{2mm}
\plotone{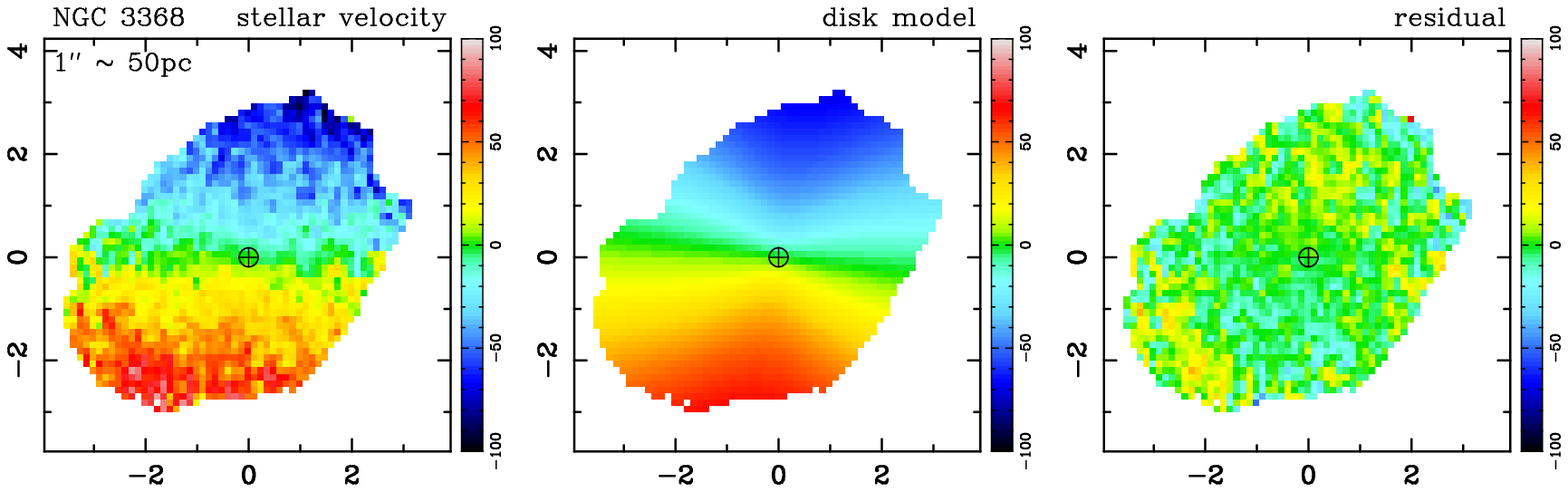}
\vspace{2mm}
\plotone{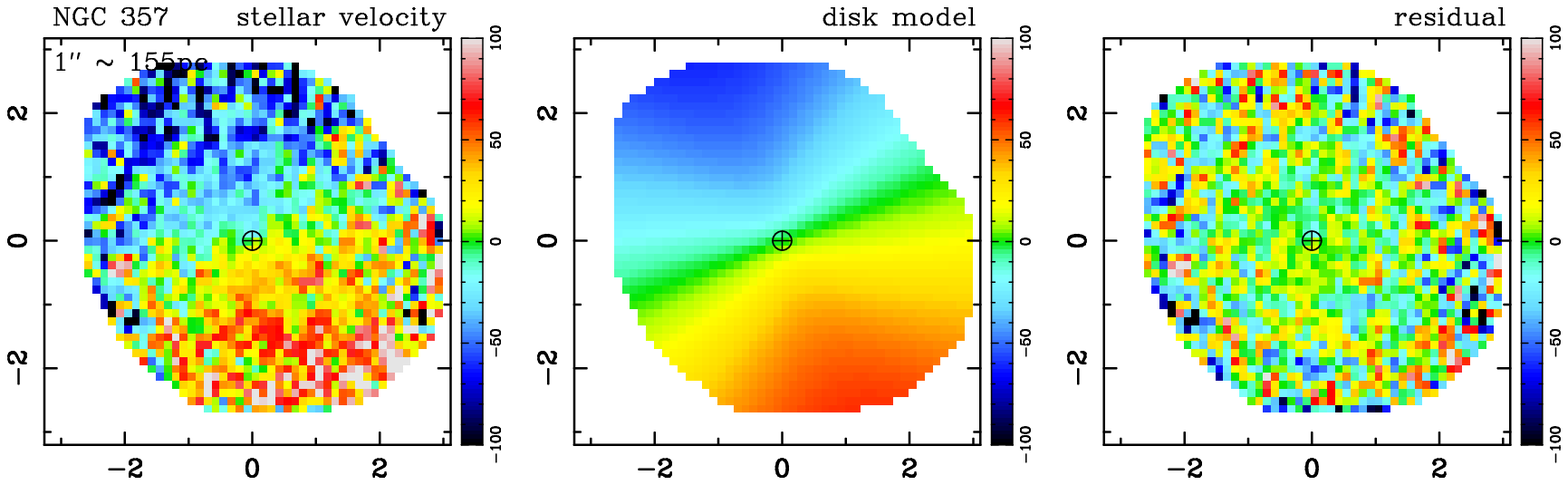}
\caption{\label{fig:starvelq}
Stellar velocity fields of the 3 inactive galaxies with measurable velocity gradients, together with simple disk models and residuals (plotted on the same velocity scale from $-$100 to $+$100\,km\,s$^{-1}$) as for Fig.~\ref{fig:starvelagn}.
The 2 inactive galaxies (IC\,5267 and NGC\,628) for which no velocity gradient could be measured, and hence for which we do not model the stellar kinematics, are not shown here; their velocity fields are presented in \citetalias{hic13}.}
\end{figure*}

\section{Disk fits to Stellar Kinematics}
\label{sec:stellardiskfits}

The principal aim of this work is to probe the non-circular motions of the molecular gas traced through the 2.1\,$\mu$m 1-0\,S(1) line, once a rotating disk component has been fitted and subtracted.
However, it is misleading to make an unconstrained fit of a disk to the H$_2$ data themselves, since the non-circular motions can be significant.
The initial step in our analysis is therefore to independently determine the host galaxy orientation.
In the Appendix we assess the large scale (i.e. the whole galaxy) orientation, and use this as the context for assessing the apparent orientation from the stellar kinematics and isophotes on small scales (i.e. the few central hundred parsecs within our field of view).
The 3 galaxies with notable circumnuclear structure in the stellar continuum are shown in Fig.~\ref{fig:starcont}.
And the position angle (PA) and inclination ($i$) for all the galaxies on large and small scales, and also the value we adopt for our analysis, are summarised in Tables~\ref{tab:pa} and~\ref{tab:inc}.
We use these to test whether a 
simple beam-smeared disk model can provide a good match to the observed stellar velocity field.
We do not claim that the stars are in such a disk; 
indeed Fig.~17 of \citetalias{hic13} shows that the circumnuclear stellar distribution traced by the 2.3\,$\mu$m CO bandheads is geometrically thick, although less so for the active galaxies than the inactive ones.
The rationale is to 
(i) determine whether the observed circular stellar velocity can be modelled with a simple prescription for the rotation curve, 
(ii) quantify any perturbations to the stellar kinematics, and 
(iii) assess whether we can reasonably apply the same technique to help interpret the more complex H$_2$ velocity fields in Sec.~\ref{sec:h2vel}.
We apply the same PA and $i$ when fitting a disk model to the gas kinematics, noting that unless there are clear indications to the contrary, and based on kinematics measured in early and late type galaxies \citep{sar06,dum07,dav11,wes12}, our working assumption is that the stellar and molecular gas kinematics are aligned.
However, we refer the reader to Section~\ref{sec:perturb} for a discussion of this point.

We make use of the DYSMAL code described in Appendix~A of \cite{dav11}, which was developed to help interpret observations of beam-smeared rotating rings and disks at arbitrary orientations.
We keep the number of free parameters to a minimum by fixing the PA and $i$ at the values given in Tables~\ref{tab:pa} and~\ref{tab:inc}, and quantifying the rotation curve in terms of a mass distribution described by a single S\'ersic function.
This requires three parameters: effective radius, S\'ersic index, and scaling.
In addition we allow the systemic velocity to vary.
The centering is fixed at the peak of the K-band continuum distribution, and the beam-smearing is applied according to the PSF size given in Table~\ref{tab:sample}.
The fitting is performed using the Levenberg-Marquardt algorithm as implemented by \cite{mar09}.

The results of the disk fitting process are shown in Fig.~\ref{fig:starvelagn} and~\ref{fig:starvelq} for the active and inactive galaxies respectively.
One should be cautious of the apparent striking visual effect that, within the field shown, the inactive galaxies are rotating faster than the active galaxies.
This is because we have not plotted -- nor modelled -- the two inactive galaxies with no measurable rotation.
We note that a comparison of the rotation curves, together with the dispersion, is discussed in \citetalias{hic13}.

In most cases the very simple rotation curve provides a fully satisfactory match to the observed velocity field, confirming that we can adopt a similar approach when analysing the H$_2$ velocity fields.
In a few cases there are measurable residuals.
For NGC\,7743, these were already compensated in Fig.~\ref{fig:starvelagn} by including a central compact mass (not necessarily a supermassive black hole, but maybe an unresolved star cluster) to make a steeper velocity gradient in the centre while maintaining also an extended mass distribution.
The residuals in NGC\,3227 are more complex.
Since these are anti-symmetric about the centre, we conclude that a more detailed prescription for the rotation curve might be more appropriate, perhaps reflecting the impact of a changing mass profile within, and either side of, the circumnuclear ring shown in Fig.~\ref{fig:nic3227}.
We have not attempted to improve on the basic fit because the residuals are at a level of 20--30\,km\,s$^{-1}$ which is only $\sim$20\% of the rotational velocity. 
In NGC\,3368, there are residuals at a similar relatively low level, suggesting that a more complex rotation curve may be required to fit the data more precisely.
However, given the low level of the residuals and the additional complexity this brings (especially to the interpretation of the H$_2$ kinematics in Sec.~\ref{sec:h2vel}), we do not feel this is justified.

\section{Molecular Hydrogen}
\label{sec:h2vel}

The H$_2$ velocity fields are much more complex than those of the stars.
As such, we adopt a similar approach to fitting disk models as described already for the stellar kinematics, with an additional iteration: initially we use all the data in the fit, but in a second iteration we mask out regions that are obviously anomalous and exclude these from the final fit.
This is to reduce any bias from non-circular motions, so that they can be better quantified in the residual once the disk model is subtracted.
We interpret the residual velocities in conjunction with the H$_2$ flux maps, and their residuals after subtracting elliptical isophotes.
We note that in some cases, it is not clear whether a simple disk velocity field is the correct model to subtract, and in 2 cases we specifically address this by looking also at velocity fields for disks with circumnuclear spirals.
By doing so, we take care that our simple approach does not lead to over-interpretation of the velocity residuals.

Table~\ref{tab:lum} lists the luminosities of the H$_2$ 1-0\,S(1) line, as well as those of the Br$\gamma$ line, extracted in a 2\arcsec\ diameter aperture.
Table~\ref{tab:h2} summarises what phenomena and structures we detect in molecular gas, via the 1-0\,S(1) lines, in these galaxies, which are described and discussed in detail in the following sub-sections.

\subsection{Active Sample}

As has been found for other nearby active and inactive galaxies, the 1-0\,S(1) distributions and kinematics can show a variety of different phenomena, and vary from being relatively simple to extremely complex (e.g. \citealt{dav09,hic09,mue09,rif09,rif10,sto10,rif11,ise13,rif13,maz14}).
In our sample the structures exhibited by the AGN typically appear to be a combination of molecular inflow and outflow, superimposed on a rotating disk.

\begin{figure*}
\epsscale{0.50}
\plotone{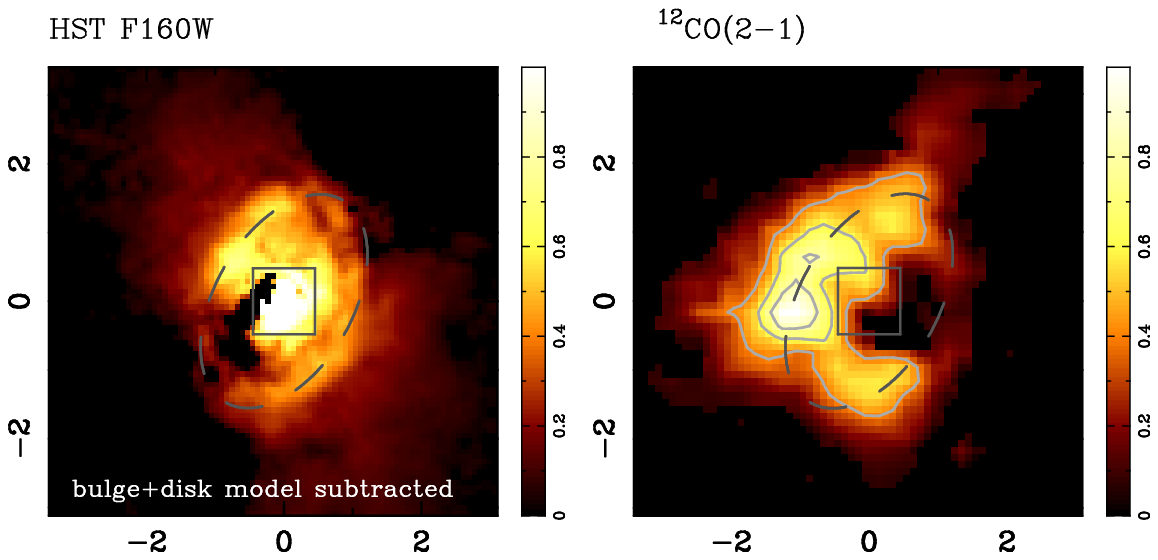}
\caption{\label{fig:nic3227}
Images of the circumnuclear ring in NGC\,3227. Left: H-band from HST (square-root scaling), after subtracting a model of the bulge and disk derived from larger scales. Right: the CO\,2-1 map from \cite{sch00}.
Reproduced from \cite{dav06}.
The central rectangle indicates the $\sim1$\arcsec\ field of view analysed in that paper, which is significantly smaller than the 6.5\arcsec\ field we use here.
In this and other figures of NGC3227, the dashed ellipse represents the location of the circumnuclear ring.
Axis scales are in arcsec; north is up and east is left.}

\vspace{3mm}

\epsscale{0.75}
\plotone{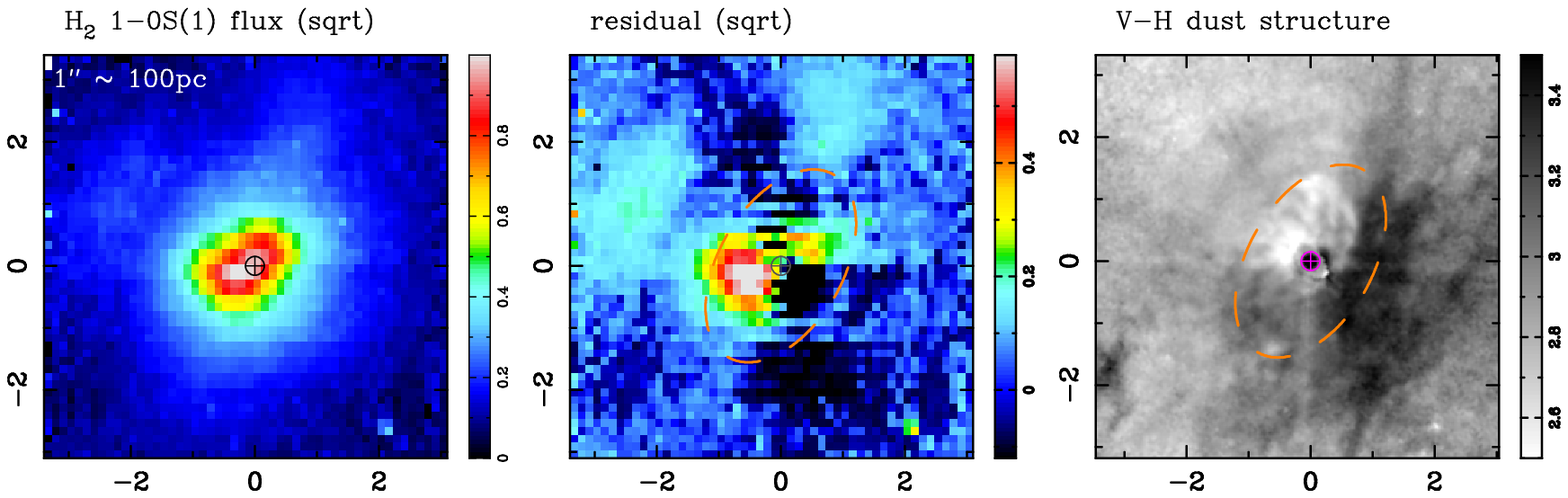}
\vspace{2mm}
\plotone{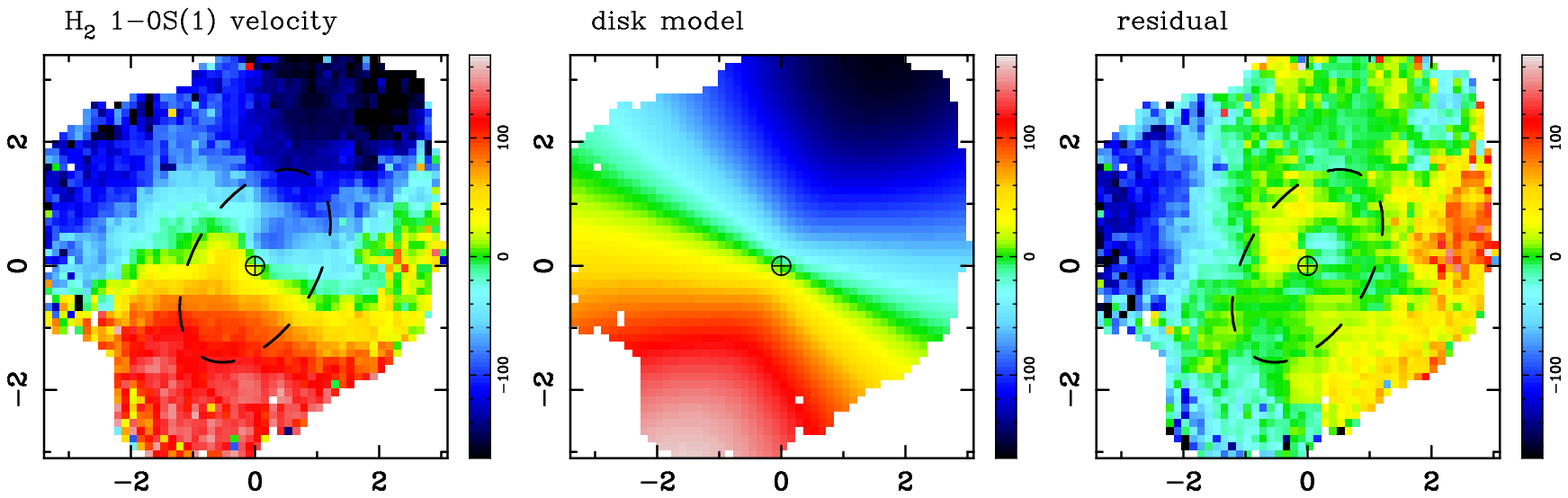}
\epsscale{0.245}
\plotone{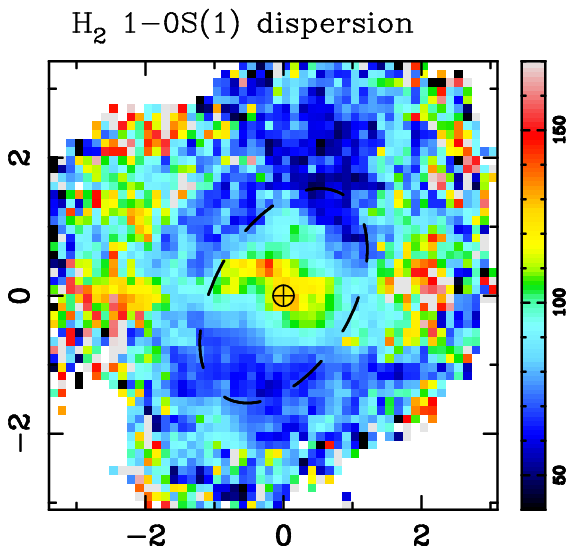}
\caption{\label{fig:ngc3227g}
Molecular gas as traced by the 1-0\,S(1) line in NGC\,3227, showing the same region as in Fig.~\ref{fig:starvelagn}.
Top row: observed flux, and the residual after subtracting elliptical isophotes centered on the nucleus; the right panel shows the V-H dust structure map reproduced from \cite{mar03}.
Bottom row: observed velocity field, disk model, and the difference between them.
The strong residuals a few arcsec from the centre suggest that a simple disk is not an appropriate model for this galaxy.
The rightmost panel is a map of the velocity dispersion.
Axis scales are in arcsec, with a conversion to parsec given; north is up and east is left.}

\vspace{3mm}

\epsscale{0.50}
\plotone{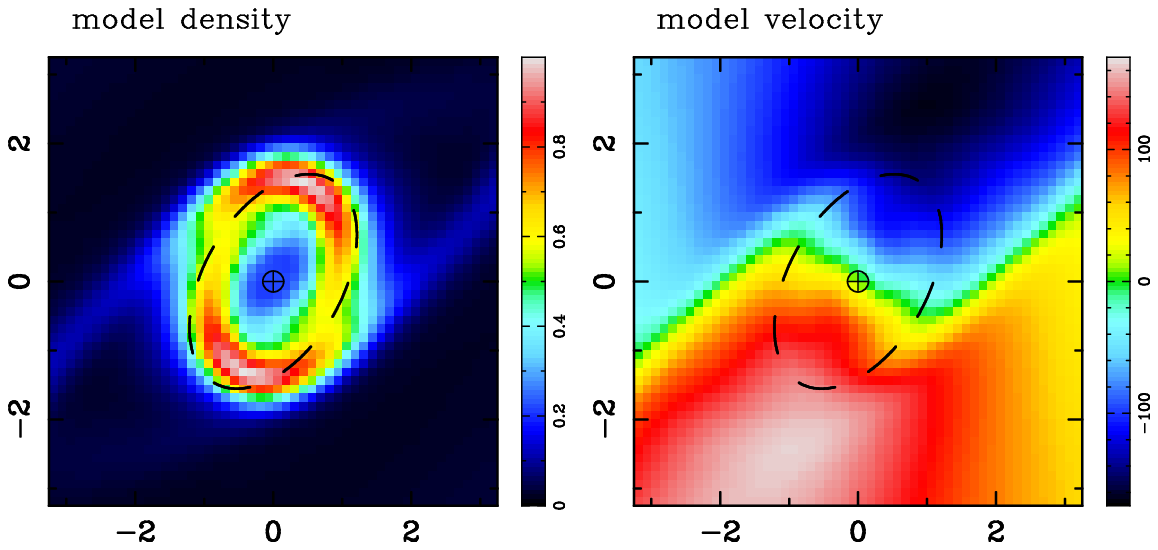}
\caption{\label{fig:ngc3227mod}
Density (left) and line of sight velocity (right) for a hydrodynamical simulation of a disk with a circumnuclear ring (spiral) generated by a large scale bar, matching model S20 from \cite{mac04}.
The bar was oriented 20\deg\ west of the line of nodes, and the model has been set at an inclination and position angle matching NGC\,3227 as given in Tables~\ref{tab:pa} and~\ref{tab:inc}.
}

\vspace{3mm}

\epsscale{0.75}
\plotone{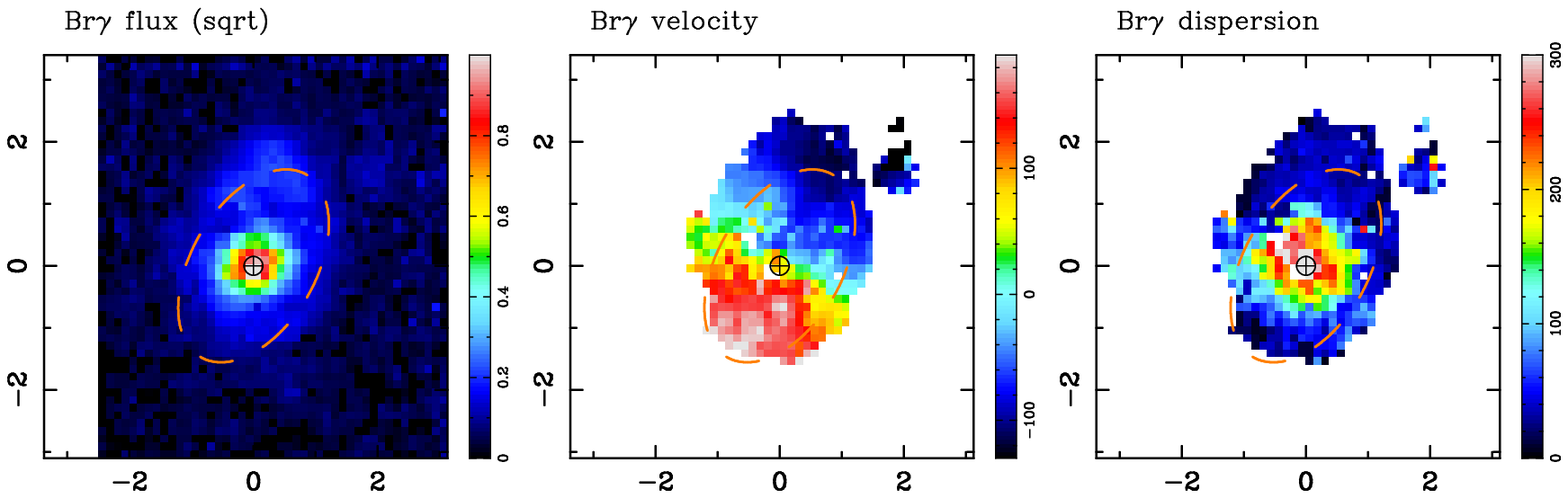}
\caption{\label{fig:ngc3227b}
Ionised gas as traced by the narrow Br$\gamma$ line in NGC\,3227, over the same field of view as shown previously.
The region of high dispersion at the centre should be interpreted cautiously, since in this region it was not possible to robustly separate the broad and narrow components.
The orange ellipse traces the stellar ring as for Fig.~\ref{fig:nic3227}.
Axis scales are in arcsec; north is up and east is left.}
\end{figure*}

\subsubsection*{NGC\,3227}

Large-scale features of NGC\,3227, which might have been caused by its
interaction with NGC\,3226, include a possible bar with bar-induced gas
flow. 
These can be traced inwards, and they affect the circumnuclear
morphology and kinematics of NGC\,3227. 
In particular, two strings of H$\alpha$ regions, in the form of two spiral arms, can be traced from a radius of 70\arcsec\ down to a radius of 6\arcsec, stretching along a PA of -45\deg\ in the inner 30\arcsec\ (see Fig.7 of \citealt{mun04}). The arms can be traced further inwards in molecular gas, where they earlier were reported as a CO bar \citep{mei90,mun95}, extending to a radius of 7--10\arcsec, and stretched along a PA of -55\deg. 
Inside the bar, \cite{sch00} report an asymmetric nuclear ring with diameter of
about 3\arcsec. 
This ring is also seen in J-K and H-K colour maps constructed from
0.15\arcsec\ resolution images \citep{cha00}. 
However, the ring does not show in the V-H HST colour map by \cite{mar03}.
This is possibly because of high extinction in the nucleus, and in that case the dust lanes in the HST V-band would be a foreground feature. 
The dust lanes in the V-H HST colour map also appear unrelated to the CO bar, which fully fits within the HST field.

Our SINFONI field covers the circumnuclear ring and the area immediately outside it (see Fig.~\ref{fig:nic3227}).
The 1-0\,S(1) flux distribution, shown in Fig.~\ref{fig:ngc3227g}, appears elongated along a PA between -45\deg\ and -50\deg, which is different from the adopted PA=-25\deg\ (Table~\ref{tab:pa}). This produces a pattern of positive and negative flux residuals, with the strongest positive residual 0.5\arcsec\ south-east of the nucleus. 
The elongation of the 1-0\,S(1) flux distribution observed in our data is consistent with that in higher resolution data for the same line transition shown in \cite{dav06}, who presented a
detailed view of molecular and ionised gas and their kinematics in the
central arcsecond of NGC\,3227. The strongest positive residual that we observe
in the \htwo\ 1-0\,S(1) emission is located inside the molecular ring reported by \cite{sch00}, and it is offset from the region of
brightest CO\,2-1 emission. This suggests that local excitation rather than
just the presence of massive molecular clouds plays a role in the excess 1-0\,S(1) line flux. However, large amounts of dense molecular gas traced by HCN(1-0) transition are reported inside the nucler ring, in the nucleus itself \citep{sch00,san12,dav12}.

The \htwo\ exhibits a very strong velocity gradient at a position angle consistent
with that of the stars. However, the zero-velocity line shows a remarkable
twist: in the innermost 2--3\arcsec\ it is almost straight at PA $\sim$40\deg,
corresponding to the line of nodes (LON) along a PA of 130\deg, slightly discrepant with the
stellar LON along a PA of 155\deg. Further out, the zero-velocity line takes
a sharp turn on both sides of the galaxy centre, assuming a PA of 120\deg,
almost perpendicular to its run in the central parts. For this reason, if we
subtract the disk model derived from stellar kinematics, then the largest
residuals, with amplitude reaching ~100 km/s, are towards the edges of our
field towards both east and west: at radii beyond 2\arcsec\ from the nucleus, roughly along the minor axis of the galaxy. Interestingly, at the same locations, the velocity dispersion also increases to $\sim$100\,km\,s$^{-1}$.

High velocity dispersion and high velocity residuals occuring at the outskirts
of the field can be caused by outflow superimposed on rotation there. On the
north-east side, these residuals are cospatial with the \htwo\ flux residual about 3\arcsec\ from
the centre, which may trace the outflowing material. This material is also
seen in H$\alpha$ emission in the F658N HST filter, with a blueshifted velocity consistent with our measurement \citep{fis13}. 
In another tracer, the [S\,III] line at 0.907\,$\mu$m, \cite{bar09} found a NE-SW velocity gradient,
with the blueshifted material towards NE extending to 3\arcsec\ from the centre, at a PA somewhat smaller than that of the H$\alpha$ emission and the \htwo\ residual.
They interpreted it as an interaction of the outflowing ejecta with the
circumnuclear gas. In the 3D geometry of the outflow in NGC\,3227 proposed by \cite{fis13}, the [S\,III], H$\alpha$ and \htwo\ emission can all highlight
portions of the biconical outflow that intersects the disk plane. This outflow
would then be also consistent with the central 2\arcsec-long, high-velocity-dispersion
feature in our H$_2$ data that is extended N-W. 
However, on small scales, in the adaptive optics integral field spectroscopic data presented by \cite{dav06}, residual (non-circular) velocities within 50\,pc of the nucleus are redshifted at a PA around 45\deg, though the 1-0\,S(1) flux is low there.
This could be consistent with the locations where the modelled bicone encloses
part of the galaxy disk, with the redshift instead of blueshift occuring
because we see the other half of the bicone in the high-resolution data of \cite{dav06}.

To summarize, we see signatures of outflow in the innermost 1--2\arcsec\ of NGC\,3227, and also in the outskirts of our field. 
However, at the location of the bend
in the zero-velocity line, these signatures are absent. In particular, on the
NE side, the \htwo\ residual emission and the H$\alpha$ emission are weak at the
bend of the zero-velocity line, and the velocity dispersion there reaches a local minimum, which counter-indicates multiple components. 
High velocity dispersion, as well as H$\alpha$ emission and \htwo\ excess are all located further out than the bend in the zero-velocity line. 
This leads us to believe that this bend is not caused by outflow superimposed on the rotation in the disk, but by gas flow
in the disk departing from circular rotation. We noted earlier in this
section that NGC\,3227 is likely barred, with gas lanes stretching throughout
the bar down to the innermost 3\arcsec, where the molecular nuclear ring is present.
Hydrodynamical models of gas flow in a bar predict strong inflow in gas lanes
that settles on the nuclear ring. Thus strong non-circular motion are expected
outside the ring, while in the ring the motion is expected to be close to
circular. 
In Fig.~\ref{fig:ngc3227mod}, we show the LOS velocity for model S20 from \cite{mac04} of gas flow in a bar, for the orientation of NGC\,3227 given in Tables~\ref{tab:pa} and~\ref{tab:inc}.
There is good qualitative agreement with the velocity field that we
observe in NGC\,3227 (both the 1-0\,S(1) H$_2$ line in Fig.~\ref{fig:ngc3227g} and, within the more restricted region over which it is detected, the Br$\gamma$ line in Fig.~\ref{fig:ngc3227b}), with the zero-velocity line bending at the outer parts of
the nuclear ring, with blueshift dominating to the east, and redshift to
the west. 
Thus the observed velocity field indicates strong inflow in the
bar that settles in the nuclear ring.

\begin{figure*}
\epsscale{0.75}
\plotone{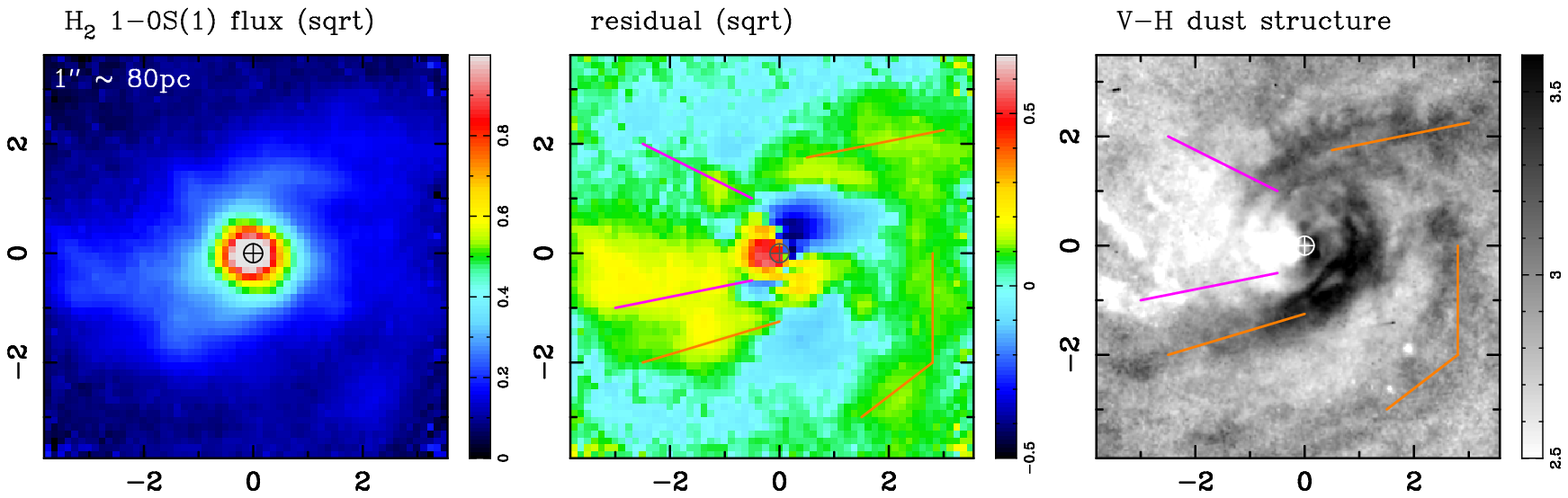}
\vspace{2mm}
\plotone{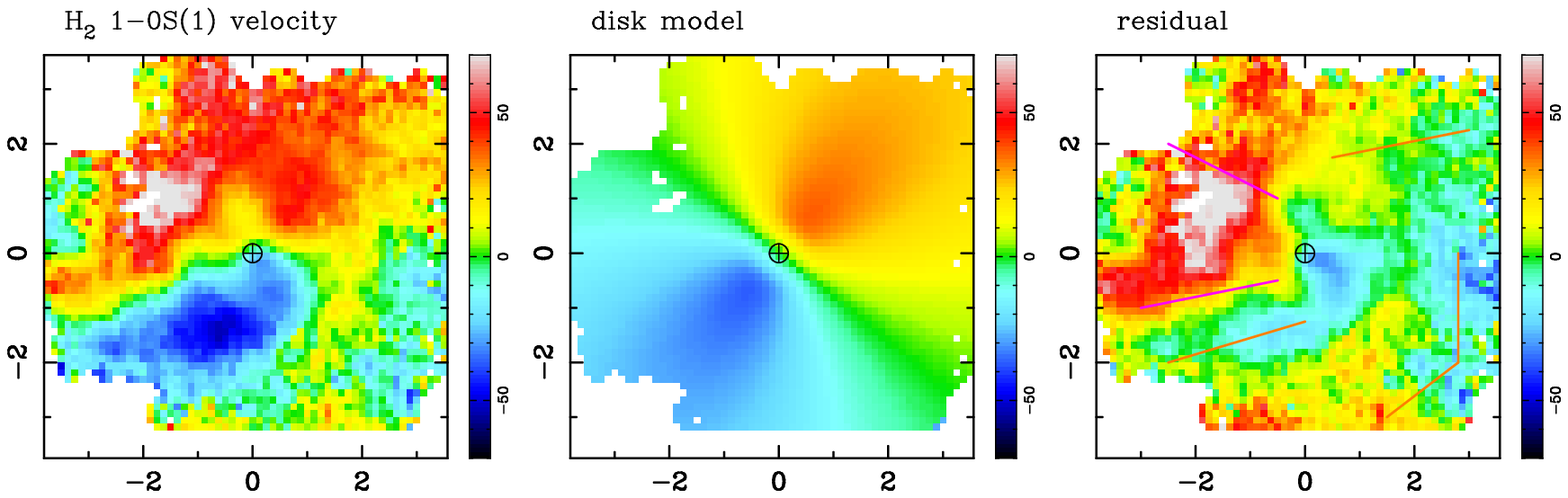}
\epsscale{0.245}
\plotone{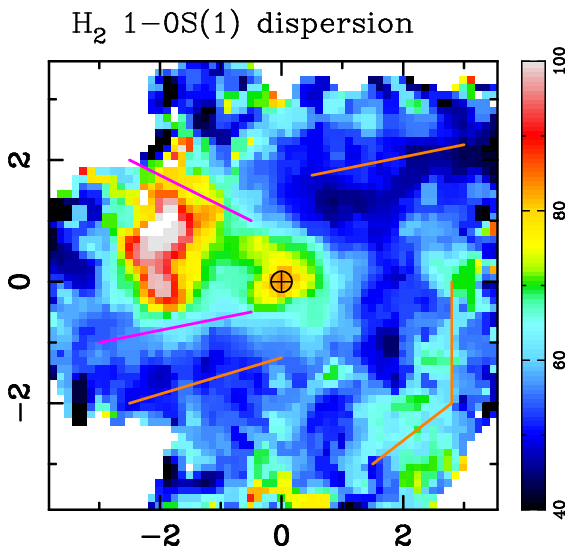}
\caption{\label{fig:ngc5643g}
Molecular gas as traced by the 1-0\,S(1) line in NGC\,5643, showing the same region as in Fig.~\ref{fig:starvelagn}.
Top row: observed flux distribution, and the residual flux after subtracting elliptical isophotes centered on the nucleus; the right panel shows the V-H dust structure map reproduced from \cite{mar03}.
Bottom row: observed velocity field, disk model, and the difference between them showing the strong velocity residuals in the central few arcsec. 
The rightmost panel is a map of the velocity dispersion.
Magenta lines denote the extent of the eastern outflow as measured from the 1-0\,S(1) dispersion and residual velocity maps; these also trace the outline of the low extinction region in the dust structure map.
The orange lines trace the 3 main absorption features in the dust structure map, although we note that the feature to the southwest may infact be associated with (opposite) outflow rather than inflow.
Axis scales are in arcsec, with a conversion to parsec given; north is up and east is left.}


\vspace{3mm}

\epsscale{0.50}
\plotone{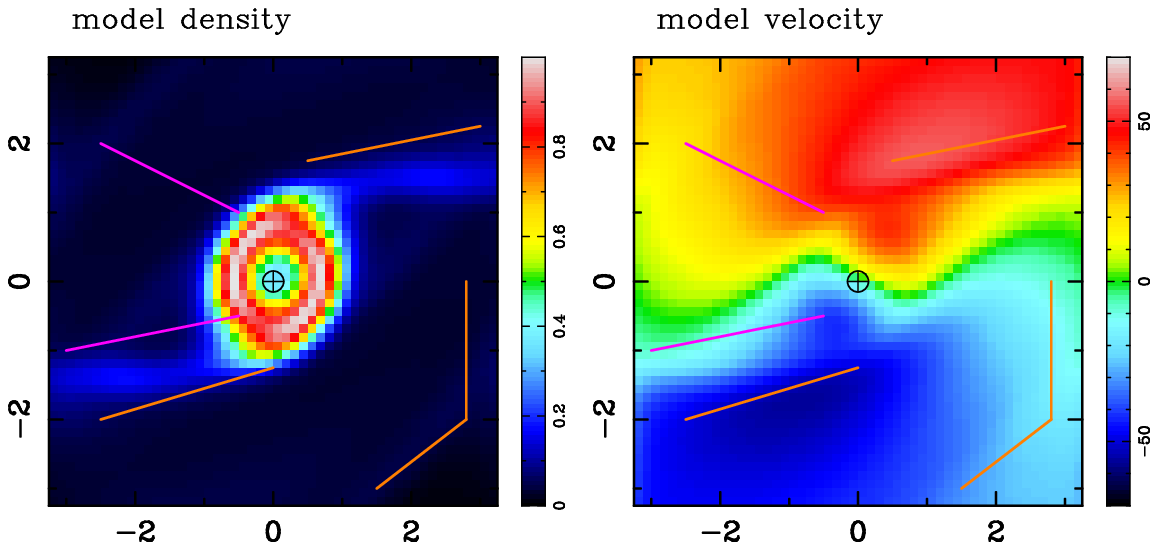}
\caption{\label{fig:ngc5643mod}
Density (left) and line of sight velocity (right) for a hydrodynamical simulation of a disk with a circumnuclear ring (spiral) generated by a large scale bar, matching model S20 from \cite{mac04}.
The bar was oriented 60\deg\ from the line of nodes in the disk plane, and the model has been set at an inclination matching NGC\,5643 as given in Table~\ref{tab:inc}, but at a PA of -30\deg, 10\degr{} different to that in Table~\ref{tab:pa}.
}

\vspace{3mm}

\epsscale{0.75}
\plotone{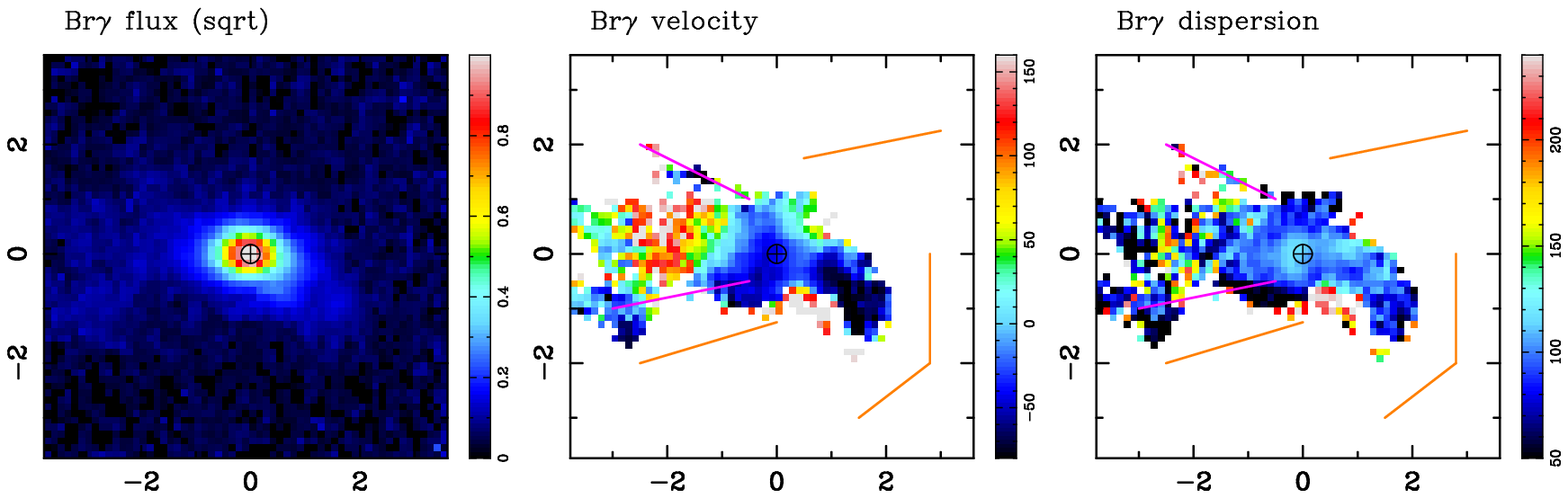}
\caption{\label{fig:ngc5643b}
Ionised gas as traced by the Br$\gamma$ line in NGC\,5643, 
showing the same region as shown previously.
Magenta and orange lines are as for Fig.~\ref{fig:ngc5643g}.
Axis scales are in arcsec; north is up and east is left.}
\end{figure*}

\subsubsection*{NGC\,5643}

The H$_2$ 1-0\,S(1) distribution and kinematics are rather complex in this object.
Because much of the field of view is dominated by non-circular motions, a disk model is hard to fit;
and to do so, the PA needs to be fixed beforehand.
As such, when subtracting the best fitting disk model, the kinematic structure is qualitatively little different from the direct velocity map.
On the other hand, the morphological structures do become much clearer when elliptical isophotes are subtracted from the flux map.

Fig.~\ref{fig:ngc5643g} confirms that NGC\,5643 has a 2-arm spiral which is clearly seen in both the dust structure map and the H$_2$ distribution.
In these regions only, the kinematics can be modelled by rotation with modest residuals.
As for NGC\,3227, in Fig.~\ref{fig:ngc5643mod}, we show the LOS velocity for model S20 from \cite{mac04} of gas flow in a bar for a geometry appropriate to NGC\,5643.
There is reasonable qualitative agreement with the observed velocity field, and also with the dust and molecular spiral structure.

About 2\arcsec\ to the north east of the nucleus, the kinematics are highly anomalous, with a redshift of 80\,km\,s$^{-1}$ and the dispersion reaching 100\,km\,s$^{-1}$.
At the same location, the dust structure map shows a lighter colour in a conical shape towards the east.
This indicates less dust, consistent with the reduced 1-0\,S(1) flux at the same location, suggesting there is little molecular gas in this region.
The simplest interpretation is in terms of outflowing molecular gas excited around the edge of an ionisation cone -- for which a sketch of the geometry is given in Fig.~22 of \cite{fis13}.
Such a structure extending up to 8\arcsec\ to the east, is already known from the H$\alpha$ and [O\,III] emission lines \citep{sim97}, as well as X-ray imaging at energies $<3$\,keV \citep{bia06}.
The Br$\gamma$ map in Fig.~\ref{fig:ngc5643b} also traces this ionisation cone, with velocities exceeding 150\,km\,s$^{-1}$ -- somewhat faster than the molecular gas.
Based on the [O\,III] image, and using H$\alpha$ and [O\,III] kinematics extracted from a single 0.2\arcsec\ wide slit at a PA of -128\deg, \cite{fis13} found that the eastern ionisation cone had an opening half-angle of 35--40\deg, and was oriented 65\deg\ away from the line-of-sight but only 42\deg\ away from the disk plane.
As such, it bisects the disk plane.
With this geometry, the emission in the 1-0\,S(1) line must trace the far side of the eastern ionisation cone, and plausibly originates from ambient molecular gas that was in the disk plane and has been swept up and pushed aside by the outflow.

NGC\,5643 has a third molecular structure to the west of the nucleus, extending for approximately 5\arcsec\ at a distance of 3\arcsec\ west and south of the nucleus.
This structure can be seen as a faint arm in the dust structure map.
Its presence could be attributed to a transient third arm, which structures have been seen in high resolution simulations of the circumnuclear region (Emsellem et al. in prep).
However, it is associated with a higher dispersion than the two arms comprising the grand design circumnuclear spiral.
As such, its presence suggests either that the central few hundred parsecs of this galaxy have recently been perturbed; or it could be tracing the far side of the outflow seen more prominently to the north-east.
Given both that the geometry of the ionisation cone suggests it is bisecting the disk, and that the Br$\gamma$ emission is extended in this direction opposite to the eastern outflow, interpreting the feature in terms of the second ionisation cone is appealing.
Based on the model of \cite{fis13} one would expect to see the near side of the backward-pointing cone, with moderate blueshifted velocities out to a distance of a few arcseconds.
The Br$\gamma$ emission is consistent with this.
And the H$_2$ 1-0\,S(1) could again trace ambient gas in the disk that is being swept up and heated by the outflow.

\begin{figure*}
\epsscale{0.75}
\plotone{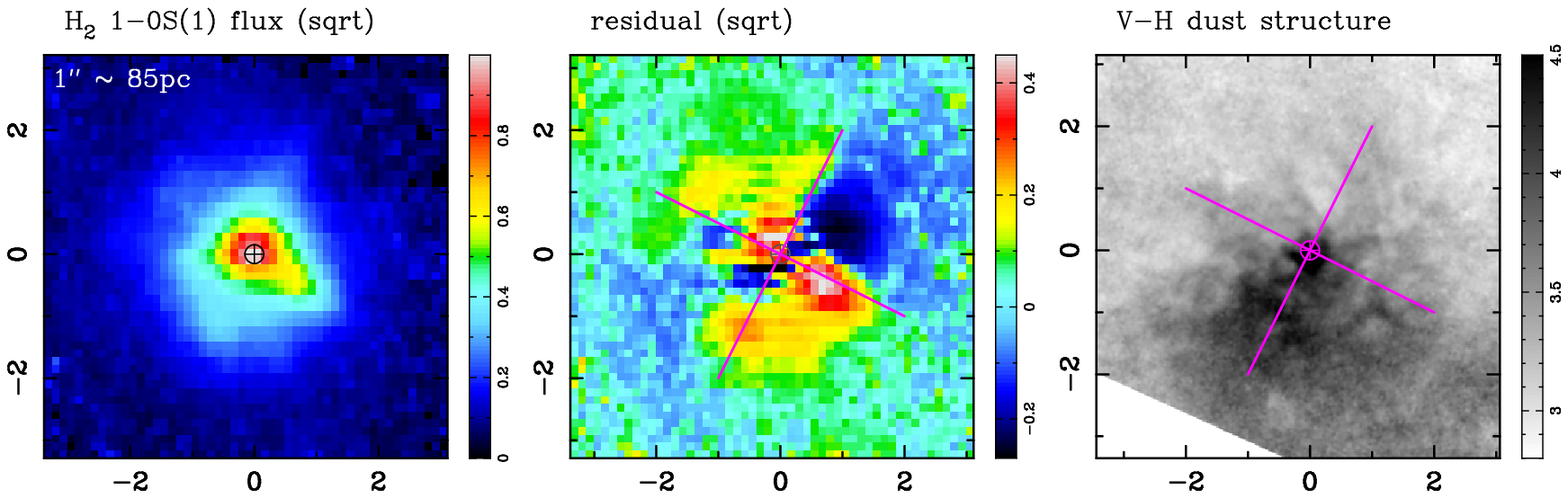}
\vspace{2mm}
\plotone{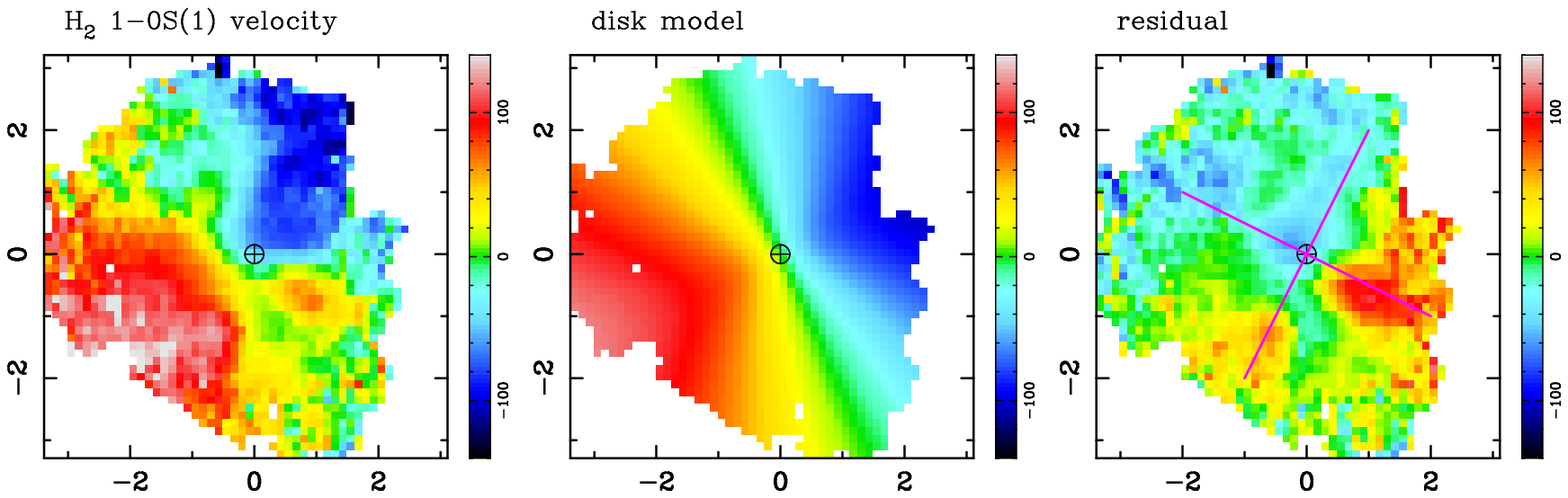}
\epsscale{0.245}
\plotone{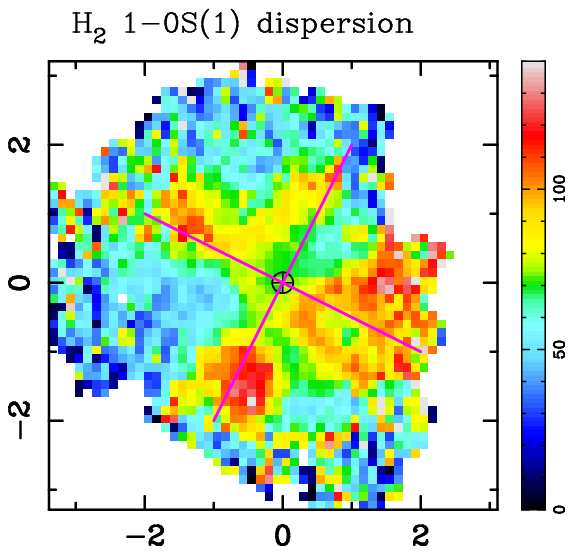}
\caption{\label{fig:ngc6300g}
Molecular gas as traced by the 1-0\,S(1) line in NGC\,6300, showing the same region as in Fig.~\ref{fig:starvelagn}.
Top row: observed flux distribution, and the residual flux after subtracting elliptical isophotes centered on the nucleus; the right panel shows the V-H dust structure map reproduced from \cite{mar03}.
Bottom row: observed velocity field, disk model, and the difference between them showing the strong velocity residuals in the central few arcsec. 
The rightmost panel is a map of the velocity dispersion.
The crossed magenta lines outline the prominent biconical feature seen in the flux residuals, that also trace regions of highest velocity residual, and also follow approximately the unusual X-shape in the dispersion map.
We note that only a very compact region (0.6\arcsec\ FWHM) of Br$\gamma$ emission was detected for this object, showing a velocity gradient along the same axis as the 1-0\,S(1) line. 
Axis scales are in arcsec, with a conversion to parsec given; north is up and east is left.}
\end{figure*}

\subsubsection*{NGC\,6300}

As for most of the other AGN in our sample, NGC\,6300 exhibits complex 1-0\,S(1) distribution and kinematics.
In this case, we argue that the simplest interpretation is an edge-on outflow superimposed on a rotating disk -- although we have detected at best only very weak Br$\gamma$ emission within the same features.

The residual flux map in Fig.~\ref{fig:ngc6300g} reveals a distinct biconical feature, aligned roughly north-south.
Especially to the south side, the flux is enhanced along the limbs of this structure.
On several of the panels in the figures, we have drawn a cross corresponding to it, which enables a better comparison of the locations of features between the various maps.
The velocity field shows a strong gradient, which is oriented similarly to that of the stars on the same spatial scales (shown in Fig.~\ref{fig:starvelagn}), as well as the large scale H$\alpha$ \citep{but01} and H\,I \citep{ryd96}: clear evidence for a circumnuclear molecular disk.
When subtracting a disk model, for which the PA was fixed to match that of the stars, the structures in the residual velocity field show a clear blueshift to the north and redshift to the south, following the arms of the cross.
A global gradient in the residual can be an indication that the PA of the disk model was incorrect.
However, when allowing this to vary freely during the fitting process, the structures and their respective velocities remain (although the scale of the residuals is reduced by a factor 2).
This suggests that they are real physical features, with a scale of 60--80\,km\,s$^{-1}$.
Outside the arms of the cross, as well as between them, the residuals are much smaller.
A similar $X$-structure is apparent in the 1-0\,S(1) velocity dispersion map, which shows significantly higher dispersion around 100\,km\,s$^{-1}$ along its arms compared to only 50\,km\,s$^{-1}$ between and around them.
The simplest interpretation that can simultaneously explain the flux and velocity residuals, as well as the dispersion map, is that they trace the edge of a biconical molecular outflow oriented roughly north-south and seen close to edge-on, with the northern side tilted slightly towards us.
The edges of the bicone are brighter because the column of hot gas along those lines of sight is greater; and the dispersion is higher in the same locations because of the range of velocities the gas is tracing.
For NGC\,6300 we conclude that there is a circumnuclear molecular disk; 
but we cannot say if there is inflow because the non-circular velocities are dominated by signatures of outflow.

In contrast to the other AGN with outflows in our sample, there does not appear to be any extreme reduction in the dust content in the outflow. 
Nevertheless, the V-H dust structure map does show that the northern side is associated with some reduction in obscuration, suggesting that the outflow is in front of the galaxy, consistent with its blueshifted residual velocity.

Our interpretation is consistent with the classification of the [O\,III] emission by \cite{fis13} as compact, because their slit was oriented at PA=90\deg, roughly perpendicular to the projected axis of the outflow.
We note finally that, if the gas kinematics in the centre of the galaxy are dominated by outflow rather than rotation, this might explain why the line-of-sight velocity of ionised gas in the nucleus differs from the galaxy's systemic velocity by 100\,km\,s$^{-1}$ \citep{but01}.

\begin{figure*}
\epsscale{0.75}
\plotone{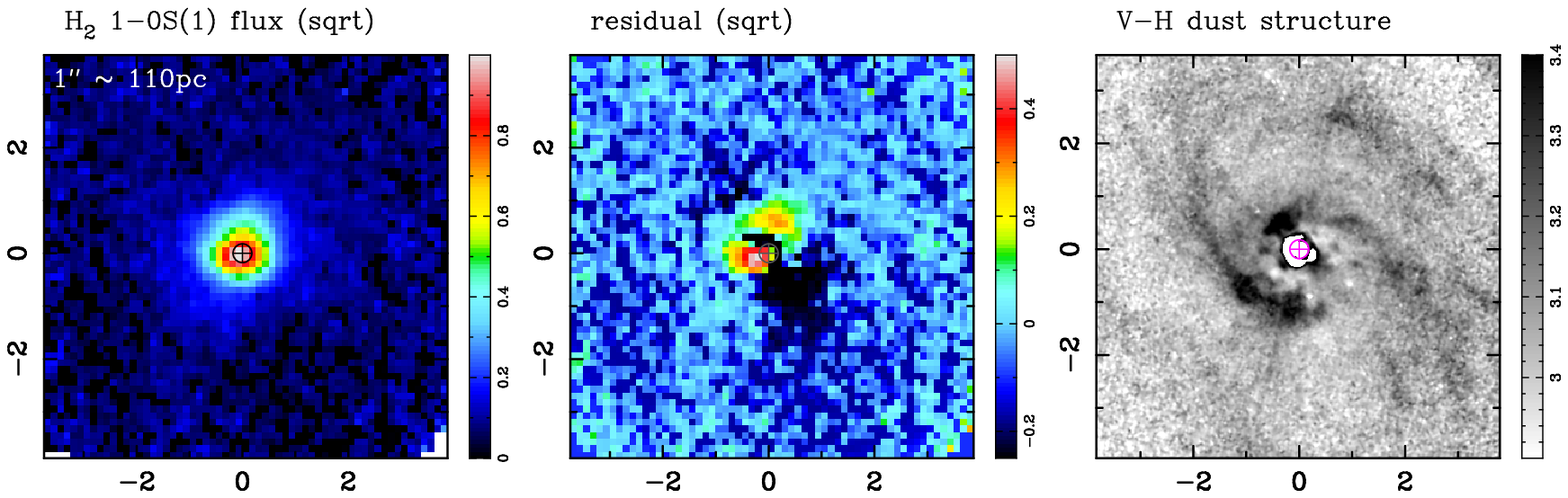}
\vspace{2mm}
\plotone{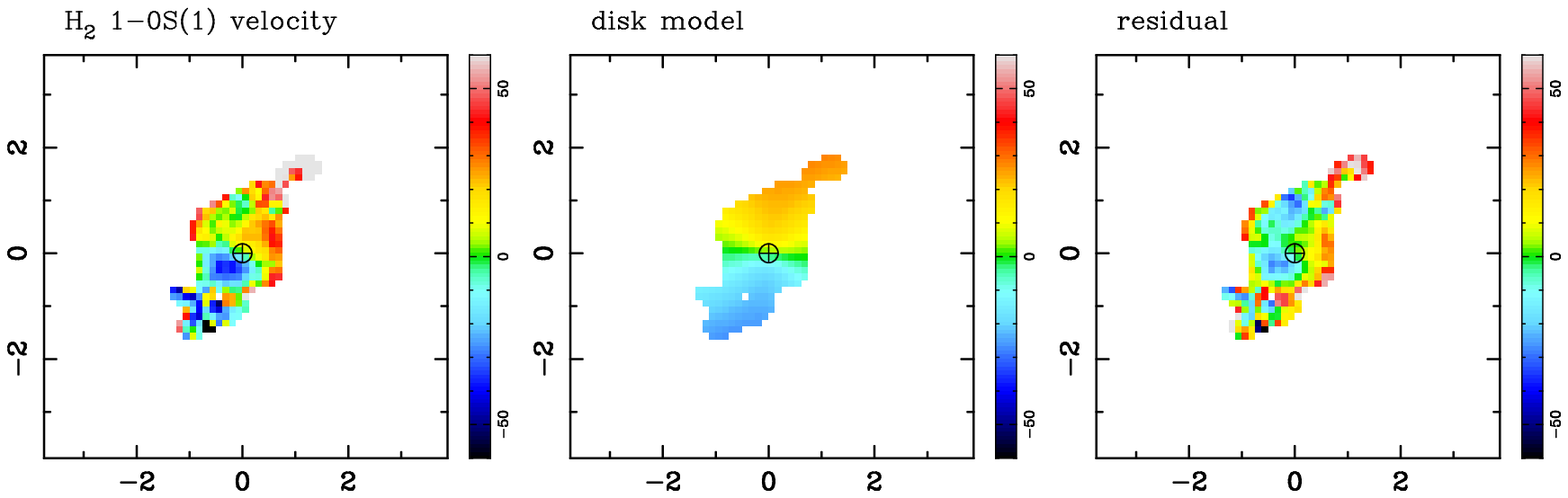}
\epsscale{0.245}
\plotone{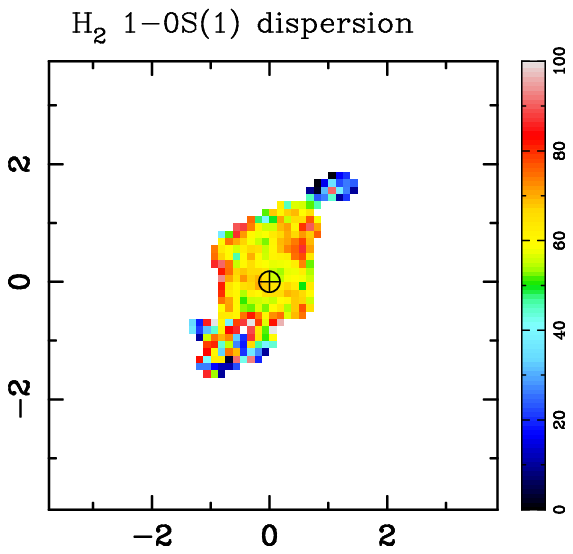}
\caption{\label{fig:ngc6814g}
Molecular gas as traced by the 1-0\,S(1) line in NGC\,6814, showing the same region as in Fig.~\ref{fig:starvelagn}.
Top row: observed flux distribution, and the residual flux after subtracting elliptical isophotes centered on the nucleus; the right panel shows the V-H dust structure map reproduced from \cite{mar03}.
Bottom row: observed velocity field, disk model, and the difference between them showing the strong velocity residuals in the central few arcsec. 
The rightmost panel is a map of the velocity dispersion.
We note that for this object, the Br$\gamma$ flux extracted from the same datacube has already been presented and analysed by \cite{mue11}.
Axis scales are in arcsec, with a conversion to parsec given; north is up and east is left.}
\end{figure*}

\subsubsection*{NGC\,6814}

In contrast to the other AGN in this sample which have extended structures, the H$_2$ flux in NGC\,6814 is rather compact, as can be seen in Fig.~\ref{fig:ngc6814g}.
The adaptive optics data analysed by \cite{hic09} and \cite{mue11} show that the intensity falls to half within a radius of 0.25\arcsec.
The gradient in the velocity field presented by these authors -- from south-east to north-west -- is matched by the seeing limited data presented here.
\cite{mue11} analysed the seeing-limited Br$\gamma$ and [SiVI] maps extracted from the same datacube, modelling their velocity fields in terms of a rotating disk with a velocity gradient oriented at $-35$\deg\ and a biconical outflow aligned towards a PA of $+34$\deg.
While the PA of the disk in this model appears different to that of the stellar kinematics on the same scales, discussed in Sec.~\ref{sec:stellardiskfits}, the small inclination of the galaxy means that the PA has a large uncertainty.
Similarly, while longslit spectroscopy on larger scales \citep{mar04} suggests that the steeper velocity gradient is at a PA of 30\deg\ rather than -30\deg, the difference is much less clear -- possibly even reversed -- on smaller scales. A slight warp of the nearly face-on disk could easily account for this change.
Neither the available H\,I \citep{ho08} or CO\,1-0 data \citep{cur01} are able to shed further light on this because of insufficient spatial resolution.

The 1-0\,S(1) velocity gradient is consistent with the $-35$\deg\ PA of the disk component modelled by \cite{mue11}, but also with our adopted $-6$\deg\ PA which was used to fit the stellar kinematics.
Due to the low flux levels beyond the central compact emission region, the data can be fit equally well with a rotating disk at either of these PAs.
As such, we cannot draw any conclusion about whether there are indications of molecular inflow or outflow.

\begin{figure*}
\epsscale{0.75}
\plotone{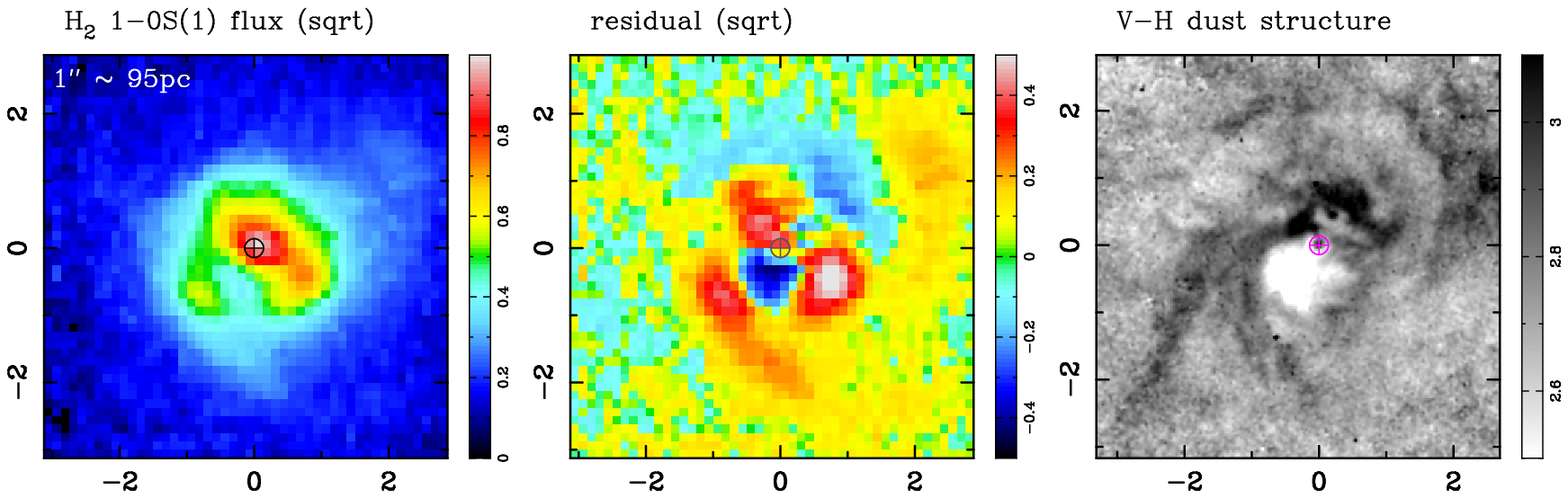}
\vspace{2mm}
\plotone{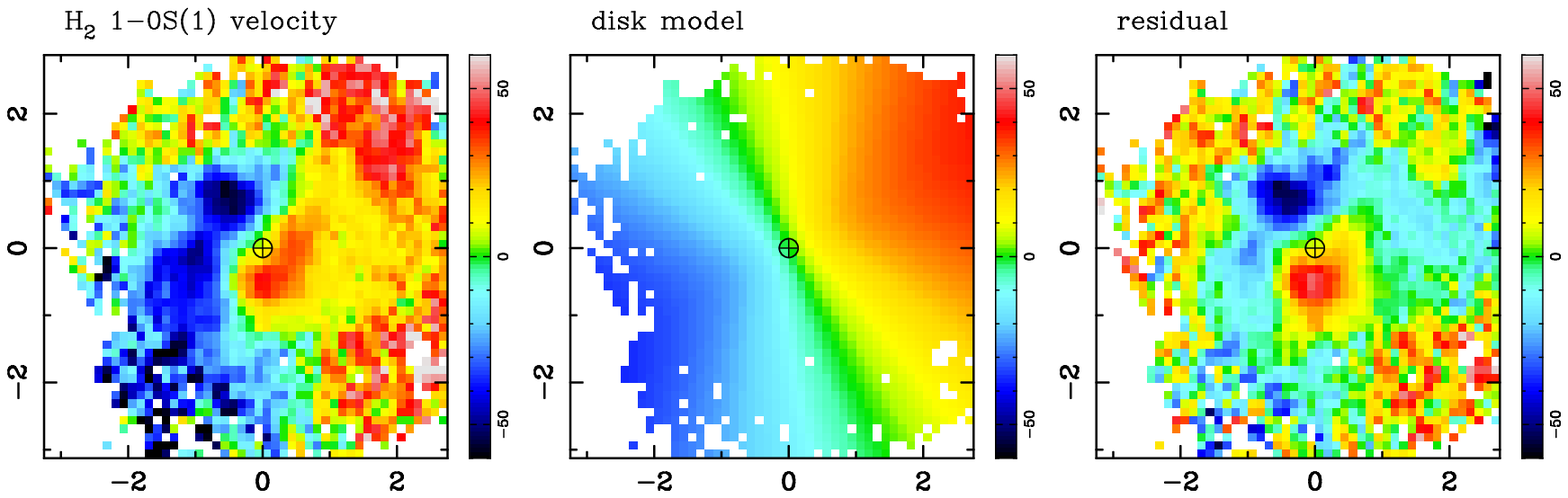}
\epsscale{0.245}
\plotone{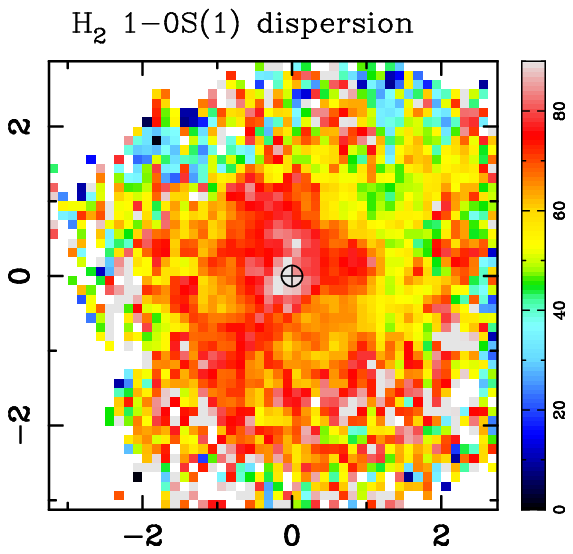}
\caption{\label{fig:ngc7743g}
Molecular gas as traced by the 1-0\,S(1) line in NGC\,7743, showing the same region as in Fig.~\ref{fig:starvelagn}.
Top row: observed flux distribution, and the residual flux after subtracting elliptical isophotes centered on the nucleus; the right panel shows the V-H dust structure map reproduced from \cite{mar03}.
Bottom row: observed velocity field, disk model, and the difference between them showing the strong velocity residuals with an arcsec of the nucleus.
The rightmost panel is a map of the velocity dispersion.
The red and blueshifted velocity residuals lie along the minor axis, and neither are spatially coincident with excess 1-0\,S(1) flux.
In contrast, the redshifted residual is associated with a `hole' in the 1-0\,S(1) flux map.
The Br$\gamma$ map for this object shows only very weak emission located within an arcsec north of the nucleus.
Axis scales are in arcsec, with a conversion to parsec given; north is up and east is left.}
\end{figure*}

\begin{figure}
\epsscale{1.0}
\plotone{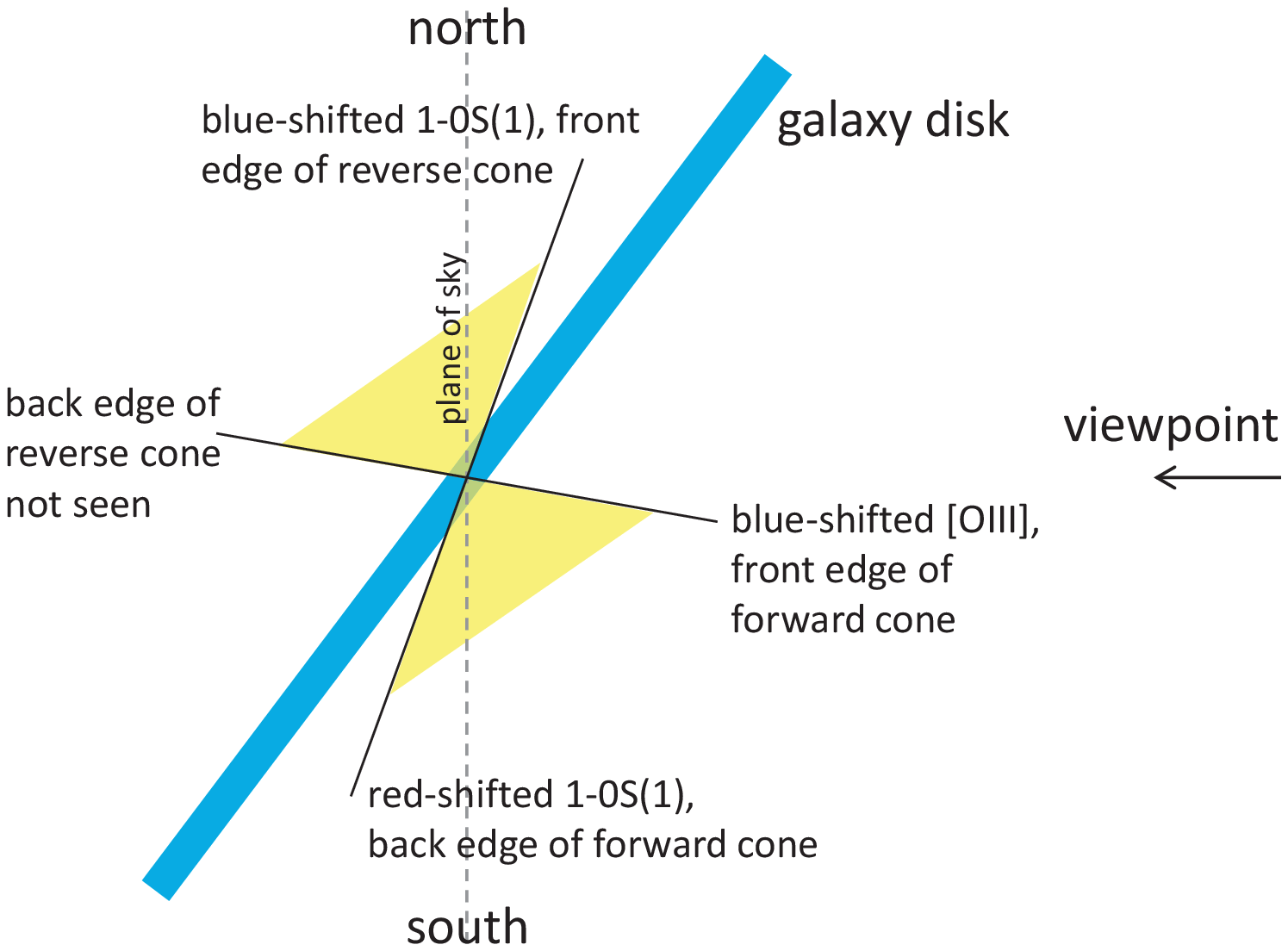}
\caption{\label{fig:ngc7743sketch}
Sketch of our proposed geometry for the outflow in NGC\,7743, as described in the text.}
\end{figure}

\subsubsection*{NGC\,7743}

The detailed structures in the 1-0\,S(1) flux map become clearer once elliptical isophotes have been subtracted. As Fig.~\ref{fig:ngc7743g} shows, this has some resemblance to the $V-H$ dust structure map from which \cite{mar03} classified the circumnuclear region of NGC\,7743 as a loosely wound spiral. Two arm-like structures, one curving towards the north-west and the other curving round to the south, are apparent in both images.
The only feature not traced by the H$_2$ emission is the arm that continues almost straight out to the south-east. Outside this spiral structure, \cite{moi04} report two dust lanes in the bar of NGC\,7743.

The velocity field also has significant structure and is dominated by non-circular motions. Two spiral arms (one blue and one red) can be traced in the velocity field: the blue arm winds from about PA$\sim0$\deg\ at small radii to PA$\sim180$\deg\ at the outskirts of the FOV, while the red arm is not so well defined. 
Such a structure should accompany a density wave of a 2-arm morphology \citep[e.g.][]{dav09}, and it is consistent with inflow driven by a circumnuclear spiral, often present inside the straight shocks in the bar, and driven by the bar \citep[e.g.][]{mac02,mac04}.

However, within the inner 1\arcsec, the line-of-sight gas velocity along the LON ($PA=-65\deg$) is almost zero, which indicates that the tangential velocity of the gas there is almost zero. Thus the velocity field at that location is dominated by a radial flow (because gas with zero tangential velocity has no angular momentum). This is inconsistent with gas flow in a bar or in a nuclear spiral, where the radial velocity component occurs in addition to the always significant tangential velocity. Moreover, within the inner 1\arcsec\ residual velocities no longer form the classical 1-arm spiral structure expected for a 2-arm photometric spiral \citep[e.g.][]{mac04,dav09,schn13}, but are instead consistent with a signature of radial flow. Thus, while at radii $\gtrsim$1\arcsec\ the dust structure and the gas kinematics are consistent with inflow in a nuclear spiral, the gradient of the line-of-sight velocity between $+50$ to $-50$\,km\,s$^{-1}$ within $\sim$1\arcsec\ of the nucleus along the minor axis is most likely caused by a radial flow.

This radial flow can either be isotropic in the plane of the galaxy disk, or can be out of the disk plane. In order to assess whether this is due to inflow or outflow, we must consider which are the near and far sides of the galaxy. In optical images of the galaxy, the spiral arms on large scales in the disk wind, from the centre outwards, in an anti-clockwise direction on sky \citep{moi04}. The circumnuclear spiral structure is wound in the same sense. Since the velocity field implies the approaching side is to the east, for the spiral arms to be trailing, the north side of the galaxy must be the near side. Thus, we can state that, if the redshifted residual to the south and the blueshifted residual to the north arise from material in the disk plane, then this material must be flowing outwards. 

That these features trace outflowing gas is also true if the material is in front of the disk to the north, and behind the disk to the south, or if it is in a feature inclined to the galaxy disk by less than 37\deg (the inclination of NGC\,7743) in such a way that the blueshifted feature is behind the galaxy disk and the redshifted feature is in front of it. This is consistent with the interpretation of the feature present in the residual velocity field in terms of an AGN driven outflow.

An outflow would be consistent with the high dispersion of $\sim70$\,km\,s$^{-1}$ in these regions.
But perhaps stronger support for this interpretation comes from the [O\,III] line, which is typically taken to be excited by an AGN and trace its ionisation cone.
In NGC\,7743, [O\,III] emission is extended along the minor axis, and shows a markedly different velocity gradient to the stars \citep{moi04,kat11}. Indeed, to the north side at a PA of 358\deg, the [O\,III] line is blue shifted, while to the south it is redshifted with respect to the stars (see Fig.~13 of \citealt{moi04} and top right panel of Fig.~2 in \citealt{kat11}). While the H$\beta$ kinematics follow the stars on scales $>$10\arcsec, at $<$5\arcsec\ they instead match those of [O\,III].
Remarkably, \cite{kat11} also note that the highest dispersion in the [O\,III] line is 1--2\arcsec\ south of the nucleus.
Using their high spectral resolution spectra, they show that here the line comprises two components, one of which is blueshifted by 300\,km\,s$^{-1}$ with respect to the primary component.
The spatial location of this component coincides with the location of the white region in the dust structure map, which shows there is almost no dust.
As is apparent for the other AGN in our sample, regions of low extinction and low molecular gas content such as this are typically associated with outflows. We note that low extinction in the cone has been observed for other AGN, e.g. NGC\,4945 \citep{mor96,mar00}. 
Putting these facts together in the context of the conical outflow models employed by \cite{fis13}, we conclude that to the south of the nucleus, the outflow is in front of the galaxy disk.
As illustrated in Fig.~\ref{fig:ngc7743sketch}, in this context:
\begin{enumerate}

\item[(i)] the blueshifted [O\,III] emission close to, but south of, the nucleus traces an edge of the cone that is oriented almost directly towards us but still blocks the direct line-of-sight to the nucleus;
\item[(ii)] the redshifted 1-0\,S(1) and [O\,III] emission south of the nucleus traces the back-edge of the cone which is tilted behind the plane of the sky, but may still be in front of the disk; we see H$_2$ emission from this edge because it is close to the galaxy disk where there is a supply of ambient molecular gas;
\item[(iii)] the blueshifted [O\,III] and 1-0\,S(1) emission to the north traces the near side of the opposite (backward-pointing) cone, which lies behind the galaxy disk but is still tilted forward from the plane of the sky;
\item[(iv)] we do not see the counter-part to (i) above because it is obscured behind the central regions of the galaxy.
\end{enumerate}
In this interpretation of the various emission components, the opening half-angle of the cone must be large, probably in the range 40--60\deg, so that one edge is angled towards us while the other is close to the galaxy disk.
We note that the 3D modelling of outflow geometries by \cite{mue11} and \cite{fis13} suggests that opening angles this wide may not be untypical.
The interpretation is also consistent with the classification of this AGN as a Seyfert~2, for which one should expect an outflow to be tilted away from our line of sight. 
This then requires the outflow to be strongly tilted with respect to the normal to the disk plane.
Such orientations may be common: there is thought to be little or no relation between host and outflow orientation, and \cite{fis13} found a significant number that were tilted by 30--50$^\circ$ from the normal to the disk.

\subsection{Inactive Sample}

Three of these objects (NGC\,4030, NGC\,628, NGC\,357) have no 1-0\,S(1) line emission detected, so there are only two objects for which we can study the H$_2$ distribution and kinematics. That both of these show severe disturbances is perhaps unexpected.
Inflow of gas from external streamers is the most promising explanation, and only a small perturbation is now required to trigger a major molecular gas inflow event and turn these objects into AGN.

\begin{figure*}
\epsscale{0.5}
\plotone{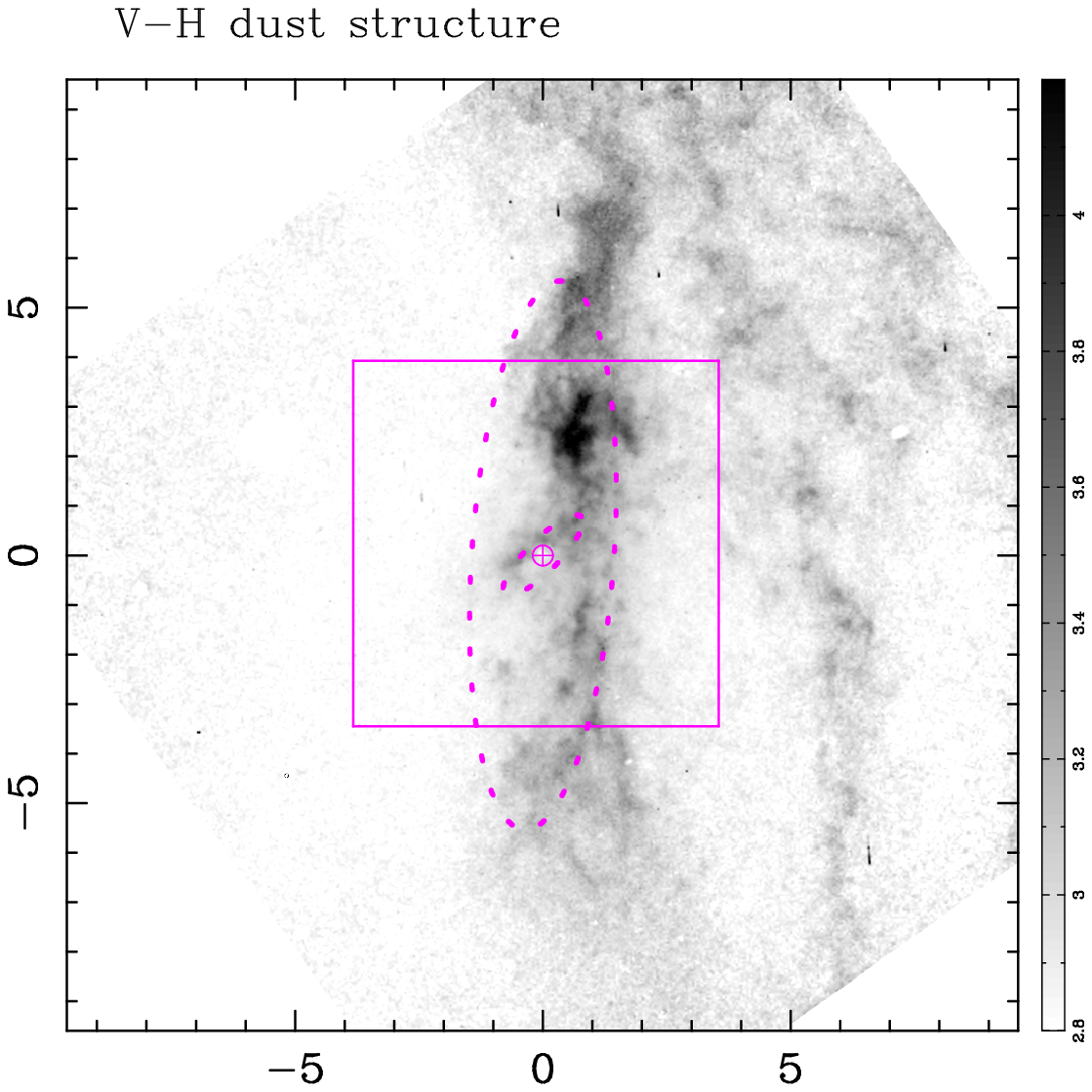}
\caption{\label{fig:ic5267d}
Dust structure (V-H) map of IC\,5267 reproduced from \cite{mar03}. The field of view of SINFONI, shown in the corresponding Fig.~\ref{fig:ic5267g}, is marked. In addition, the same ellipses as shown in some panels of that figure are overdrawn with dotted lines.
Axis scales are in arcsec; north is up and east is left.}

\vspace{3mm}

\epsscale{0.75}
\plotone{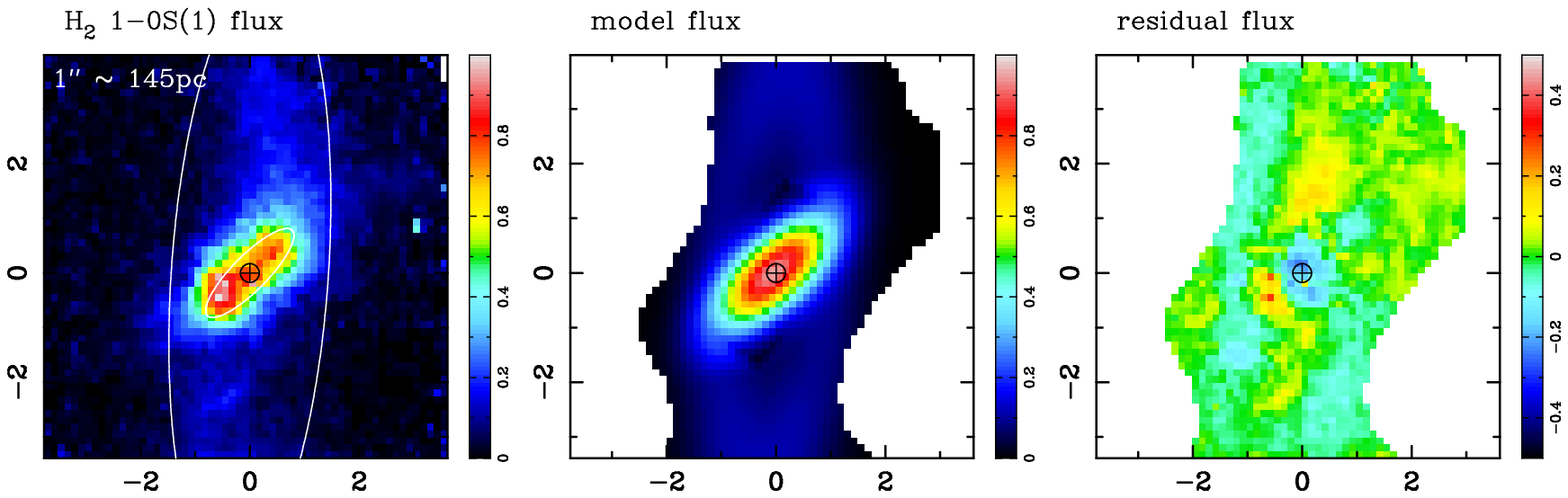}
\epsscale{0.245}
\plotone{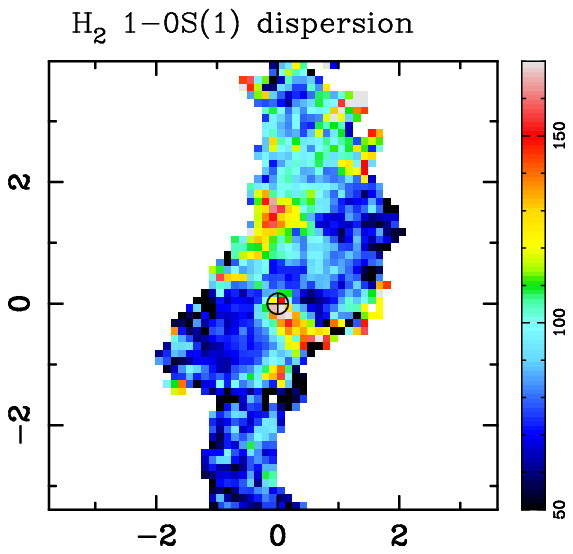}
\vspace{2mm}
\epsscale{0.245}
\plotone{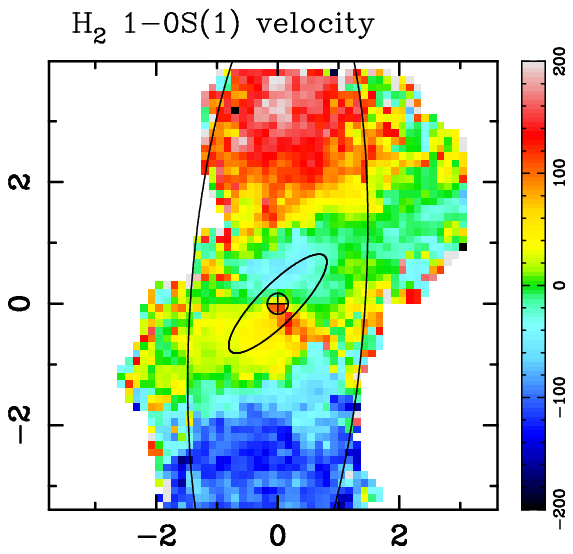}
\epsscale{0.75}
\plotone{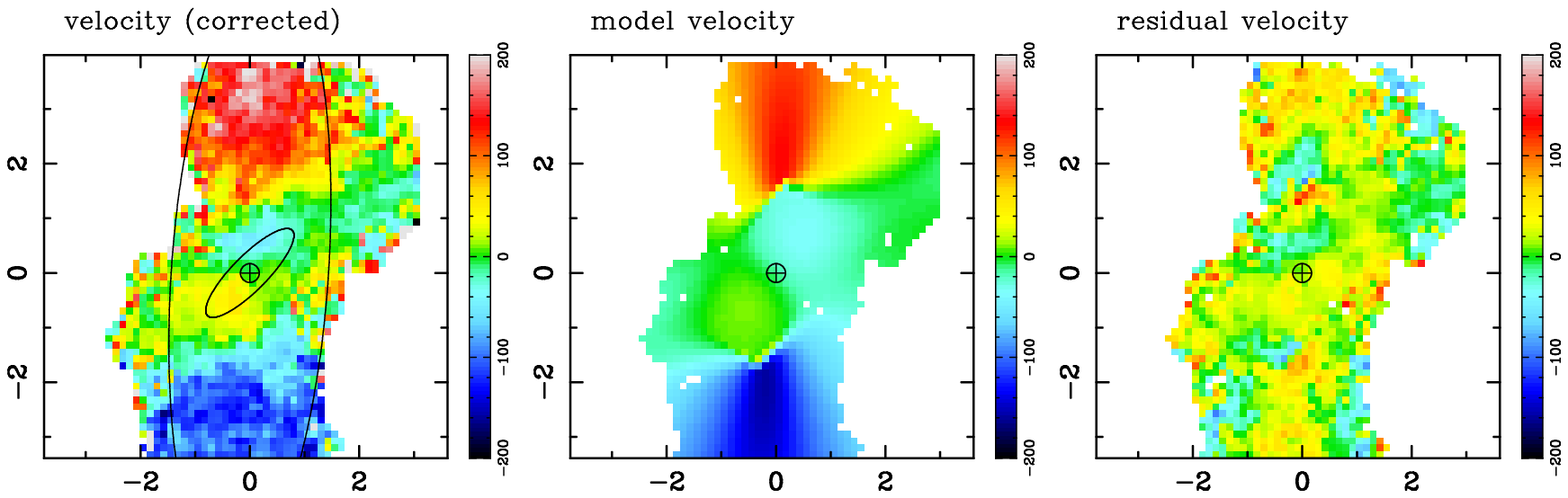}
\caption{\label{fig:ic5267g}
Molecular gas as traced by the 1-0\,S(1) line in IC\,5267.
Top row: observed flux distribution, the model comprising 2 rings, and the difference between them (note that some of the differences may be due to the model being axisymmetric while the illumination of the rings may vary between locations). On the far right is shown the velocity dispersion.
Bottom row: observed velocity field, before and after correcting for the double peaked line profile causing the finger of anomalous velocities extending south-west from the nucleus (see text for details). On the right side are the 2-ring model, and the residual velocity after subtracting this model.
The two ellipses represent the 2 rings (drawn corresponding to a size of Radius+FWHM/2, as given in Tab.~\ref{tab:ic5267}, to denote the full extent of the two structures).
Axis scales are in arcsec, with a conversion to parsec given; north is up and east is left.}
\end{figure*}

\subsubsection*{IC\,5267}

\begin{deluxetable*}{cccccccc}
\tablecaption{Properties of the two rings in IC\,5267\label{tab:ic5267}}
\tablehead{
\colhead{Radius} &
\colhead{FWHM} &
\colhead{Rel.} &
\colhead{PA} &
\colhead{i} &
\colhead{$M_{enc}$} &
\colhead{$V_{rot}$} &
\colhead{$\tau_{orb}$}\\
\colhead{(\arcsec)} &
\colhead{(\arcsec)} &
\colhead{Int.} &
\colhead{($^\circ$ E\,of\,N)} &
\colhead{($^\circ$)} &
\colhead{($M_\odot$)} &
\colhead{(km\,s$^{-1}$)} &
\colhead{(Myr)}\\
}
\startdata
3.9\phm{\tablenotemark{a}} & 3.3 & 1.0 & \phn\,-4 & 75 & $2.5\times10^9$ & 139 & 25 \\
0.2\tablenotemark{a} & 1.8 & 6.2 & 135      & 73 & $3.0\times10^7$ & \phn66 & \phn3 \\
\enddata

\tablenotetext{a}{Because the fit yields a radius of the ring smaller than its width, this can be considered a disk with a rather flat inner radial profile, extending out to 1.1\arcsec.}

\end{deluxetable*}

The H$_2$ distribution and kinematics in the central few hundred parsecs, shown in Fig.~\ref{fig:ic5267g}, appear totally unrelated to the stars, which exhibit no measurable velocity gradient \citepalias{hic13}. On scales of a few arcsec, the warm molecular gas lies in a narrow north-south band with a strong velocity gradient, spatially coincident with the band of extinction in the dust structure map of \cite{mar03}, which is reproduced in Fig.~\ref{fig:ic5267d}. This band is part of a larger dust structure most likely located in front of the galaxy disk \citep{mul97}. The much more intense H$_2$ emission within $\sim1$\arcsec\ of the nucleus comes from another elongated structure, separated from the north-south band by an abrupt change of position angle, and having a velocity gradient with the opposite sense (redshifted gas is towards the south rather than the north). This inner H$_2$ feature is spatially coincident with a much fainter dust lane that approximately shares its extent and PA. 
These kinematics are consistent with a possible moderate ($<$10\,km\,s$^{-1}$) counter-rotation observed in stars \citep{mor08}, and a stronger (30\,km\,s$^{-1}$) counter-rotation in ionised gas (L.~Morelli, priv. comm.).
If the system were close to face-on, the observed kinematics could be caused by a modest warp, so that the disk's outer part is tilted slightly behind the plane of the sky and its inner part slightly in front along the same radial direction (see e.g. \citealt{eng10,dia13}). However, the narrow dust lane in the north-south direction indicates that at least part of the system is close to edge-on, which is inconsistent with the warp scenario.

The most natural interpretation is that the system consists of at least 2 separate components: one is the north-south dust lane, together with the associated H$_2$ emission extending across the full SINFONI field, and the other is the inner H$_2$ emission of distinctly different kinematics, associated with the weaker dust lane. The increased gas dispersion at the boundary between these two components is then caused by the two components overlapping there. The first component is likely part of a streamer in front of the galaxy. If we assume that this streamer is edge-on and orbiting the galaxy nucleus, we can estimate the maximum radial distance at which it can be in order to yield the observed projected velocity of $\pm$150\,km\,s$^{-1}$ at an offset of $\pm$3\arcsec\ (440\,pc) in projection. Since the H\,I line profile has a full width of 400\,km\,s$^{-1}$ \citep{dri88}, the maximum rotation velocity is 300\,km\,s$^{-1}$ for an inclination of 42$^\circ$. The maximum distance of the streamer from the center is then $\sim$900\,pc.
The second component is a disk or a ring that extends radially to about 250\,pc, with a PA and inclination that are different to the streamer, but closer to those of the galaxy disk on large scales.
Such a scenario, with gas orbiting in 2 different planes, can only be quasi-stable and would not require much to perturb it.

In order to test this interpretation further by attempting to reproduce the observed 1-0\,S(1) flux map and velocity field, we model this streamer, and also the more compact structure in the 1-0\,S(1) map, as two rings of gas. Using DYSMAL, we have created a model, shown in Fig.~\ref{fig:ic5267g}, which comprises of two massless rings with circular orbits.
The best fitting model was obtained by varying the enclosed mass together with the ring radius, width, PA and inclination for both rings in order to obtain the best fit to the flux and velocity maps for all pixels above a given 1-0\,S(1) flux threshold (the extent of the mask can be traced in the figure). The parameters of the best fitting model are summarised in Table~\ref{tab:ic5267}, and indicate that both rings are close to edge-on but nearly counter-rotating, as they differ in PA by 140\deg. In this model, the outer ring is broad, extending from 330\,pc to 810\,pc. The radius of the inner ring is formally 30\,pc, but its large FWHM indicates that it is effectively a disk, whose intensity drops to half of the maximum value at 160\,pc, and is also slightly less than the maximum at radii smaller than 30\,pc. Thus the inner component has a relatively flat inner luminosity distribution (we have modelled it as a ring rather than an axisymmetric centrally peaked disk only to better match the dip in 1-0\,S(1) flux at the centre). The enclosed masses are well-defined. The inner ring yields an upper limit (excluding the impact of dispersion) on the mass of a central black hole of $3\times10^7$\,M$_\odot$. 
This is more than 1.2 dex below the M$_{BH}-\sigma$ relation, given the stellar velocity dispersion in the bulge of 200\,km\,s$^{-1}$ \citep{mor08}. 
The outer ring suggests there is a considerable mass of stars in the central few hundred parsecs. 
There is a chance for gas clouds to collide where the orbital planes of the two structures intersect.
This will occur if either the gas in the rings diffuses inwards or outwards, or if there is cool gas already distributed through the orbital planes that we cannot trace with the 1-0\,S(1) line.
If a sufficient mass of gas begins then begins to fall inwards, the galaxy would become an AGN -- within a relatively short time given the orbital timescales for both rings.

One obvious discrepancy between the observed and modelled velocity fields, is the zero velocity line (ZVL) of the inner structure: in the data this is at a PA of 90$^\circ$ while the model puts it at $45^\circ$ matching the $135^\circ$ orientation of the LON. One way to account for this difference is to assume that a radial flow is superimposed on the circular motion. However, matching the observed ZVL and LON requires outflow at a velocity of 0.3 times the rotation speed. Alternatively, the difference may be the result of gas moving on elliptical orbits (see the Appendix of \citealt{maz14}) instead of being in circular motion in the inner structure. For the observed PA of the flux distribution in the inner structure of 135\deg, the observed axial ratio of the structure of about 0.3, and the PA of the ZVL of 90\deg, only one PA of the LON is possible: PA=140$^\circ$. This is exactly the PA of the large-scale galactic disk. The inclination is unconstrained and is tied only to the intrinsic ellipticity of the orbits. If we assume that the inner structure is in the galactic disk, then at $i=40$$^\circ$, we get an intrinsic axis ratio b/a=0.4, implying a rather eccentric flow. Moreover, such gas in the inner structure being in the disk plane would rotate in the opposite direction than the large-scale H\,I \citep{dri88}. On the other hand, the elliptical flow is roundest, with b/a=0.77, when the inclination is close to $i=73$$^\circ$, which is the inclination of the inner ring in the model above that assumes circular flow.

One final anomaly in the 1-0\,S(1) data is the finger of high velocity extending about 1\arcsec\ to the south-west from the nucleus. This feature appears also in the dispersion map, and is the result of a second peak in the emission line, redshifted by about 200\,km\,s$^{-1}$ with respect to systemic. Separating the two components in the flux and velocity maps (Fig.~\ref{fig:ic5267g}, bottom panels) suggests this extra feature is spatially unresolved and centered close to the nucleus. Thus there may be a third component in our data, located within 150\,pc of the nucleus (in projection), and moving at 200\,km\,s$^{-1}$ relative to it.

In summary, these two discrepancies (the position angle of the ZVL, and the finger of high dispersion) shed light on the details of the inner disk structure in IC\,5267, but do not detract from the interpretation that the molecular gas is confined to 2 independent planes rotating at different orientations.

\begin{figure*}
\epsscale{0.75}
\plotone{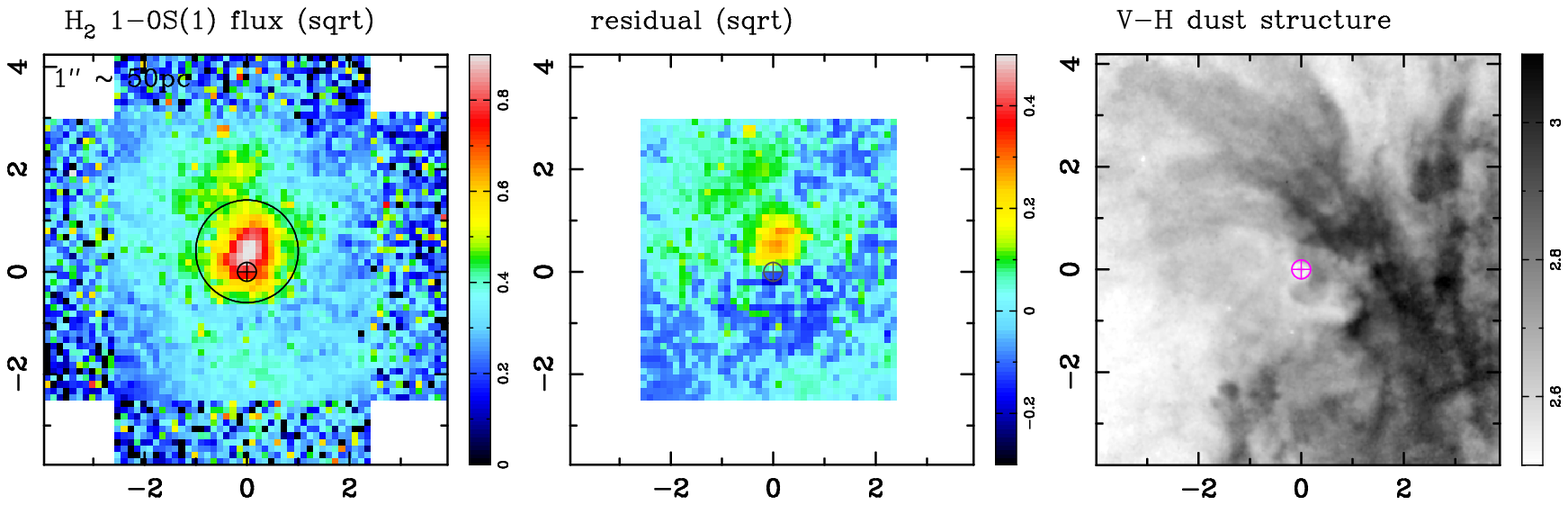}
\vspace{2mm}
\plotone{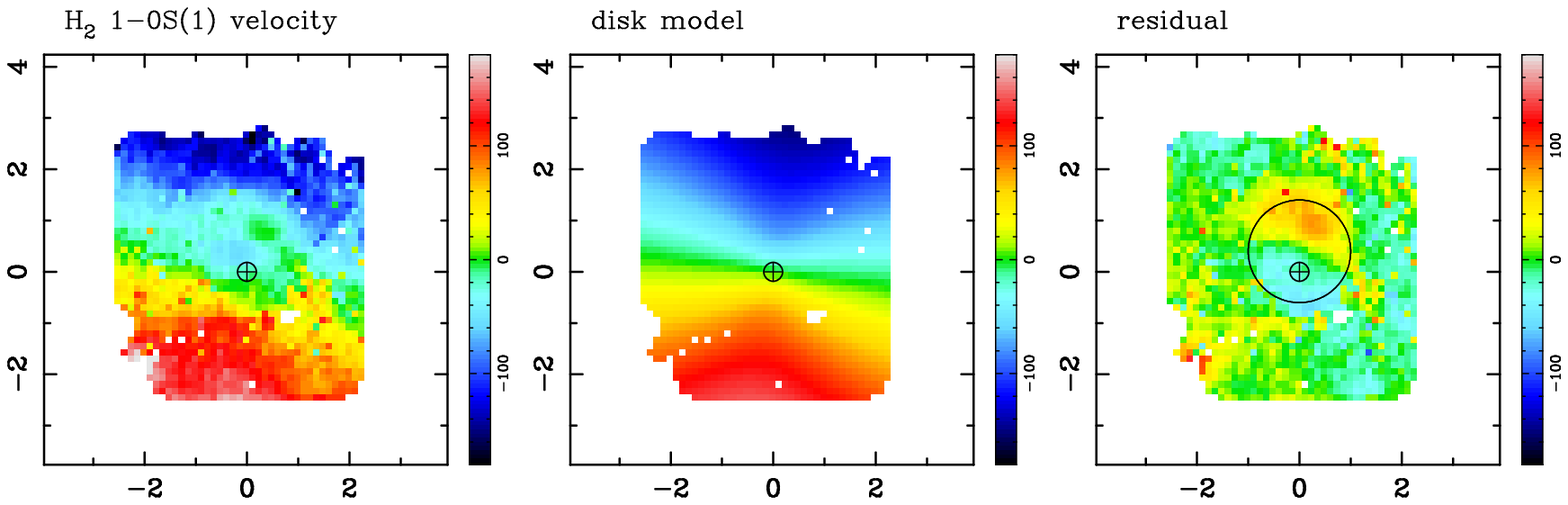}
\vspace{2mm}
\plotone{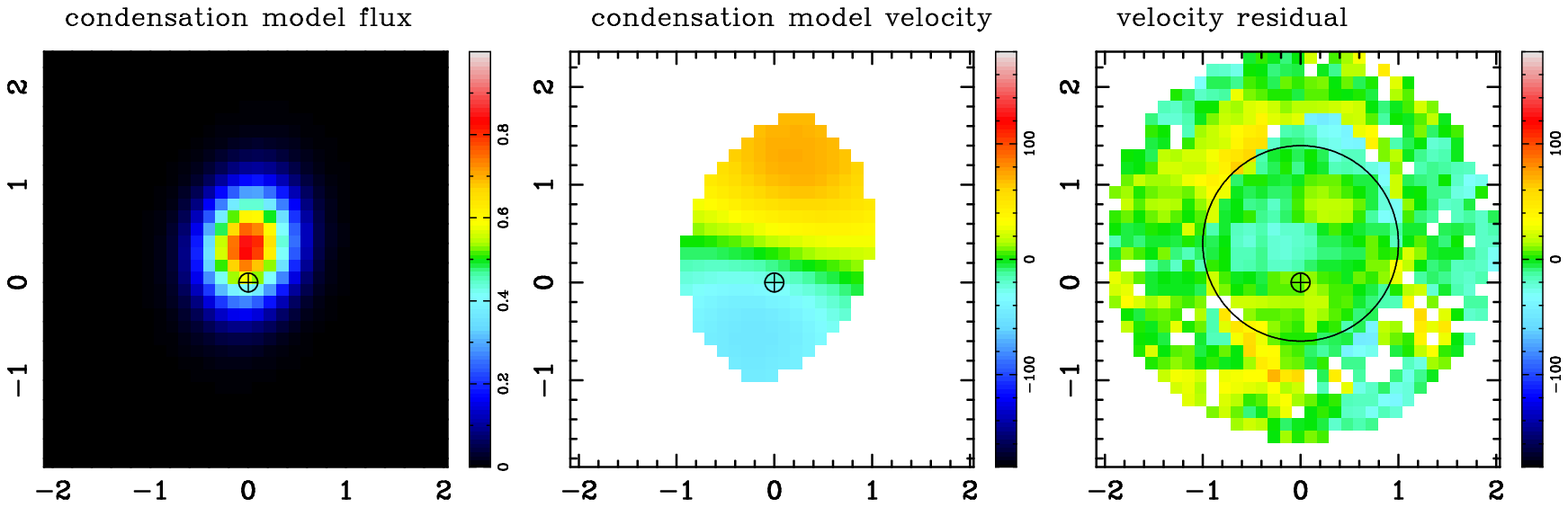}
\caption{\label{fig:ngc3368g}
Molecular gas as traced by the 1-0\,S(1) line in NGC\,3368, showing the same region as in Fig.~\ref{fig:starcont} and Fig.~\ref{fig:starvelq}.
Top row: observed flux distribution, and the residual flux after subtracting elliptical isophotes centered on the nucleus; the right panel shows the V-H dust structure map reproduced from \cite{mar03}.
Middle row: observed velocity field, disk model, and the difference between them showing the clear counter-rotation centered about 0.4\arcsec\ north of the nucleus.
Bottom row: zoomed-in view of the flux, velocity and velocity residual (with respect to the right-most panel in the middle row) of the counter-rotating condensation model.
The circle is centered 0.4\arcsec\ north of the nucleus (near the peak of the 1-0\,S(1) emission in the top left panel) and has a radius of 1\arcsec. Although the flux of this off-centre knot is compact, the impact of its velocity field can be traced out to $\sim$1.5\arcsec.
Axis scales are in arcsec, with a conversion to parsec given; north is up and east is left.}
\end{figure*}

\begin{figure}
\epsscale{0.95}
\plotone{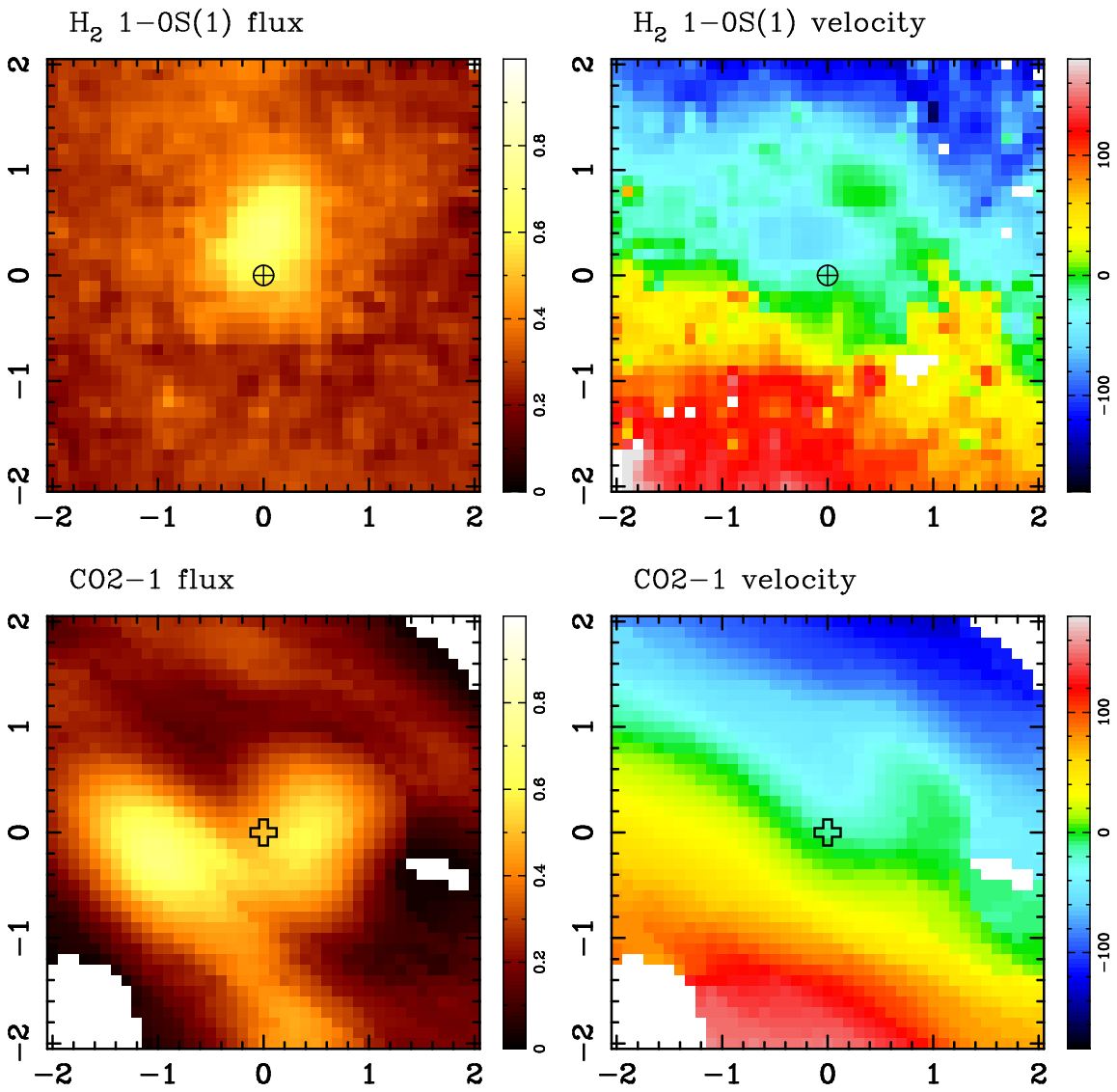}
\caption{\label{fig:ngc3368co}
Molecular gas as traced by the 1-0\,S(1) and CO\,2-1 lines in the central $4\arcsec\times4\arcsec$ of NGC\,3368.
Top row: flux and velocity of the 1-0\,S(1) line. The crossed circle denotes the centre as measured from the K-band peak.
Bottom row: flux and velocity of the CO\,2-1 line. The hollow plus denotes the dynamical centre derived by \cite{haa09}.
The flux maps of the 2 tracers are remarkably different and so we have registered them by assuming that the dynamical centre of the CO map is located at the K-band continuum peak.
When this is done, the non-circular velocities to the north-west of the nucleus match up between the near-infrared 1-0\,S(1) line and the 1.3\,mm CO\,2-1 line.
Axis scales are in arcsec; north is up and east is left.}
\end{figure}

\subsubsection*{NGC\,3368}

The H$_2$ distribution is not centered on the nucleus of this galaxy, but peaks about 0.4\arcsec\ north, as can be seen in Fig.~\ref{fig:ngc3368g}.
Some fainter structure is also apparent, and can be seen more clearly in the second panel of the figure, which shows the residual flux after fitting and subtracting elliptical isophotes.
The offset of the brightest H$_2$ emission was reported by \cite{now10}, and their adaptive optics data resolved it into two knots that appeared to be moving at different, unrelated velocities, decoupled from the disk rotation.

Our larger scale velocity field also shows structure near the nucleus, although overall it is dominated by rotation -- with a velocity of 180\,km\,s$^{-1}$ at a radius of 2\arcsec.
We see no signature of gas inflow in bar, despite a double bar being reported in NGC\,3368 by \cite{erw04}, with the inner bar confined within our SINFONI field (semimajor axis of 4--5\arcsec) and present in the continuum data. The LON of the stellar and gas velocity field is consistent with the LON determined on larger scales. 
Fitting and removing the global rotation with a disk model reveals an extended structure remarkly similar to that expected for coherent counter-rotation, centered about 0.4\arcsec\ north of the nucleus.
This velocity structure can be traced beyond a radius of 1\arcsec\ from its centre, and as such filled most of the field of view observed by \cite{now10} -- which made their H$_2$ velocity field difficult to interpret.
It is also apparent in the velocity map of the 1.3\,mm CO\,2-1 line, at a similar spatial resolution to our data \citep{haa09}, the central part of which is reproduced in Fig.~\ref{fig:ngc3368co}.
This argues against the idea that the H$_2$ emission is simply two knots moving rapidly in opposite directions with respect to the galaxy disk.
Instead, a simple explanation could be that the two knots are bright spots in a self-gravitating molecular condensation (i.e. cloud or cloud complex) that is also traced partly by the CO line emission.

However, it is difficult to reconcile the vastly different distributions of the two molecular gas tracers in this context: the 1-0\,S(1) line emission to the north of the nucleus is between the brightest CO\,2-1 emitting knots which lie to the north-west and north-east of the nucleus.
This must be related to the gas temperature, since the former traces gas at 1000-2000\,K while the latter typically traces gas at 10--100\,K.
Normally, these are expected to trace broadly similar distributions.
However, on these scales, where the resolution corresponds to only 30\,pc, local heating effects play a major role in the flux distribution (see, for example, the relative distributions of CO\,2-1 and 1-0\,S(1) in the central 300\,pc of NGC\,1068, \citealt{mue09}).

In order to pursue the idea of a molecular condensation, and to provide an approximate constraint on its properties, we have modelled it as a thin rotating disk with a mass distribution described by a S\'ersic function.
We fit the model to the residual 1-0\,S(1) velocity field (but not to the flux distribution) as shown in the bottom panels of Fig.~\ref{fig:ngc3368g}).
While the model can only be representative rather than highly constrained, 
it suggests that the condensation is slightly inclined, has a rotation velocity of 70\,km\,s$^{-1}$ at a radius of 1\arcsec,
and that the mean velocity is offset from that of the disk by more than 10\,km\,s$^{-1}$.
It is possible that such a gas condensation has arisen internally, perhaps similar to the $10^7$\,M$_\odot$ star cluster in the cirumnuclear region of M\,83 \citep{piq12}, which also exhibits internal rotation unrelated to the host galaxy but without any significant systematic velocity offset.
On the other hand, as discussed in Section~\ref{sec:environ}, there is unambiguous evidence from observations of intergalactic H\,I that there is a bridge of gas leading to NGC\,3368 \citep{mic10}.
As such, external accretion of gas is a natural, and simple, explanation for the condensation.
Furthermore, because molecular gas moving anomalously in the disk plane would rapidly lose angular momentum, it seems likely that the condensation lies out of the disk plane, and hence may be moving on an orbit unrelated to the disk.
If it is in front of the disk plane, then it would be falling towards the disk.
Such an orbit, combined with the paucity of gas in the disk, could explain why the condensation has survived long enough to be observed.
Although we cannot fully rule out an origin from secular processes, the anomalous orbit argues rather for accretion of gas from a streamer as its origin.

Ignoring velocity dispersion, the dynamical mass of the structure is $\sim6\times10^7$\,M$_\odot$.
This estimate is close to the minimum since, even if the modelled disk were edge-on, the dynamical mass would decrease only a little.
We have tested the plausibility of this dynamical mass by performing an independent estimate of the gas mass in the two brightest CO\,2-1 knots based on their luminosities.
Adopting a CO-to-H$_2$ conversion factor of 
$\alpha_{CO} = 4.3$\,M$_\odot$\,(K\,km\,s$^{-1}$\,pc$^2$)$^{-1}$ \citep{bol13} yields a mass in the right and left knots of 
$6\times10^7$\,M$_\odot$ and $7\times10^7$\,M$_\odot$ respectively.
These two knots comprise about 40\% of the CO luminosity, and hence mass, in the 4\arcsec\ field shown in Fig.~\ref{fig:ngc3368co}, and a little under 20\% of the mass in the CO `core' reported by \cite{sak99} (once their adopted distance to this object is matched to ours).
Thus, both the CO luminosity of the right-hand knot and the dynamical mass derived from the 1-0\,S(1) residual velocities, are consistent and point to a substantial mass in this condensation.
In the Appendix and Fig.~\ref{fig:starcont} we showed that there was residual stellar continuum just east of the nucleus.
The stellar mass associated with this was in the range 1--5$\times10^6$\,M$_\odot$.
As such, the gas fraction of the condensation is very high, indicating that it should not be considered as a dwarf galaxy, but as associated with the gas streamers.

The mass of the condensation is very significant compared to that in the galaxy at these scales.
Taking a stellar velocity dispersion of 120\,km\,s$^{-1}$ \citepalias{hic13}, one can estimate the dynamical mass of the galaxy within 0.4\arcsec\ of its nucleus as $2\times10^8$\,M$_\odot$, only a factor 3--4 greater.
As such, when the condensation does merge with the galaxy nucleus, it will have a marked impact.
Interestingly, the model of the condensation requires a steep rise in the rotation curve, reaching a maximum of nearly 90\,km\,s$^{-1}$ at a radius of 0.3\arcsec.
This could account for the high anomalous velocities of the two bright knots of H$_2$ emission noted by \cite{now10}: they light up the most rapidly moving part of the condensation, perhaps through thermal or shock heating since the shear at this point would be a maximum.

\section{Recent versus On-Going Nuclear Star Formation}
\label{sec:starform}

To assess the age of the stellar population in the sample of galaxies, we estimate the equivalent width of the Br$\gamma$ emission.  As can be seen in Table~\ref{tab:lum}, within a 2\as\ aperture (corresponding to a radius of typically 100\,pc) Br$\gamma$ emission is detected in 4 of the 5 active galaxies, the exception being the lowest luminosity Seyfert NGC\,7743.  No Br$\gamma$ emission is detected in any of the inactive galaxies.  The equivalent width is calculated using the stellar {\em K}-band luminosity, which is determined from the total {\em K}-band luminosities given in Table~\ref{tab:lum} corrected for any AGN contribution.  The fraction of the {\em K}-band luminosity attributable to non-stellar AGN continuum is estimated from the dilution of the CO bandheads, which are expected to have a constant (to within 20\%) intrinsic equivalent width over a wide range of star formation histories and ages \citep{dav07}.  The dilution factor for each galaxy was determined from an integrated spectrum within the same 2\as\ aperture and, consistent with our results presented in \citetalias{hic13}, only three of the galaxies have signs of significant AGN contribution to the continuum.  These are NGC\,3227, NGC\,6300, and NGC\,6814, which have a dilution factors of 0.41, 0.59, and 0.55, respectively.  It is this correction factor that is used to determine the stellar {\em K}-band luminosity and to calculate Br$\gamma$ equivalent width 
(i.e. $L_{stellar}$ = L$_{K}$$\times$f$_{dilution}$).  
As shown in Table~\ref{tab:lum}, 3$\sigma$ upper limits on the Br$\gamma$ luminosities and equivalent widths are also given for those galaxies in which no Br$\gamma$ emission is detected within the integrated 2\as\ aperture spectra.

The low Br$\gamma$ equivalent widths, ranging from 0.1-1.4\AA (Table~\ref{tab:lum}), suggest that there is no on-going star formation in our sample on the scales measured.  These equivalent width values are characteristic of a stellar population that is at least 10\,Myr old and potentially as old as a few 100\,Myr, depending on the assumed star formation history (e.g. \citealt{dav07}).  Although the measured Br$\gamma$ equivalent widths rule out on-going star formation, they do not exclude the possibility of relatively recent star formation within the circumnuclear region measured.

\cite{esq14} find evidence for recent star formation in 45\% of their sample of Seyfert galaxies using 11.3\,$\micron$ polycyclic aromatic hydrocarbons (PAH) emission.  They measure the star formation activity in the vicinity of the AGN on scales similar to those probed with our sample, as well as on larger scales of 100s of parsecs, and find that the activity detected is centrally concentrated to scales of tens of parsecs.  In the fraction of the sample in which the PAH emission was not detected there is either no star formation occurring or the population has aged such that B stars have evolved off the main sequence.  Although PAH emission is a tracer of young stars, it is believed that the emission is more sensitive to B stars than O stars (Peeters et al. 2004).  Diaz-Santos et al. (2010) verify this by showing that the PAH/Pa$\alpha$ ratio increases for lower Pa$\alpha$ equivalent width.  Since the equivalent width of Pa$\alpha$ is an indicator of age this implies that the PAHs trace older stars than the hydrogen recombination lines sensitive to O stars.  Therefore, a population with strong PAH emission is not likely to have on-going star formation but instead will have formed stars recent enough to still contain B stars, which is less than approximately 100\,Myr.  This is consistent with earlier work \citep{dav07} and with our conclusion that there is recent, but not on-going, star formation with an age since the last active star formation of less than a few 100\,Myr.

\begin{deluxetable*}{lccccc}
\tabletypesize{\scriptsize}
\tablecaption{Properties of Outflowing Gas\label{tab:outflow}} 
\tablewidth{0pt}
\tablehead{
\colhead{Galaxy} &
\colhead{$v_{out}$} &
\colhead{$\sigma_{out}$} &
\colhead{$L_{out}^{1-0S(1)}$} &
\colhead{$R/\langle \cos{\theta}\rangle$} &
\colhead{$\dot{M}_{out}$} \\
\colhead{ } &
\colhead{km\,s$^{-1}$} &
\colhead{km\,s$^{-1}$} &
\colhead{$10^4$\,L$_\odot$} &
\colhead{pc} &
\colhead{M$_\odot$\,yr$^{-1}$} \\
}
\startdata

NGC 5643 & 60--80   & 90--100   & 3.0 & 250 & 10 \\
NGC 6300 & 60--90   & 100--120  & 5.3 & 170 & 30 \\
NGC 7743 & $\sim40$ & 70--90    & 0.8 & 115 &  5 \\

\enddata

\tablecomments{
Col 1: object name;
col 2: the range or typical value, in the residual velocity field associated with the outflow;
col 3: the range, or typical value, in the dispersion map associated with the outflow;
col 4: luminosity in 1-0\,S(1) line associated with the outflow (locus is based on $v_{out}$ and $\sigma_{out}$ except for NGC\,6300 where it is based on the residual flux directly);
col 5: root-mean-square distance from nucleus to locus associated with outflow, adjusted to take accout of the unknown outflow orientation;
col 6: outflow rate calculated as $MV/R$.
}

\end{deluxetable*}


\section{Molecular Outflows}
\label{sec:outflow}

Three of the five active galaxies in our sample (NGC\,7743, NGC\,5643, and NGC\,6300) show spatially resolved molecular gas flowing out from the AGN, and a fourth (NGC\,3227) shows strong disturbances such as high dispersion associated with the ionised outflow although it is unclear whether the H$_2$ itself is outflowing.
In two of these cases, the outflow exhibits the classical bi-cone morphology.
For the fifth AGN, NGC\,6814, it is not possible to say whether or not there is molecular outflow due to the limited field over which we can detect 1-0\,S(1) line emission.
However, we note that the 3D geometry of the ionised outflow \citep{mue11} shows that it intersects the disk plane to the south west, so that, based on the discussion below and with higher signal-to-noise data, one might expect to see a molecular outflow also in this galaxy.

A property they all have in common is that the orientation of the outflow is such that it intersects, or has its edge close to, the galaxy disk.
Our sketch in Fig.~\ref{fig:ngc7743sketch} of the geometry for NGC\,7743 shows that the (redshifted) far edge of the southern outflow and (blueshifted) near edge of the northern outflow are close to the galaxy disk.
The geometrical model of \cite{fis13} for NGC\,5643 indicates that the (redshifted) far edge of the approaching outflow to the east is close the disk, as is the (blueshifted) near edge of the western outflow.
The outflow in NGC\,6300 appears almost edge-on.
Because it is aligned nearly along the minor axis of the host galaxy, it too must intersect the host galaxy disk.
The ionised outflow in NGC\,3227 also encloses part of the galaxy disk, in such a way that, if gas clouds in the disk were accelerated outwards, these would be blueshifted to the west and redshifted to the east \citep{fis13}.
While the non-circular velocities of only $\sim\pm20$\,km\,s$^{-1}$ in the central 1--2\arcsec\ of NGC\,3227 are modest, they are consistent with this geometry.

These results suggest that a minimum requirement for a molecular outflow must be that there is sufficient ambient molecular gas to be swept up by the wind from the AGN;
and for a Seyfert nucleus residing in a disk galaxy, the most abundant source of ambient gas is the disk itself.
It therefore seems likely that for a Seyfert galaxy to drive molecular gas outwards, the outflow must be tilted over sufficiently far that it is close to, or intersects, the disk plane.
This may occur rather commonly, given that outflows are oriented randomly with respect to the host galaxy \citep{cla98,nag99,kin00} and that they have a finite opening angle.
Adding together the $\theta_{max}$ and $\beta$ parameters derived by \cite{fis13} yields the angle between the normal to the disk plane and the edge of the ionisation cone in the objects they modelled.
In 9 of 17 cases, the cone does indeed intersect the disk, and another 2--4 (allowing for a tolerance of 10--20\deg) come close.
Thus we may expect more than half of Seyfert galaxies to drive molecular outflows.
The key questions for these are how much mass is entrained, and how fast are they flowing?

To address these, we have identified regions in the 3 galaxies where outflowing H$_2$ is seen based on dispersion and residual velocity (except for NGC\,6300 where it is based on the residual flux), and measured the 1-0\,S(1) luminosity in these regions as well as their {\sc rms} distance from the nucleus.
The typical values for these parameters are reported in Table~\ref{tab:outflow}.
We have derived a characteristic outflow speed as 
$V \sim \sqrt{v_{out}^2+\sigma_{out}^2}/\langle \sin{\theta}\rangle$
taking $\langle \sin{\theta}\rangle = \langle \cos{\theta}\rangle = 0.7$ to account for the unknown orientation of the outflowing H$_2$ (note that this correction is also applied to the distance of the outflowing gas from the AGN).
This gives values in the range 130--190\,km\,s$^{-1}$.
To estimate the mass of molecular gas, we have derived the mass of hot gas from the luminosity of the 1-0\,S(1) line in the relevant regions of the residual flux maps, assuming the molecules are thermalised at 2000\,K and applying a scaling factor based on M$_{cold}$/M$_{hot}$.
On large scales in star forming galaxies, this ratio is typically $10^{6.3}$ to within a factor of a few \citep{dal05,mue06}.
We note that estimating a total gas mass from the residual 1-0\,S(1) luminosity in NGC\,3368 using this ratio yields a mass of 1--$2\times10^7$\,M$_\odot$ consistent with our other estimates in Sec.~\ref{sec:h2vel}.
On scales of a few hundred parsecs or less, this ratio may be smaller, in the range $10^{5.5}$--$10^{6.2}$ \citep{maz13};
and there are indications that this may also be the case for Seyferts \citep{dal05}.
As such, in order to avoid over-estimating the total gas mass, we have adopted a ratio of $10^{5.5}$.
This highly uncertain step yields masses in the range 5--$25\times10^6$\,M$_\odot$.
A final step gives the outflow rate as $\dot{M}_{out} = MV/R$, yielding rates of order 10\,M$_\odot$\,yr$^{-1}$.

Based on these order-of-magnitude calculations, we conclude:
\begin{itemize}

\item
The H$_2$ outflow speeds of around 150\,km\,s$^{-1}$ are lower than the outflow speeds of ionised gas, which are mostly in the range 200--2000\,km\,s$^{-1}$ \citep{mue11,fis13,sto14}.
This difference may be because, as gas clouds are accelerated outwards, an increasing fraction of the clouds are ablated and ionised, in which form the gas is more easily accelerated (large filling factor, lower density), which is why one sees larger velocities for the ionised gas.
We have estimated the escape velocities of these 3 galaxies from the radial distances given in Table~\ref{tab:outflow}, deriving the galaxy mass from the H-band magnitude given in \citetalias{hic13}. 
We find $V_{esc} \sim 1500$\,km\,s$^{-1}$. 
This demonstrates that, unless it is continuously accelerated, the outflowing molecular gas cannot escape and will instead fall back, 

\item
The H$_2$ outflow rates, although rather uncertain, could be sufficiently high to have an impact on the local environment depending on how long the central black hole is active. For an activity timescale of 1--10\,Myr, the AGN could drive $10^7$--$10^8$\,M$_\odot$ of gas outwards. However, as argued above, this is likely to fall back again; and is certainly not enough to impact the large scale properties of the host galaxy.

\item
The median outflow speed of molecular gas (based on the 119\,$\mu$m OH line) reported by \cite{vei13} of $\sim200$\,km\,s$^{-1}$ for ULIRGS and QSOS is similar to that derived above for these few Seyferts.
However, it is not clear whether one can compare the outflows from ULIRGS/QSOs to those from Seyferts due to the very different nature of the host system.
And it is notable that the molecular outflow rates derived by \cite{stu11} for a small number of those ULIRGs/QSOs are 1--2 orders of magnitude greater than the rates we derive.

\end{itemize}

\begin{deluxetable*}{lcccccl}
\tabletypesize{\scriptsize}
\tablecaption{Summary of H$_2$ phenomena, dust structures, and environment\label{tab:h2}} 
\tablewidth{0pt}
\tablehead{
\colhead{Galaxy} &
\colhead{rotating} &
\colhead{circumnuclear} &
\colhead{perturbed} &
\colhead{outflow} &
\colhead{dust\tablenotemark{a}} &
\colhead{environment\tablenotemark{b}} \\
\colhead{} &
\colhead{disk} &
\colhead{spiral} &
\colhead{} &
\colhead{} &
\colhead{} &
\colhead{} \\
}
\startdata

NGC 3227 & X & X &   & \phm{\tablenotemark{c}}(X)\tablenotemark{c} & C  & group of 13--14 \\
NGC 5643 & X & X &   & X                         & GD & isolated \\
NGC 6300 & X &   &   & X                         & C  & undisturbed, group of 9\\
NGC 6814 & X &   &   &                           & GD & isolated \\
NGC 7743 & X & X &   & X                         & LW & undisturbed \\
\\                                                    
IC 5267  & X &   & X &                           & C  & group of 11--14\\
NGC 4030 &   &   &   &                           & TW & undisturbed; group of 4? \\
NGC 3368 & X &   & X &                           & CS & group of 9--14 \\
NGC  628 &   &   &   &                           & N  & undisturbed; group of 7? \\
NGC  357 &   &   &   &                           & N  & undisturbed; group of 6? \\

\enddata

\tablecomments{Summary of whether the following phenomena have been observed in the H$_2$ line distribution and kinematics within the central few hundred parsecs: the presence of a rotating disk, signs of inflow driven by a circumnuclear spiral, evidence for perturbing events, and detection of outflowing molecular gas.
The lack of a mark indicates only that a particular structure is not traced by the 1-0\,S(1) H$_2$ line in our data; it may still be apparent from maps of other tracers such as Br$\gamma$ line or dust structure.}

\tablenotetext{a}{Circumnuclear dust structure as classified by \cite{mar03}: GD = grand design spiral, TW = tightly wound spiral, CS = chaotic spiral,  C = chaotic, N = no dust structure.}

\tablenotetext{b}{Group membership as discussed in Sec.~\ref{sec:environ} (primarily from \citealt{gar93} and \citealt{cro07}; see text for full details of references). A question mark indicates where there is some uncertainty about membership (e.g. because of the different definitions used to classify groups).}

\tablenotetext{c}{The H$_2$ is disturbed (high dispersion) as a result of the outflow, but we do not directly see signs that the H$_2$ itself is outflowing (non-circular velocities are modest).}

\end{deluxetable*}


\section{Gas Inflow to Galactic Nuclei}
\label{sec:perturb}

In this section we discuss the link between gas inflow mechanisms that might fuel AGN, the properties of the host galaxy, and the environment. 

\subsection{Circumnuclear Structures in Our Sample}
\label{subsec:structures}

\subsubsection{Two modes of gas inflow to the central tens of parsecs}

When analyzing individual galaxies in our sample in Sec.~\ref{sec:h2vel}, we found gas kinematics consistent with inflow caused by bars and nuclear spirals in all 3 active galaxies in which inflow can be separated from outflow. 
Numerical models show that such inflow can be sustained for long timescales in a quasi steady state (e.g. \citealt{mac04}). 
However, we also found that in both galaxies classified as inactive in which we detected warm molecular gas, a counter-rotating gas component is present. 
Couter-rotation requires reversion of the angular momentum vector from that of gas co-evolving with the galaxy disc, and this implies a dramatic change in velocity. 
For this reason, physical processes internal to the galaxy are unlikely to provide enough energy to bring about this change for large quantities of gas. 
On the other hand, accreted gas can naturally account for counter-rotation if its intrinsic angular momentum is significantly different from that of the disc. 
Therefore counter-rotation likely points at external accretion. 
External accretion likely occurs in the form of stochastic events, which happen on shorter timescales than the internally driven gas inflow that we observe in the active galaxies. If AGN can be fed by external accretion, then there are two major consequences of this difference for the detectability of such process. Firstly, there is a higher chance to see an AGN fed by internally driven gas inflow than by external accretion; and secondly, galaxies with external accretion onto the nucleus, which are likely to become active within a short timescale, may currently be classified as inactive.

We postulate that there are two distinct kinds of processes that drive gas inwards to feed an AGN: internal secular inflow that is quasi-continuous for long timescales, and stochastic external accretion events. 
It is likely that these have different duty cycles, and manifest in different types of galaxies and environments. 
In particular, external accretion is unlikely to occur when there is no supply of cold gas external to the galaxy. 
This may apply either to isolated galaxies, or to galaxies in clusters or in large, evolved groups with X-ray cores. 
Also, a large reservoir of native gas in a galaxy may mask the signature of external accretion by reducing its impact and also influencing how easily it can be seen -- this may be happening in at least one of our targets (NGC\,3227). 
With a plentiful internal supply of gas, secular processes can dominate the inflow \citep{mor06,you11}. 
Thus, the prevalent mechanism of accretion onto the nucleus is related primarily to the environment and gas content of the galaxy, rather than to the galaxy classification. 
Despite this, the morphological classification is important, because it broadly reflects whether the host galaxy is likely to have internal gas, and it roughly correlates with the environment.
We note, however, that the correlations are coarse and, for example, there is a significant population of blue early type galaxies with substantial gas reservoirs \citep{kan09,wei10}.

\subsubsection{Possible feeding of AGN by externally accreted gas}

It is well known that a significant fraction of gas in early type galaxies is of external origin \citep{sar06,tdav11}. 
The idea that the nuclear activity in early type galaxies is sustained by externally accreted gas has been proposed by \cite{sim07}, who showed that for early type galaxies, all active galaxies have circumnuclear dust structures while only a small fraction (26\%) of the inactive galaxies do. This is in contrast to the late type galaxies in their sample, all of which, active and inactive, have dust structures. This result indicates that circumnuclear dust structures (tracing gas) are a necessary, but not sufficient, condition for fueling a central massive black hole. \cite{sim07} postulated that for early type galaxies the gas might be accreted from external sources, in contrast to late type galaxies where the large scale disk provides an internal source of gas. 
\cite{mar13} pursued this hypothesis, looking at the spectral energy distributions and dust masses of 38 early type galaxies. They argued that although there is evidence for circumstellar dust around evolved stars, this could not account for the total dust mass in galaxies that exhibit dust structures. They proposed that for these objects, the dust is seeded by external accretion and continues to form in the accreted cold gas over long timescales, accounting for the observed gas masses. 

Our observations show that if gas counter-rotating with respect to the stellar component (and to the native gas) implies its external origin, then such external counter-rotating gas can reach the innermost tens of parsecs of galaxies. 
This is seen in two galaxies (IC\,5267 and NGC\,3368) of our sample. 
Gas in the nucleus appears to form dynamically unstable configurations which are bound to result in an eventual rapid loss of angular momentum, with gas falling onto the galaxy centre. 
Thus it is likely that gas inflow to the nucleus will follow the accretion event. Depending on the timescale over which this occurs, and also on whether these are isolated accretion events, these inactive galaxies may either become AGN in the future (i.e. are `pre-AGN') or, due to rapid variability of inflow on the smallest scales, are currently in a short `off' cycle of an on-going AGN phase.
In assessing which of these scenarios is more likely, we recall that while the analysis of our targets in \citetalias{hic13} revealed systematic differences between the active and inactive galaxy samples, there was not such a clean separation for individual sources.
Thus, even in the circumnuclear region where the dynamical timescale is 2--3\,Myr, we may not have fully overcome the problem of timescale matching.

We note that rapid (i.e. 1--10\,Myr) variability particularly affects inactive control samples, and a mismatch of the timescale between this and other factors that may influence AGN activity has a major impact on how such samples should be interpreted.
This can most easily be realised with a simple thought experiment in which one hypothesises that some aspect of the host galaxy or environment is responsible for triggering inflow. This may influence the galaxy for more than 1\,Gyr; the resulting accretion onto the central black hole will occur at some point during this phase, but only on-and-off within a $\sim100$\,Myr period.
If one then takes a snapshot to look at which galaxies are active and inactive, one will see only a random subset, which will be different for another snapshot.
Based on any of these snapshots, although the host galaxy or environmental factor will accompany AGN fueling, the conclusion from the control sample would be that one does not lead to the other, which contradicts the initial hypothesis of the thought experiment.

\subsubsection{Minor mergers versus intergalactic gas}
\label{subsubsec:hi}

In the literature, external accretion is usually associated with dwarf satellites falling onto the host galaxy and releasing their gas to the host. 
However, in the particular case of NGC\,3368, there is an intergalactic H\,I bridge leading to NGC\,3368 \citep{mic10}. 
Because of the lack of non-uniformities in the circumnuclear stellar kinematics, we postulate that accretion of gas rather than stellar systems -- i.e. streamers rather than (minor) mergers -- is likely to play a role in this particular case, and that streamers may provide an additional source of external gas, besides the commonly considered mergers.

There is some evidence from H\,I observations to support this.
\cite{kil09} found that the H\,I mass function in groups is flatter than in the field, and that there is a lack of high H\,I mass galaxies in groups.
They suggested this may be due to tidal stripping of gas from disk galaxies through interactions; rather than the result of ram pressure stripping, which occurs in clusters but is not so important in groups.
\cite{wil09} argued that galaxies falling into groups are experiencing the impact of a dense environment for the first time, and that the transformation of galaxies and the inter-galactic medium (often referred to as pre-processing) may take place in the group environment prior to infall of the group into a cluster.
There is, however, a maximum group size in which such pre-processing occurs.
\cite{hes13} found that as group membership increases from 5 to 20, two things happen: the fraction of H\,I galaxies decreases, and those with H\,I tend to be less centrally concentrated.
Their results suggest that groups with 5--20 members are an ideal place for H\,I to be stripped away from the outskirts of galaxies into the IGM, and hence be available (still as neutral gas) to accrete onto other galaxies in the group.
While this could provide a source of atomic gas in the local environment, further work is needed to investigate whether the origin of counter-rotating molecular gas in IC\,5267 and NGC\,3368 is related to stripping of H\,I from the outskirts of galaxies. We note that intragroup molecular gas has been observed, but only in strongly interacting groups such as Stephan's Quintet \citep{lis02}.

\subsubsection{Trends in our sample}
\label{subsubsec:trends}

When assessing the conditions that are favorable to internally versus externally driven circumnuclear inflow, we use a combined sample containing all our galaxies, rather than separating them according to whether they are specifically active or inactive at the current time. We relate the H$_2$ morphology and kinematics reported in Sec.~\ref{sec:h2vel} to the environment discussed in Section~\ref{sec:environ} and the circumnuclear dust structures reported by \cite{mar03}. 
These properties are all summarised in Table~\ref{tab:h2}, and enable us to make the following points:
\begin{enumerate}
\item
There are three galaxies without 1-0\,S(1) detections: NGC\,4030, NGC\,628, and NGC\,357. They are either isolated or in small groups with $<$10 members, and two show no circumnuclear dust structure. All three undisturbed galaxies without circumnuclear molecular disks are inactive.
\item
All four galaxies with chaotic dust structure (designated C or CS) reside in groups with $\sim10$--15 members. They are two inactive galaxies IC\,5267 and NGC\,3368, and two AGN NGC\,3227 and NGC\,6300. 
The chaotic dust structure may either reflect the H$_2$ distribution and kinematics which are perturbed by external accretion (leading to counter-rotating gas structures), or be superimposed on an internal mechanism revealed by the H$_2$ such as bar driven inflow.
\item
The remaining three AGN are isolated or have undisturbed large scale disks: NGC\,5643, NGC6814, and NGC\,7743. All three of these undisturbed AGN have circumnuclear dust structures classed as spirals (GD, LW, TW) as well as circumnuclear molecular disks.
\end{enumerate}

These point to a scenario in which 
(i) a circumnuclear spiral in an undisturbed galaxy with a plentiful gas supply is a sign of secular internal inflow of gas to the nucleus; and
(ii) chaotic circumnuclear dust structures indicate external accretion: they are preferentially happening in galaxies that are part of a moderate-size group with 10--15 members, and in Sec.~\ref{subsubsec:hi} we showed that in groups of similar size, H\,I stripped from the outskirts of galaxies could provide a source of intergalactic gas.

\subsection{Comparison to Other Samples}

The small number of targets limits the statistical robustness of our results because it cannot properly sample the full range of galaxy hosts (Hubble type, and presence/absence of a bar). 
As such, we now extend our analysis to look at other samples of matched pairs of active and inactive galaxies that were observed using integral field spectroscopy (IFS) so that the stellar and gas kinematics are spatially resolved.
We examine first the samples of \cite{dum07}, in which nearly all galaxies are late type, and of \cite{wes12}, in which the majority of the galaxies are early type.
For these two samples, which between them provide more than 30 additional sources, the spatial scales probed extend out to beyond 1\,kpc at resolutions of 100--600\,pc.
In this part of the analysis, we focus on the kinematic evidence for a relation between the inflow mechanism and host galaxy type.
Because there is no sharp dividing line that separates host galaxy types into two distinct groups, we use the terms `early type' and `late type' loosely.
Specifically, for the purpose of this analysis, we take galaxies classed as Sa or earlier to be `early', and Sab or later to be `late'.

We do not attempt a detailed study of the kinematics, but look at whether the gas and stars are co-rotating or counter-rotating, how this is related to the host type, and if it indicates whether external accretion or secular evolution is at work.
We then turn to the sample of \cite{maz14} in which all galaxies are late types, but several of them are in a cluster environment. 
We take a brief look at whether this has an impact on their circumnuclear structures.

\subsubsection{Secular internal inflow}

When secular evolution is driving gas inflow, the gas and stars are expected to be co-rotating. On the other hand, in the external accretion scenario, the accreted material should show no preference for co- or counter-rotation with respect to the pre-existing stellar population. 
In their analysis, which included 7 pairs of matched active/inactive galaxies, \cite{dum07} reported the difference in mean position angle between the ionised gas and stars.
Interestingly, the galaxy with the largest discrepancy between these is the AGN NGC\,2655. 
This S0/a galaxy is known to have off-planar gas and complex dust structures \citep{erw03}, and is in a group with 4--7 members \citep{gar93,cro07}.
In contrast, the galaxies in the other 6 matched pairs are all morphological type Sab to Sbc.
None of the 6 inactive late types showed evidence for misalignments between the stellar and gas kinematics: in all of them, the gas and stars are co-rotating.
And while 3 of the 6 active late types did have misalignments, their range of 25--55\deg\ indicates the gas and stars are still co-rotating.
Thus, interactions and external accretion might play a role here, but it is also possible that internal processes such as bar driven inflow (NGC\,4151 and NGC\,4579 are both barred and have circumnuclear rings) could generate these moderate misalignments.
We note in the case of NGC\,5194, the [O\,III] line from which the gas kinematics were derived, appears to be influenced by the outflow \citep{cec88,ter01}; the H$\beta$ kinematics are more closely aligned with the stellar kinematics.

We recall that in our sample, there were 3 AGN with undisturbed hosts, circumnuclear molecular disks, and circumnuclear spiral dust structures.
Of these, NGC\,6814 and NGC\,5643 are classified as Sbc and Sc; NGC\,7743 is an S0.
NGC\,3227, an interacting Sa galaxy with a chaotic dust structure, also exhibits unambiguous spiral driven inflow.
We have shown that the presence of a bar can play a key role in generating these structures.
We argue below that the presence of a bar and a plentiful supply of gas in a large scale disk may be more important than the host type in understanding such galaxies.
Archetypal examples of this sort of AGN host may be Narrow Line Seyfert 1s (i.e. type 1 AGN with a `narrow' broad line region with FWHM $<$2000\,km\,s$^{-1}$).
In terms of bulge properties, these AGN appear to lie at the extreme end of the range covered by type~1 AGN in general, having pure pseudo-bulges \citep{orb11,mat12}.
The implication is that they have evolved in relatively isolated environments without major mergers.
\cite{orb11} also brought together studies in the literature which show that these are associated with strong bars, enhanced circumnuclear star formation, and grand design circumnuclear spirals: all indications of secular evolution bringing gas inwards from the large scale disk.
In these objects, the bar drives gas into the circumnuclear region as well as is stimulating the formation of the circumnuclear spiral; and it is the circumnuclear spiral that brings gas to smaller scales, enabling it to fuel the AGN.
For these AGN, gas inflow is internally driven, and does not appear to result from external accretion.

One important question is what inflow rates can internal secular evolution drive?
An example we can turn to is NGC\,1097, which has a strong bar, a pseudo-bulge, and circumnuclear spiral structure.
H\,I observations indicate the only interaction is with its dwarf companions NGC\,1097A which has M$_{HI}<10^6$\,M$_\odot$, or NGC1097B which has M$_{HI}=5\times10^6$\,M$_\odot$ \citep{ond89,hig03}.
The inflow rate to the central 10--20\,pc has been estimated to be of order 0.05\,M$_\odot$\,yr$^{-1}$ \citep{dav09}.
Given that much of this gas turns into stars at that radius, and inflow to smaller scales is still likely to be inefficient, this is unable to power a luminous Seyfert.
And indeed, the AGN's 2--10\,keV luminosity of $6\times10^{40}$\,erg\,s$^{-1}$ \citep{lut04} is moderate.
Thus, while secular processes can provide the gas inflow required, the inflow rate, and hence accretion rate onto the black hole, is modest.
In contrast, much higher inflow rates of 0.1--1\,M$_\odot$ have been reported by \cite{sto14} for a variety of LINER and Seyfert galaxies.
It is not yet clear whether these can be achieved through spiral driven inflow alone, or if some perturbation (internal or external) is required to temporarily enhance the inflow rate, as appears to have happened in the luminous Seyfert NGC\,1068.
In this galaxy, \cite{mue09} showed that in the central $\sim100$\,pc, clouds inflowing along the circumnuclear bar collide with clouds swept up at the edge of an expanding cavity, leading to gas falling almost directly towards the AGN at a high rate.

\subsubsection{Stochastic external accretion}

Stochastic external accretion is most easily detected in early type galaxies, as on average they contain less gas than the late types. 
It may provide a natural explanation for why in the \cite{wes12} sample, emission from ionised gas was detected in all 10 early type active galaxies but was not (or only weakly) detected in 4 of the 5 early type inactive galaxies: the inactive galaxies are inactive simply because they have not recently accreted gas and do not have their own internal supply. 
This sample includes 21 galaxies (14 active and 7 inactive) with morphological classifications, of which 15 (10 active and 5 inactive) are classified as S0 to Sa.\footnotemark[1]
\footnotetext[1]{Applying a cutoff at type S0, as was done by \cite{mar13}, has no impact on the result of the analysis except to reduce the numbers of objects.}
Stellar kinematics could be measured in 11 of these 15 early type galaxies, and 7 of these, show clear signs of stellar rotation.
Of these, which are all AGN, 5 have co-rotating H$\alpha$ and 2 have counter-rotating H$\alpha$.
In contrast, all 4 of the late type (Sab or later) galaxies in which stellar kinematics could be measured, show clear stellar rotation with co-rotating H$\alpha$ (we note that, possibly due to their shorter integration times, in neither of the inactive late-type galaxies could stellar kinematics be measured; and neither had good H$\alpha$ detections).
While the numbers involved are still small, these results suggest that the \cite{wes12} sample upholds the following observational signatures of external accretion (of intergalactic gas, or minor mergers) as a source of the gas fueling AGN: 
(i) the paucity of gas in most inactive early type galaxies contrasting with the presence of gas in all active early types, and (ii) the existence of at least some early type galaxies with counter-rotating gas.  The fraction above ($\sim$30\%) of early type galaxies in which the gas is counter-rotating with respect to the stars is similar to that reported by \cite{sar06} for the SAURON sample of galaxies, and also by \cite{tdav11} for the larger and complete ATLAS$^{3D}$ sample.
We note that both of these samples cover the central few kiloparsecs at a sampling of 100--200\,pc.

\subsubsection{Cluster versus group environment}
\label{subsubsec:clusters}

\cite{tdav11} show that for the ATLAS$^{3D}$ sample of early type galaxies, there are few misalignments for galaxies in the Virgo Cluster while there are numerous misalignments in `field galaxies' (i.e. including isolated galaxies as well as those in groups).
We have argued above that misalignments due to external accretion occur in {\em group} environments, rather than {\em cluster} environments.
The reason was already given by \cite{tdav11}: that intra-cluster gas is heated and so the source of external atomic or molecular gas is curtailed.
These authors find that the kinematic misalignments occur exclusively in galaxies residing in environments whose density is below a critical local density limit, corresponding to a number density of $\sim$20\,Mpc$^{-2}$ -- which includes moderately dense groups.

\cite{maz14} present an analysis of the circumnuclear H$_2$ properties of 6 late type (Sab or later), mostly barred, galaxies. While half of them lie in groups, the other half is in the core of the Virgo cluster and hence provides an opportunity to assess circumnuclear properties in the cluster environment. Below, we briefly look at all 6 of their galaxies in the context of environment.

NGC\,3351 and NGC\,3627 are both in the Leo\,I group (which has 9--14 members, \citealt{gar93,cro07}), and have large neighbours 60--130\,kpc away.
Despite this, the former appears undisturbed on large scales, while the latter is interacting.
NGC\,4536, while formally belonging to the Virgo cluster, is in a sub-group located about 3\,Mpc from the cluster centre.
Like the first 2 galaxies, it has a similar sized companion 120\,kpc distant, and a small blue irregular galaxy 40\,kpc away.
The circumnuclear molecular kinematics for all 3 of these galaxies show circular or elliptical motions in their disk planes, which are consistent with being internally driven.
NGC\,3351, which is undisturbed and has the circular motions, is an inactive galaxy with a limit on its 0.3-8\,keV hard X-ray luminosity of $<10^{38}$\,erg\,s$^{-1}$ \citep{gri11}, suggesting that no inflow is occurring.
For the other 2, with elliptical motions, inflow may be happening at a low level:
they are both AGN, although only NGC\,3627 has a measured 2--10\,keV luminosity, which is $3\times10^{39}$\,erg\,s$^{-1}$ \citep{bri11}.\footnotemark[2]
\footnotetext[2]{The instrinsic 2--10\,keV luminosities given in \cite{bri11} have been corrected to the distances adopted by \cite{maz14}.}

The remaining 3 galaxies -- NGC\,4569, NGC\,4501, and NGC\,4579 -- are all within 600\,kpc of the Virgo cluster centre, with different numbers of close companions, and varying signs of ram pressure stripping.
All 3 are classified as AGN.
Both NGC\,4579 and NGC\,4501 have 2--10\,keV luminosities of $10^{41}$\,erg\,s$^{-1}$ within the range of our AGN sample, while NGC\,4569 has a more modest hard X-ray luminosity of $3\times10^{39}$\,erg\,s$^{-1}$ \citep{bri11}.\footnotemark[2]
The circumnuclear kinematics were all found to show streaming motions by \cite{maz14}, although for both NGC\,4569 and NGC\,4579 they were disordered, with complex kinematics and no indications of underlying ordered rotation.
Only NGC\,4501 showed circular rotation with streaming superimposed.
Based on the linear features in the dust structure map \citep{sim07}, that is possibly of external origin.

Our conclusion from the \cite{maz14} sample is two-fold: 
(i) even in a group environment, gas inflow in late type galaxies may still be driven by internal mechanisms, which can be understood if the mass of any accreted gas is much less than the gas mass already in the galaxy disk; and
(ii) the circumnuclear molecular gas kinematics of galaxies in clusters may be very complex, presumably because the extreme environment can have an impact even deep in the potential well of a galaxy.

Putting this together with the observations of H\,I in galaxy groups (Sec.~\ref{subsubsec:hi}) and the trends seen in our sample (Sec.~\ref{subsubsec:trends}), these results suggest that, in the context of fueling AGN, external accretion appears to be a process that requires specific circumstances: a dense enough environment (i.e. a group with 10--15 members) to make streamers and/or minor mergers sufficiently likely, but not so dense ($>$20 members) that gas in the inter-galactic medium is predominantly shock-heated.

\section{Conclusions}
\label{sec:conc}

We present an analysis of the spatially resolved stellar and molecular gas kinematics and distributions for a sample of 5 pairs of matched active and inactive galaxies, covering their central few hundred parsecs. 
This has enabled us to study molecular inflows and outflows, with the aim of relating these to the large scale host galaxy properties and the environment.
We see a link between the local environment, the circumnuclear dust structures (which may also be caused by dust superimposed along the line of sight), and the circumnuclear H$_2$ structures/kinematics.
The number of galaxies in our sample is too small for robust statistical conclusions, and so some of the main findings below require further confirmation:
\begin{itemize}
\item
Circumnuclear molecular disks are seen in all 5 active galaxies but only 2 inactive galaxies, suggesting they are required for fueling Seyfert luminosity AGN. 
\item
There are two modes of inflow feeding AGN: secular processes and external accretion. The former is quasi-continuous for long timescales, and takes place in galaxies with a plentiful internal supply of gas. The latter occurs in stochastic events, and may bring in significant gas masses in a short time. It is revealed by counter-rotation in gas in the central few hundred parsecs that leads to dynamically unstable configurations.
\item
In addition to minor mergers, externally accreted gas may be coming from streamers that form intergalactic filaments and bridges in groups observed in H\,I. In such cases, moderately dense groups (with 5--20 members) provide an environment conducive to external accretion, because of efficient stripping of gas from galaxies into the IGM there.
\item
Circumnuclear spiral structures are associated with relatively isolated galaxies, and indicate that the large scale disk is the source of gas. On the other hand, chaotic circumnuclear dust structures appear to be associated with external accretion in groups. This difference is driven primarily by environment; the relation to galaxy type is a reflection of the intrinsic gas content of galaxies. Early type galaxies tend not to have a plentiful supply of internal gas and so the impact of external accretion is more easily seen.
\item
For an AGN fuelled by external accretion one can expect (i) paucity of gas in inactive galaxies vs presence of gas in active galaxies, (ii) the existence of counter-rotating gas in the circumnuclear region of early type (i.e. intrinsically gas poor) galaxies, and (iii) moderately dense local inter-galactic environments.
These characteristics appear to be supported by matched active/inactive galaxy samples presented here, as well as in \citealt{dum07} and \citealt{wes12}, and by early type samples \citep{sar06,tdav11}.
\item
Inactive control samples, need to be assessed in the context of the relevant timescales. 
This is a particular concern if there is a mismatch between the timescale of AGN accretion events (which occur on sub-pc spatial scales) and the timescale of large scale phenomena (on kpc to Mpc spatial scales). The role of environment also needs to be considered (perhaps via proxies such as host type) when selecting and analysing matched active and inactive samples.
\item
There cannot be on-going star formation in the central $\sim$100\,pc of either the active or inactive galaxies in our sample. We note this is consistent with the detection of PAH features in the centres of AGN, if PAH emission is a signature of stars older than the ionising stars traced by H-recombination lines.
\item
Spatially resolved molecular outflows are found in 3 or 4 of the 5 active galaxies, but none of the inactive galaxies. In these galaxies, the outflow (ionisation) cone is tilted over far enough that it intersects the host galaxy disk, which we argue is the source of ambient molecular gas being expelled.
\item
The molecular outflow speeds of $\sim$150\,km\,s$^{-1}$ are less than the 200--2000\,km\,s$^{-1}$ typical of ionised gas, and well below the escape velocity of the galactic potential so that the gas is likely to fall back. 
The outflow rates are of order 10\,M$_\odot$\,yr$^{-1}$.
While this is high enough to have an impact on the circumnuclear inter-stellar medium, it is not enough to impact the large scale properties of the host galaxy.
\end{itemize}

\acknowledgments

The authors are grateful to those who generously agreed that we could make use of their data: Paul Martini for sharing his HST data for each of the galaxies in our sample; Sebastian Haan for sharing his CO\,2-1 and CO\,1-0 data for NGC\,3368; Leslie Hunt and Matt Malkan sharing their data and analysis of Seyfert, LINER, starburst, and normal galaxies.
And we thank the referee for a thorough reading of the paper, and for making a variety of suggestions that have helped to improve it.
This research has made use of the NASA/IPAC Extragalactic Database (NED) which is operated by the Jet Propulsion Laboratory, California Institute of Technology, under contract with the National Aeronautics and Space Administration.  E.K.S.H. acknowledges support from the National Science Foundation Astronomy and Astrophysics Postdoctoral Fellow- ship under award AST-1002995, as well as support from the NSF Astronomy and Astrophysics Research Grant under award AST-1008042.

{\it Facilities:} \facility{VLT}

\appendix

\section{Orientation and Large Scale Structural Properties of Individual Galaxies}
\label{sec:orient}

In this Appendix we derive the kinematic major axis position angle (PA) and inclination ($i$) for each galaxy.
The aim is to find the appropriate values to adopt for our analysis.
On small scales, we use two methods.
The first makes use of kinemetry \citep{kra06} to quantify the stellar kinematics. We apply this method globally to the CO\,2-0 velocity and dispersion simultaneously, to all radii over which at least 75\% of an ellipse can be measured, and adjust the PA in order to minimise the absolute value of the $A_1$ parameter (see \citealt{kra06}).
We note that, as expected, this yields very similar values to those derived using the method described in Appendix~C of \cite{kra06} since it is based on the same concepts and differs only in the detail of its implementation.
The second method is based on the stellar distribution, fitting elliptical isophotes to a map of the CO\,2-0 bandhead flux.
For this, we made use of the multi-Gaussian expansion method \citep{ems94}, using the fitting method described in \cite{cap02} to parametrize galaxy surface brightness profiles.
As for the kinematics, the aim was to find single global estimates for the PA and $i$ based on the data within our field of view.
However, there can be large uncertainties in the derived PA and $i$, and also large differences between the kinematic and isophotal PA or $i$.
In order to provide crucial guidance in such cases, we first assess the large scale properties of each galaxy.
The notes below compare the large and small scales, providing a rationale for the values adopted in our analysis, which are given in the second column of Tables~\ref{tab:pa} and~\ref{tab:inc}.

\subsection{NGC\,3227}

Determining the global orientation of NGC\,3227 with precision is difficult, partly because the presence of a large, strong bar means that gas (and possibly stellar) kinematics may be somewhat non-circular in much of the visible disk, and also because it is interacting strongly with its elliptical-galaxy neighbour NGC\,3226 and with a nearby gas-rich dwarf galaxy (which may in fact be a tidal dwarf resulting from the interaction, \citealt{mun95,mun04}). Nevertheless, a combination of published 2D gas and stellar kinematics implies a consistent PA for the line of nodes and weaker constraints on the galaxy inclination. \citet{mun95} found from their analysis of VLA \hi{} data that the main gas disk was ``really quite undisturbed'', with a kinematic PA of $158 \pm 2\arcdeg$; they estimated an inclination of 56\arcdeg{} from the shape of the \hi{} isophotes (excluding the extended plumes to the north and south which are probably tidal features due to the interaction). The tilted-ring analysis of $^{12}$CO data by \citet{sch00} indicated a kinematic PA of $160 \pm 2\arcdeg$ and an inclination of $56\arcdeg$, almost identical to the \hi{} analysis. Finally, \citet{dum07} found a stellar-kinematic PA  of $160 \pm 5\arcdeg$ from their analysis of a SAURON stellar-velocity field; this is consistent with the stellar-kinematic PA of 157\arcdeg{} reported by \citet{bar06} based on higher-resolution, smaller-scale Gemini/GMOS IFS data. Taken all together, this suggests a line of nodes with PA $\approx 159\arcdeg$, and a somewhat less certain inclination of $\sim 56\arcdeg$. The position angle of the bar is very similar to that of the disk; we estimate it at $\sim 155 \pm 5\arcdeg$, based on the $K$-band image of \citet{mul97} and an archival Spitzer IRAC1 image \citep{gal10}.

On small scales, our data show a strong stellar velocity gradient, so that the kinematic PA is well determined at $155\arcdeg$.
Because it is comparable to the PA on large scales, we adopt it.
It is also consistent with that reported by \cite{dav06} for the circumnuclear ring at a radius of 1.7\arcsec\ (see Fig.~\ref{fig:nic3227}), which is completely within our field of view and is visible in our K-band data, as shown in the top panels of Fig.~\ref{fig:starcont}.
Assuming the ring is intrinsically round, its axis ratio suggests an inclination of 47\degr{} rather than the 56\degr{} characteristic of the large scales, and so we adopt that value.
Although this assumption may not be strictly correct (the average deprojected ellipticity of such rings is in the range 0.1--0.2, \citealt{com10}), an error in inclination has only a small impact on our modelled velocity fields, for the same reason that deriving an inclination from a velocity field is associated with a large uncertainty.
We note that the ring was also reported by \cite{bar06}, who found its location corresponds to low stellar dispersion as measured in the Ca\,II triplet -- an effect that can also be seen in the stellar dispersion measured from the CO\,2-0 bandhead (Fig~4 of \citetalias{hic13}).
We have overdrawn an ellipse with a semi-major axis of 1.7\arcsec, and our adopted PA and $i$ of 155\deg\ and 47\deg.
The slightly different PA derived from the stellar isophotes could be due to a secondary bar which \cite{bak00} and \cite{sch00} proposed may exist on scales of 10--15\arcsec.

\subsection{NGC\,5643}

Various estimates based on the shape of the outer isophotes of this
barred spiral galaxy suggest a global inclination of $\sim 23$--25\arcdeg{}
\citep{mor85,gar04,li11}. Measurements of the position angle
of the outer isophotes yield estimates of $\sim 121$--128\arcdeg{}
\citep{mor85,li11}, roughly consistent with the kinematic major axis
of 137\arcdeg{} determined by \citet{mor85} from their Fabry-Perot
H$\alpha$ velocity field. 
Thus on large scales we adopt a compromise position angle of
129\arcdeg{} (or equivalently -51\arcdeg), while noting this is somewhat uncertain, and an inclination
of 24\arcdeg. Ellipse fits to the $K$-band image of \citet{mul97} yield
a bar position angle consistent with their measurement of $\approx
85\arcdeg$.

In the circumnuclear region, we measure a difference in the stellar kinematic and isophotal PA and $i$, which may partially be explained by the bar.
However, the rather circular isophotes also indicate that the stellar distribution is thick rather than disky on small scales, as would be expected from the $\sim$70\,km\,s$^{-1}$ stellar velocity dispersion \citepalias{hic13}.
We therefore adopt the stellar kinematic values for PA and $i$, which are well constrained due to the large velocity gradient, bearing in mind that the small scale kinematics may be influenced by the bar.
We note that the apparent excess of stellar continuum seen in Fig.~\ref{fig:starcont}, especially to the east of the nucleus, may be due to reduced extinction in that region resulting from an outflow, similar to the situation reported for NGC\,4945 by \cite{mor96} and \cite{mar00}.
This is discussed more in Sec.~\ref{sec:h2vel}.

\subsection{NGC\,6300}

This barred spiral galaxy has been extensively studied by Buta and
collaborators \citep{but87,ryd96,but01}. The data presented in those
papers, including multiple long-slit H$\alpha$ rotation curves, \hi{}
mapping, and Fabry-Perot H$\alpha$ maps, all suggest a consistent
kinematic orientation with position angle $\sim 109\arcdeg$ and
inclination $\sim 50\arcdeg$; these values are also consistent with
those derived from the shape of the outer isophotes. 
Analysis of an archival Spitzer IRAC1 image (proposal ID 80072) suggests a position angle of $\approx 63\arcdeg$ for the bar.

The stellar kinematic and isophotal PA and $i$ in the central few hundred parsecs are consistent with each other and also with those on large scales.
As such, we simply adopt the values derived from our kinematic analysis of this region.

\subsection{NGC\,6814}

This galaxy is very close to face-on; the presence of strong spiral
structure makes determining the orientation from the outer isophotes
difficult, since the measured position angles and ellipticities will be
dominated by variations in local spiral arm strength. Based on a VLA \hi{} velocity field, \cite{lis95} found a kinematic position angle for the optical disk ($r \la 90\arcsec$) of 10--20\degr; at larger radii, the \hi{} disk appears to be warped. 
Unfortunately, they did not attempt to determine an
inclination from the velocity field, but noted that it was probably less than 20\degr.
Due to the uncertainty in PA and $i$ from the isophotes and also due to the warp in the H\,I disk, we take the large scale inclination to be $i=21$\degr{} as reported in the Third Reference Catalogue of Bright Galaxies (RC3) by \cite{vau91}, along with 15\degr{} as an estimate for the position angle.

On small scales, the modest stellar velocity gradient and fairly circular isophotes suggest that the inner region of this galaxies is also relatively close to face-on.
Given the uncertainty of the PA on both large and small scales, the values are not vastly dissimilar, and so we simply adopt that derived from the small scale kinematics.
The small scale stellar kinematic and isophotal estimates of $i$ also differ, although this is again due to the significant uncertainties arising from the face-on nature of the galaxy.
As such, we adopt $i$ from large scales, which is similar to our isophotal value.

\subsection{NGC\,7743}

This galaxy has a prominent bar with visible box/peanut structure \citep{erw13} and a luminous lens surrounding the bar with a
position angle of $\sim 80$--90\degr, consistent with the RC3 position
angle. However, SDSS $g$, $r$, and $i$ images, as well as the deeper
$R$-band image of \cite{erw03}, show faint outer isophotes
with a different orientation. This outer disk has a position angle of $\sim
110\degr$ and an ellipticity of $\sim 0.22$, suggesting an inclination
of $i \sim 39$--40\degr. This agrees quite well with the stellar
kinematic analysis of \cite{kat11}, who analyzed a SAURON IFS
stellar velocity field and three long-slit stellar velocity profiles and
derived a kinematic position angle of $\sim 120\degr$ and an inclination
of $i = 40\degr \pm 2\degr$. Our compromise estimate is a position angle
of $\sim 115\degr$ (or equivalently $\sim -65\degr$) and an inclination of 40\degr.

In the central few arcsec, we measure a PA and $i$ from both stellar kinematics and isophotes that are remarkably consistent with those on large scales -- supporting the conclusion in Sec.~\ref{sec:environ} that it is an undisturbed galaxy.
The stellar velocity gradient is significant enough that the kinematic PA is well defined, and so we adopt this value.
The inclination has some uncertainty, although this is not enough to have a significant impact on the modelled velocity field; and so again we simply adopt the value derived from the kinematics.

\subsection{IC\,5267}

The position angle of this galaxy is well-defined: we obtained a mean
position angle of $\sim 139$\degr{} from our ellipse fits to the
isophotes from a Spitzer IRAC1 image \citep{she10}, consistent with
the mean value of 137\arcdeg{} found by \citet{li11} from their $I$-band
image. The inclination is somewhat less well defined; the ellipticity of
the IRAC1 isophotes varies from a low of $\sim 0.15$ to a maximum of
$\sim 0.30$ (corresponding to inclinations of $\sim 33$--47\arcdeg). The
highest value is found in a region of partial spiral structure, and so
may be affected by local star formation. We note that both
\citet{lau10} and \citet{li11} suggest inclinations of
34--38\arcdeg, the former from a 2D bulge/disk decomposition of a
$K$-band image and the latter as a mean from $I$-band ellipse fits.
There is no evidence for a bar in this galaxy; the spiral structure is
tightly wrapped and flocculent, with two narrow, elliptical star-forming
rings or pseudo-rings (semi-major axes $a = 78$ and 147\arcsec) identified from
H$\alpha$ imaging \citep{gro10} and also apparent in UV images \citep{gil07}.
The axis ratio of these rings also suggests an inclination around $\sim40\degr$.

The circumnuclear stellar velocity field shows almost no velocity gradient \citepalias{hic13}, resulting in a highly uncertain kinematic position angle.
Similarly, the circumnuclear stellar isophotes are rather circular, indicating that the inner region of the galaxy is face-on and/or that the stars in the central few hundred parsecs are predominantly in a spheroidal (or at least geometrically thick) distribution with little ordered rotational component.
The high stellar dispersion of $\sim$160\,km\,s$^{-1}$ reported in \citetalias{hic13} supports the latter option.
For our purposes, the difference between the inclination measured on large and small scales is not critical because the H$_2$ velocity field clearly shows two distinct components, and we can model these directly without needing to subtract an underlying disk.
As such, we adopt the PA from the circumnuclear stellar isophotes, and $i$ from the large scales.

\subsection{NGC\,4030}

This unbarred Sbc spiral has a moderately well-defined orientation, with
literature estimates of the position angle and inclination ranging from
20--38\arcdeg{} and from 37--44\arcdeg, respectively, in the compilation
of \cite{gar04}; they adopt mean values of PA = 32\arcdeg{} and $i =
40\arcdeg$ from their Fourier analysis approach. The SAURON stellar and
gas kinematics presented in \cite{gan06} are very regular and
consistent with this overal orientation (e.g. inspection of their Fig.~5
suggests a position angle of $\sim 35\arcdeg$ for the stellar kinematic
line of nodes).

In the central few arcsec, the stellar kinematics and isophotes yield similar PA and $i$, which are also not very different from those on large scales.

\subsection{NGC\,3368}

\citet{now10} undertook a detailed analysis of the orientation of this
nearby, double-barred \citep{erw04} spiral. From the outer-disk
isophotes they derived a position angle of $\sim 172\arcdeg$,
consistent with kinematic estimates from \hi{} data
\citep{sch89,sak99} and Fabry-Perot H$\alpha$ data
\citep{sil03}. The shape of the isophotes suggested an inclination
of $\sim 53\arcdeg$, roughly consistent with the Fourier analysis of
\citet{bar04} and with an inversion of the $K$-band Tully-Fisher
relation using the observed Cepheid distance to the galaxy (see \citealt{now10} for details). The position angles of primary and secondary
bars are 115\arcdeg{} and 129\arcdeg, respectively \citep{erw04}.

In the central few hundred parsecs, we find a large discrepancy between the stellar isophotal PA and the stellar kinematic PA.
This is due to its small scale bar \citep{kna03,erw04,now10} for which such an offset is typical.
However, the kinematic PA is effectively the same as that at large scales, so we adopt it.
Because of this, and due to the uncertainty in $i$ as derived from kinematics, we also adopt $i$ from large scales.
We note also that Fig.~\ref{fig:starcont} appears to indicate a slight excess of stellar continuum to the west of the nucleus.
That there is no equivalent feature in the stellar kinematics could be because it contributes only 20--25\% of the total stellar continuum at that location.
It could result from either a stellar population associated with the gas condensation that we discuss in Section~\ref{sec:h2vel}, or a distortion of the central stellar population by that condensation.
In either case we can estimate an associated stellar mass from the luminosity of the excess, L$_K = 2\times10^6$\,L$_\odot$, using mass-to-light ratios for different stellar populations.
For continuous star formation over an age of 0.1--1\,Gyr we find that the ratio of mass currently in stars (i.e. excluding both non-luminous remnants and gas that has been returned to the ISM) is 1--2.5$\times10^6$\,M$_\odot$.
For an older 1--10\,Gyr population that is no longer forming stars, the mass still in stars would be $\sim5\times10^6$\,M$_\odot$.

\subsection{NGC\,628}

The VLA \hi\ mapping of \cite{kam92} showed regular
rotation in a nearly face-on main disk ($r \la 380\arcsec$), with a
position angle of 25\arcdeg{} and an assumed inclination of 6.5\degr,
along with evidence for warping of the \hi{} disk at larger radii. The
more recent analysis of {\em The H\,I Nearby Galaxy Survey} (THINGS) \hi{} data
by \cite{tam08} yielded $i = 7\degr$ and a PA of 20\degr{} for the
main disk, consistent with the position angle of $19\degr \pm 7\degr$ found
from analysis of Fabry-Perot \ha{} data by \cite{fat07}. Since isophotal
analysis of galaxies this close to face-on can be easily confused by the distorting
effects of spiral arms, the best estimate is that from the gas kinematic
observations: $i = 7\degr$ and PA = 20\degr.

The lack of stellar rotation in the central few arcsec of this galaxy are consistent with its nearly face-on appearance at large scales.
The low stellar dispersion ($\sim$30\,km\,s$^{-1}$, \citetalias{hic13}) suggests that this is due to its orientation rather than because the stars are in a spheroidal distribution.
As such, the PA and $i$ remain uncertain.
But because there is no measurable stellar velocity gradient, and also no 1-0\,S(1) line emission, we do not model this galaxy at all and so do not adopt any specific PA or $i$ for the circumnuclear region.

\subsection{NGC\,357}

Although there are no published 2D kinematics for this double-barred galaxy, the outer disk is symmetric and apparently undisturbed. The deep $I$-band image of \citet{agu05} yields a consistent position angle of 20\arcdeg{} (or equivalently $\sim -160\degr$) and an inclination of 37\arcdeg. The position angles for the primary and secondary bars are 120\arcdeg{} and 45\arcdeg{} respectively \citep{erw04}.

In the central few hundred parsecs, the stellar kinematic PA matches that at large scales closely.
However, the isophotal PA differs, suggesting a triaxial system.
As such, we have adopted $i$ from large scales.


\end{document}